\documentclass[12pt]{iopart}
\usepackage{iopams}  

\usepackage{setstack}
\usepackage{epsf}
\usepackage{rotating}
\usepackage{multirow}

\usepackage{graphicx,epsfig}% Include figure files
\usepackage{dcolumn}% Align table columns on decimal point
\usepackage{bm}% bold math

\def\cD{{\cal D}}
\def\cN{{\cal N}}
\def\cL{{\cal L}}
\def\cP{{\cal P}}
\def\cI{{\cal I}}
\def\cQ{{\cal Q}}
\def\cE{{\cal E}}

\def\cZ{{\cal Z}}
\def\cM{{\cal M}}
\def\CH{{\cal H}}

\newcommand{\req}[1]{Eq.~(\ref{#1})}
\newcommand{\fig}[1]{Fig.~\ref{#1}}
\newcommand{\tab}[1]{Table \ref{#1}}
\newcommand{\avg}[1]{\langle #1\rangle}

\newcommand{\cut}[1]{{\!\!}}
\def\cC{{\cal C}}
\def\cB{{\cal B}}
\def\cS{{\cal S}}
\def\mnk{{n^\ast_k}}

\def\snk{{\{n_k\}}}
\def\nk{{n_k}}
\def\sfe{{\mbox{\sf e}}}

\def\sP{{\cal P}}
\def\li{\lambda^{({\rm I})}}
\def\lii{\lambda^{({\rm II})}}
\def\xei{x^{({\rm I})}}
\def\xeii{x^{({\rm II})}}
\def\kinjmi{{k\in {\cal L}_j\backslash\{i\}}}

\def\jini{{j\in {\cal L}_i}}
\def\linjmi{{l\in {\cal L}_j\backslash\{i\}}}

\def\linjmik{{l\in {\cal L}_j\backslash\{i,k\}}}
\def\cst{{$c$ state} }
\def\bst{{$b$ state} }
\def\sst{{$s$ state} }
\def\Cst{{$C$ state} }

\def\Sst{{$S$ state} }

\def\psiji{\psi^{j\rightarrow i}}

\def\psikj{\psi^{k\rightarrow j}}
\def\ecji{\eta_c^{j\rightarrow i}}
\def\esji{\eta_s^{j\rightarrow i}}
\def\egji{\eta_g^{j\rightarrow i}}
\def\eskj{\eta_s^{k\rightarrow j}}
\def\egkj{\eta_g^{k\rightarrow j}}

\begin{document}

\title[Optimal Location of Sources in Transportation Networks]
{Optimal Location of Sources in Transportation Networks}

\author{C.~H.~Yeung and K.~Y.~Michael Wong}
\address{Department of Physics, The Hong Kong University
of Science and Technology, Hong Kong, China}

\begin{abstract}
We consider the problem of optimizing the locations of 
source nodes in transportation networks.
A reduction of the fraction of surplus nodes induces
a glassy transition.
In contrast to most constraint satisfaction problems involving
discrete variables,
our problem involves continuous variables which lead to 
cavity fields in the form of functions.
The one-step replica symmetry breaking (1RSB) solution involves
solving a stable distribution of functionals,
which is in general infeasible.
In this paper,
we obtain small closed sets of functional cavity fields and 
demonstrate how functional recursions are converted 
to simple recursions of probabilities,
which make the 1RSB solution feasible.
The physical results in the replica symmetric (RS)
and the 1RSB frameworks are thus derived and the 
stability of the RS and 1RSB solutions are examined.
\end{abstract}
%\pacs{02.50.-r, 05.20.-y, 89.20.-a}
% 02.50.-r: Probability theory, stochastic processes, and statistics
% 05.20.-y: Classical statistical mechanics
% 89.20.-a: Interdisciplinary applications of physics

\maketitle

%%%%%%%%%%%%%%%%%%%%%%%%%%%%%%%%%%%%%%%%%%%%%%%%%%%%%%%%%%%%%%%%%%%%%%%%%%

\section{Introduction}

Constraint satisfaction problems (CSPs),
which are highly relevant to many applications such as
electronic circuit design 
and frequency assignment in cellular mobile networks,
have been studied in the fields
of applied mathematics,
computer science and engineering.
Despite their usefulness,
many CSPs are NP-complete problems \cite{garey1979} 
associated with algorithmic hardness.
It is thus important to understand the physical origin of their hardness
and map out the easy and hard regimes for typical instances.
Physicists approach the problem 
by making analogy between CSPs and spin glasses
\cite{mezardBook,nishimoriBook}.
Objective functions are mapped to spin glass Hamiltonians,
enabling the analysis of CSPs using statistical physical techniques.
Successful examples are found in the $K$-satisfiability 
problem \cite{mezard2002},
graph coloring \cite{mulet2002},
and vertex cover \cite{weigt2000, weigt2001, zhou2005}.
They suggest a rich physical picture of CSPs corresponding to
the glassy phase in spin glasses.

In this paper we study a problem having a wide range of applications 
and sharing the characteristics of CSPs. 
Specifically,
we consider the optimal locations of source nodes 
in transportation networks.
Transportation networks consist of nodes with either
surplus or deficiency of resources,
and an important problem is to distribute them 
so as to achieve a networkwide satisfaction 
with a minimum transportation cost \cite{wong2006, wong2007, yeung2009a}.
This problem is important in load balancing in computer 
networks \cite{shenker1996}
and network flow of commodities \cite{rardin1998}.
Progress has been made in generalizing the message-passing technique 
of discrete variables to the passing of cavity energy functions 
in terms of the continuous 
current variables \cite{wong2006, wong2007, yeung2009a}.

Subsequent work considered networks in which shortages are allowed but
cost penalty is imposed \cite{yeung2009b}.
This models applications such as communications networks
where shortages are detrimental to the performance 
of the nodes.
Their effects were modeled by step-like shortage costs.
This high nonlinearity gives rise to unique behavior 
and a physical picture absent in the previous models.
When the shortage cost is comparable to the transportation cost,
the total cost may be optimized either by 
saving the transportation cost feeding a poor node while
sacrificing the satisfaction of the node,
or by saving the shortage cost 
while spending more on the transportation cost.
The picture in reminiscent of 
the learning of noisy examples in perceptrons,
where the field distribution of the examples consist of the bands,
corresponding to the learned and sacrificed examples respectively 
\cite{wong1990, wong1993, whyte1995, luo2001}.
As a result,
frustration arises from competition for resources among connected nodes.
Numerous metastable states emerge,
leading to typical glassy behavior.

The problem of optimal source location in this paper 
addresses an even more general
and practically relevant issue in network design and optimization.
Compared with \cite{yeung2009b} 
where some nodes remain unsatisfied in the optimized state, 
this paper moves one step forward 
and considers the situation in which the {\it location}
of the source nodes can also be optimized, 
and all nodes are satisfied. 
The source location problem has wide applications
in the design of optimized transportation networks.
For example, the optimal locations of access points 
in wireless networks can be determined
by balancing the signaling cost of the access points 
and the power and bandwidth limitations
of the channels linking the mobile subscribers
(which can be expressed as the transportation cost).

As demonstrated in \cite{wong2006, wong2007, yeung2009a},
the resource allocation problem involves passing messages 
of continuous variables.
When the cost function includes nonlinear terms,
the messages generally become extremely complicated.
However,
as will be described in this paper,
there are phases where the space of continuous messages 
can be replaced by small closed sets of cavity energy functions,
and their recursions can be converted to simple recursions of probabilities.
In the context of the source location problem,
this takes place when the consumer nodes form small clusters
surrounded by source nodes.
When the ratio of the installation cost of the source nodes and the
transportation cost changes,
regimes with different maximum cluster sizes are observed,
resembling the Devil's staircase observed in the circle map and other dynamical
systems \cite{devaney1989}.
We will show that the use of small closed sets of cavity energy functions
is particularly successful in the {\it singlet} regime
where the consumer nodes are isolated (clusters of size 1) and,
when the cost ratio is commensurate,
in the {\it doublet} regime where the consumer nodes can be paired or isolated.

\begin{figure}
\centerline{\epsfig{figure=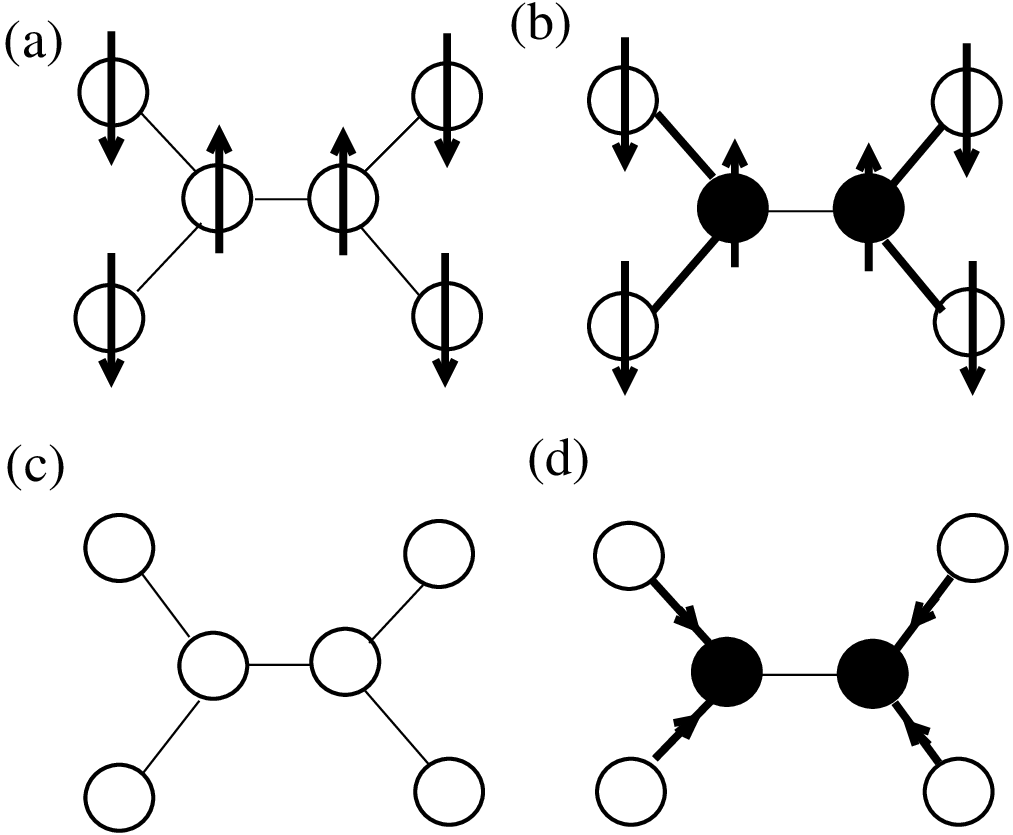, width=0.45\linewidth}}
\caption{
(a)-(b) RFIM of two spins and four fixed boundary spins.
Symbols: filled (unfilled) circles -up (down) spins,
arrows - direction of random fields of magnitude $h$,
thick segments - frustrated couplings of magnitude $J$.
When $h/J$ increases from below 2 to above,
pattern (a) changes to (b) to form a cluster
demarcated by a domain wall of frustrated couplings. 
(c)-(d) The source location problem defined by \req{eq_energy} 
with two nodes and four fixed boundary nodes,
and $\Lambda_i=-1$ for all $i$.
Symbols: filled (unfilled) circles - consumer (source) nodes,
arrows - current flows.
When $u^{-1}$ decreases from above $\sqrt{2}$ to below,
pattern (c) changes to (d) to form a cluster 
demarcated by a ``domain wall'' of inward current flow.}
\label{gr_RFIM}
\end{figure}

In fact,
clusters formed by similar energetic considerations have 
been found to play an important
role in disordered systems such as the random field Ising model (RFIM) 
\cite{imry1975},
as illustrated in \fig{gr_RFIM}.
Indeed, 
domain sizes in RFIM are determined by the interplay between the random
field energy and the domain wall energy,
giving rise to the so-called Griffiths singularities 
and cascades of phase transitions \cite{bruinsma1983, bruinsma1984}.
Analogously,
cluster sizes in the source location problem are determined 
by the balanced between the installation
cost of the source nodes and  the transportation costs.

%%%%%%%%%%%%%% Figure 1 %%%%%%%%%%%%%%
%%%%%%%%%%%%%%%%%%%%%%%%%%%%%%%%%%%%%%%%%%%%%%%%%%

Another class of problems that exhibits similar cluster behaviors 
is the lattice glass models~\cite{biroli2002, rivoire2004, krzakala2008}.
In the lattice glass models,
each site of a network can be occupied or empty,
but the number of nearest neighbors cannot exceed a maximum.
Analogously, 
the energetics of cluster formation in the source location problem implies that
two neighboring consumer nodes have effective repulsions,
since their contiguity prevents them from drawing resources
from more neighboring source nodes.
However,
the present model is richer in behavior,
since the cluster energy depends on the current distribution 
in non-trivial ways,
rather than merely counting the number of neighbors.
When the particle density
in a lattice glass increases,
configurations of clusters are formed,
causing the system to undergo dynamical
freezing transitions preceding the ideal glass transition.
Similar transitions will be reported in the source location problem.

The vertex cover problem \cite{weigt2000, zhou2005},
one of the NP-complete problems 
in computational complexity theory \cite{garey1979}
that attracted recent attention from physicists,
also shares similar cluster behaviors.
Drawing the analogy of assigning guards in a museum \cite{weigt2000},
each site of a network can be covered or uncovered,
but none of the links can have both vertices uncovered.
Hence the uncovered nodes form a configuration 
with an effective repulsion among them.
Indeed,
an equivalent configuration of consumer nodes can be found 
in the singlet regime of the source 
location problem,
since for neighboring consumer nodes,
each has one less link to draw resources from the 
neighboring source nodes.
When the average connectivity of the network increases,
we will show that it undergoes a glassy transition analogous to that
in the vertex cover problem.

The origin of these interesting phenomena 
can be traced to the presence of {\it frustrations},
which refer to the conflicts between competing interaction energies 
in the system \cite{toulouse1977}.
This connects our problem with a broad class of network CSPs in which 
frustrations are inherent.
When the system consists of numerous states,
the replica symmetry-breaking (RSB) solution is applicable,
but the distribution of the cavity energy functions is in general 
infeasible to solve.
Nevertheless, 
with the small closed set of cavity energy functions
introduced in this paper,
the one-step RSB (1RSB) solution
becomes computationally feasible.
The physical results based on the replica symmetric (RS) ansatz 
and the 1RSB configurational entropy are thus derived.

The paper is organized as follows.
We introduce our problem in Section \ref{sec_model},
presenting simulation results of the optimal behaviors.
In Section \ref{sec_RS},
we discuss the general RS formalism
and derive the piecewise quadratic ansatz of 
the cavity energy functions.
In Section \ref{sec_RSresult},
we obtain a small closed set of cavity energy functions in the singlet regime
and demonstrate how functional recursions are converted to recursions
of probabilities.
The average energy,
the fraction of soft nodes and the RS-RSB phase diagram
are derived.
In Section \ref{sec_RSB},
we apply the small closed set of cavity fields to 
the 1RSB formalism and  
obtain results of the configurational entropy.
The conclusion is given in Section \ref{sec_conclusion}.
In \ref{sec_closeSetDoubleSat},
we describe how the small closed set of cavity energy functions can be applied
to the doublet regime in the commensurate case,
and the corresponding RS and 1RSB results are derived.

%%%%%%%%%%%%%%%%%%%%%%%%%%%%%%%%%%%%%%%%%
\section{The Model}

\label{sec_model}
\subsection{Model Formulation}

We consider a network of $N$ nodes, 
labelled $i=1\dots N$.
Each node $i$ is connected randomly to a set $\cL_i$ of 
$K$ neighbors.
Each node $i$ has {\it capacity} $\Lambda_i$;
nodes with positive and negative values of $\Lambda_i$ correspond to 
{\it surplus} and {\it deficient} nodes respectively.
The capacities $\Lambda_i$ are 
randomly drawn from a distribution of $\rho(\Lambda_i)$.
With network applications in mind,
we consider a bimodal distribution in which 
$\Lambda_i=A(\gg 1)$ with probability $\phi_s\ge 0$ 
and $\Lambda_i=-1$ with probability $\phi_d\equiv 1-\phi_s$. 
Naturally, the surplus nodes serve as {\it source nodes} 
providing resources to the {\it consumer nodes}. 
However, to minimize cost functions that include transportation costs, 
it is often desirable to convert some deficient nodes 
into source nodes as well. 
Hence in general, 
the task is to optimally locate these {\it extra} source nodes 
so as to minimize the total cost function.
The relevant glossary used in this paper 
is summarized in \tab{tab_glossary}.

%%%%%%%%%%%%%%%%%%% Table 1 %%%%%%%%%%%%%%%%%%%%%%%%
\begin{table}\centering
\begin{tabular}{|c|c|c|c|}
\hline
$\Lambda_i$ & before optimization & $\xi_i$       & after optimization \\
\hline
            &                     &               &                    \\
$\ge 0$     & surplus node        & $\ge 0$       & source node        \\
            &                     &               & (resource	       \\
\cline{1-3}
            &                     &               & providing)         \\
            &                     & $< 0^\dagger$ &                    \\
            &                     &               &                    \\
\cline{3-4}
$<0$        & deficient node      &               &                    \\
            &                     & $\ge 0$       & consumer node      \\
            &                     &               & (consuming)        \\
            &                     &               &                    \\
\hline
\end{tabular}
\leftline{}
\leftline{$^\dagger$The deficient nodes with $\xi_i<0$ after optimization 
are converted to source nodes.}
\caption{A summary of glossary used in this paper.}
\label{tab_glossary}
\end{table}
%%%%%%%%%%%%%%%%%%%%%%%%%%%%%%%%%%%%%%

We first consider the minimization of the cost function in \cite{yeung2009b}, 
whose optimization variables are the currents $y_{ij}\equiv -y_{ji}$ 
of real values from node $j$ to node $i$,
\begin{equation}
\label{eq_energy}
	 E=\frac{u^2}{2}\sum_i \Theta(-\xi_i)
	 +\sum_{(ij)}\frac{y_{ij}^2}{2}.
\end{equation}
$\xi_i\equiv \Lambda_i + \sum_{\jini}y_{ij}$ 
is the final resource of node $i$, 
and $\Theta(x)=1$ when $x>0$, 
and $0$ otherwise. 
The link connecting nodes $i$ and $j$ is denoted as $(ij)$. 
The first term corresponds to the unsatisfaction cost imposed on nodes 
with negative final resource. 
The second term is the transportation cost.
This cost function models load balancing situations 
in which insufficient provision of resources to a deficient node 
produces detrimental effects on it 
(irrespective of the magnitude of insufficiency).

The key to applying the cost function in \req{eq_energy} 
to optimize the location of source nodes 
is to note that once the final resource of a deficient node is negative, 
the unsatisfaction cost remains the same 
even when its resources are maximally drawn by other nodes of the network. 
Hence the deficient node effectively becomes a resource provider. 
Ref. \cite{yeung2009b} contains many such examples. 
If we consider the coefficient $u^2/2$ to be the installation cost 
of a source node, 
then we can solve the optimal source location problem 
by first minimizing the cost function in \req{eq_energy}, 
then identifying the deficient nodes whose final resources are negative, 
and converting them to source nodes.

Formally, in the optimal source location problem, 
we introduce the state variables $s_i=\pm 1$ for deficient nodes 
when node $i$ is a consumer or a source node respectively. 
The cost function is then
\begin{equation}
\label{eq_optsource}
	 E=\frac{u^2}{4}\sum_{i\in\cN_D}(1-s_i)
	 +\sum_{(ij)}\frac{y_{ij}^2}{2},
\end{equation}
subject to $\xi_i\ge 0$ for $s_i=1$. 
No constraints are imposed on nodes with $s_i=-1$, 
since an arbitrary amount of resource can be provided 
when they are converted to source nodes.
$\cN_D$ is the set of deficient nodes.

To check the equivalence between the cost functions 
in Eqs.~(\ref{eq_energy}) and (\ref{eq_optsource}), 
we can easily see that when $s_i=1$, 
the installation (or unsatisfaction) cost 
vanishes in both cost functions. 
When $s_i=-1$, we only have to consider the case $\xi_i<0$, 
and the installation (or unsatisfaction) cost 
is $u^2/2$ in both cost functions. 
This is because when $\xi_i\ge 0$, 
we can set $s_i=1$ to minimize the total cost.

Note that the cost function of the optimal source location problem 
is identical to that in~\cite{yeung2009b},
but the interpretation is far more relevant to network applications.
All previous results on networks with nodes of negative capacity 
can be directly mapped to networks 
whose unsatisfied nodes are replaced by source nodes.
For example,
the single-sat regime studied in \cite{yeung2009b} corresponds to the 
case that each consumer node is surrounded by source nodes,
since the installation cost is low compared with the 
transportation cost. 
When the installation cost is gradually raised,
resource provision is achieved with less source nodes,
but optimization requires the consumer nodes to be 
located in clusters surrounded by source nodes,
forming the clusters observed in \cite{yeung2009b}.

To formulate an algorithm,
we introduce the constraints $s_i \xi_i\ge 0$ for each deficient node.
These contraints are not applied to surplus nodes as they are always
satisfied.
Introducing Lagrange multipliers $\mu_i$ for the resource constraint, 
we minimize the Lagrangian
\begin{equation}
\label{eq_lagr}
	L=
	\frac{u^2}{4}\sum_{i\in \cN_D}(1-s_i)
	+\sum_{i\in \cN_D}\mu_i s_i \xi_i+\sum_{(ij)}\frac{y_{ij}^2}{2}
\end{equation}
with the K\"uhn-Tucker conditions $\mu_i s_i \xi_i=0$ 
and $\mu_i\le 0$.
Optimizing $L$ with respect to $y_{ij}$,
one obtains
$y_{ij}=\mu_j s_j-\mu_i s_i$ and
$\mu_i =\min[0, (\Lambda_i+\sum_\jini\mu_j s_j)/s_i c]$.
Given a particular set of $\{s_i\}$, 
we iterate these equations to find the corresponding set of $\{\mu_i\}$.
The set of optimal $\{s_i\}$ is found by an approach similar to the
the GSAT algorithm \cite{selman1996},
by comparing the Lagrangian in \req{eq_lagr} for each choice of $\{s_i\}$.
In each step of this algorithm,
a cluster of $N_{\rm flip}$ nodes is randomly selected.
The network energies involving the different configurations of this cluster 
are compared,
and the cluster configuration is updated to the one 
that yields the lowest network energy \cite{yeung2009b}.

\subsection{Major Simulation Results}
\label{sec_simulation}

%\subsubsection{The General Physical Picture and the Devil's Staircase}

%%%%%%%%% Figure 2 %%%%%%%%%%%%%%%%%
\begin{figure}
\centerline{\epsfig{figure=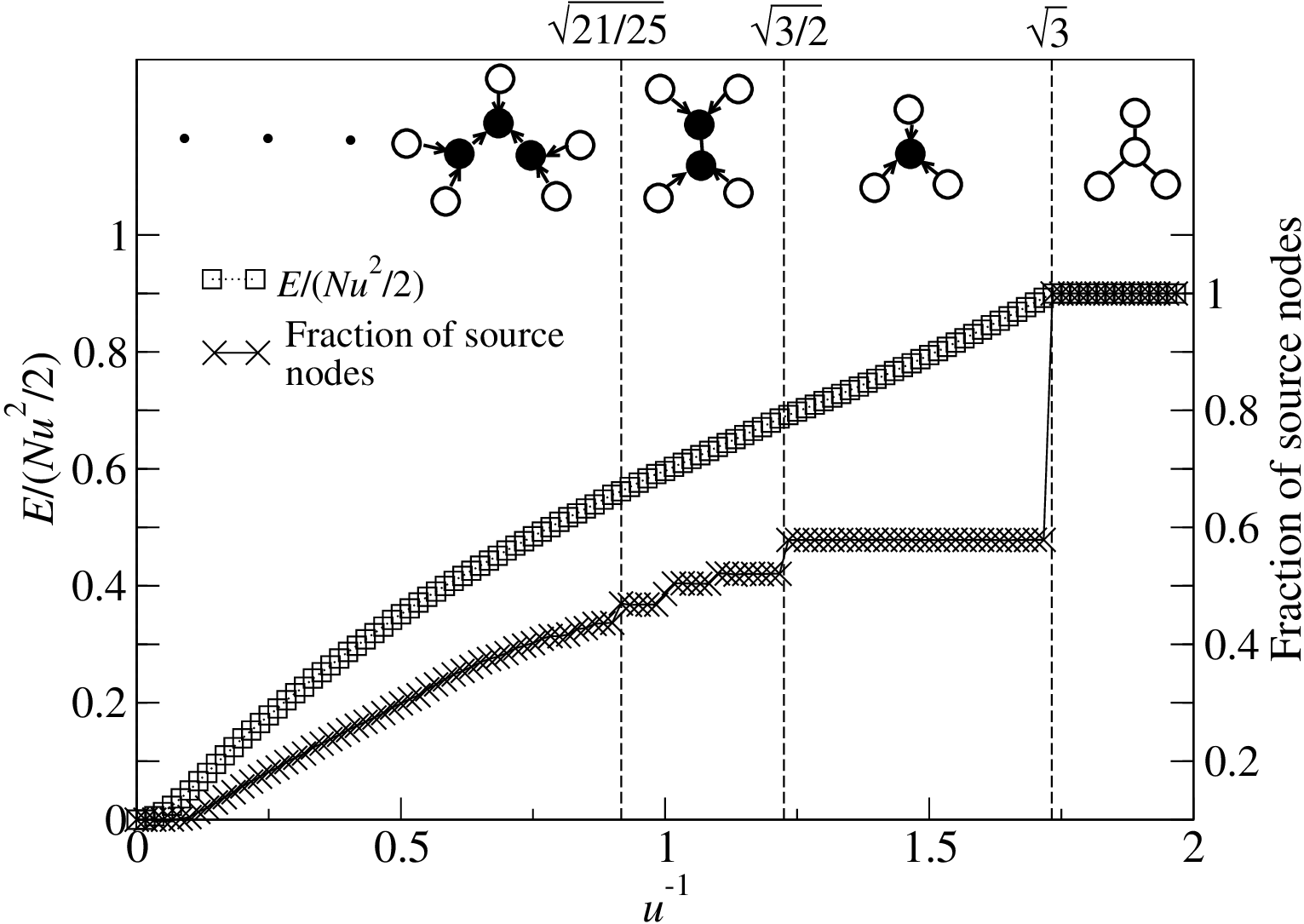, width=0.65\linewidth}}
\caption{Simulation results of average energy per node and 
the fraction of network nodes acting as source nodes. 
Parameters:  $K=3$, $\phi_d=0.9$, $N=100$, $N_{\rm flip}=4$, 100 samples
and 1000 flips.
New clusters formed on increasing $u^{-1}$ are sketched at the top,
with filled and unfilled circles 
representing consumer and source nodes respectively.
}
\label{gr_energy}
\end{figure}
%%%%%%%%%%%%%%%%%%%%%%%%%%%%%%%%%%%%

As shown in \fig{gr_energy} for $K=3$, two phases can be identified:
(1) {\it all-source} phase for $u^{-1}\ge\sqrt{3}$,
in which all nodes are assigned to be source nodes due to the very high 
transportation cost;
(2) {\it partial-source} phase for $0<u^{-1}<\sqrt{3}$,
in which only some nodes are assigned to be source nodes.
(In \cite{yeung2009b} we also identified a phase transition at $u^{-1}=0$ 
to an all-consumer phase.)

The fraction of source nodes is a discontinuous function of $u^{-1}$,
showing abrupt jumps at threshold values of $u^{-1}$.
The step size of the curve decreases as $u^{-1}$ increases,
and gradually becomes unresolvable by the numerical experiments.
This resembles the {\it Devil's staircase} observed in the circle map
and other dynamical systems \cite{devaney1989}.
These threshold values of $u^{-1}$ mark the positions at which
certain configurations of the source and consumer nodes become 
energetically stable.
Similar features are observed in RFIM due to the formation of 
ferromagnetic clusters resultant from the competition between
the strengths of couplings and random fields \cite{bruinsma1983, bruinsma1984}.
Except for a shift of the average energy per node, 
these features are qualitatively similar 
to the simulation results of \cite{yeung2009b} 
which correspond to the case $\phi_d=1$, 
if the latter is reinterpreted 
from the perspective of the source location problem.

%%%%%%%%%%%%% Figure 3 %%%%%%%%%%%%%%%%%%%%
\begin{figure*}
%\centerline{\epsfig{figure=singleSat2_current.eps, width=0.38\linewidth}}
%\centerline{\epsfig{figure=singleSat2_current.eps, width=0.38\linewidth}}
%\centerline{\epsfig{figure=doubleSat3_current.eps, width=0.48\linewidth}}
%\centerline{\epsfig{figure=tripleSat2_current.eps, width=0.48\linewidth}}
\includegraphics[width=0.1\linewidth]{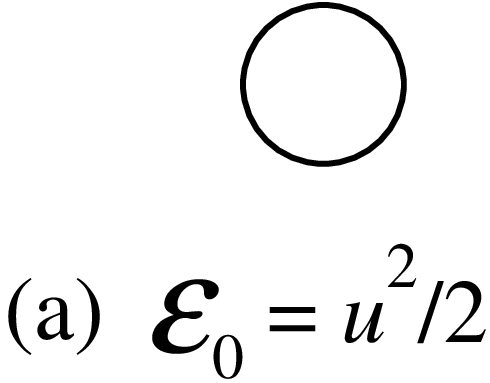}~
\hspace{1cm}
\includegraphics[width=0.17\linewidth]{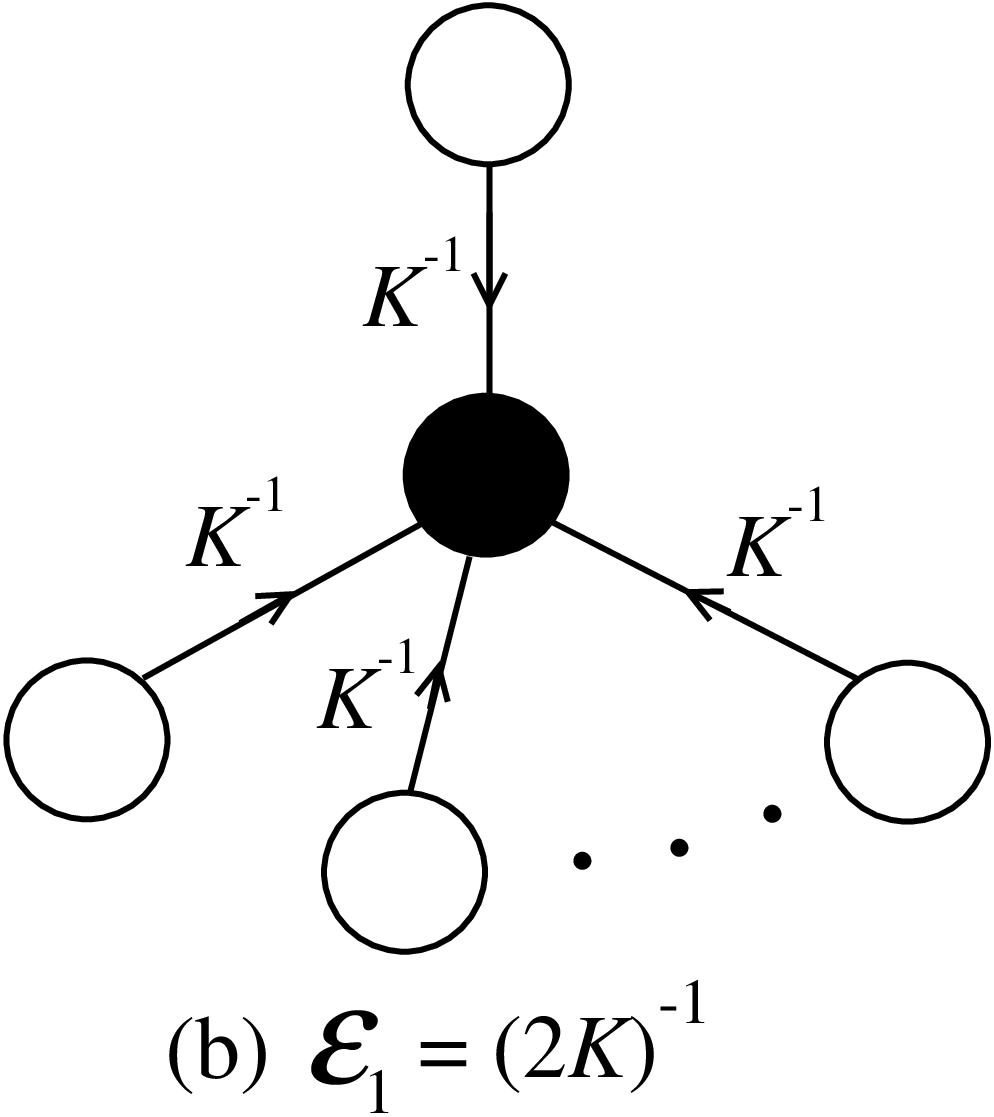}~
\includegraphics[width=0.27\linewidth]{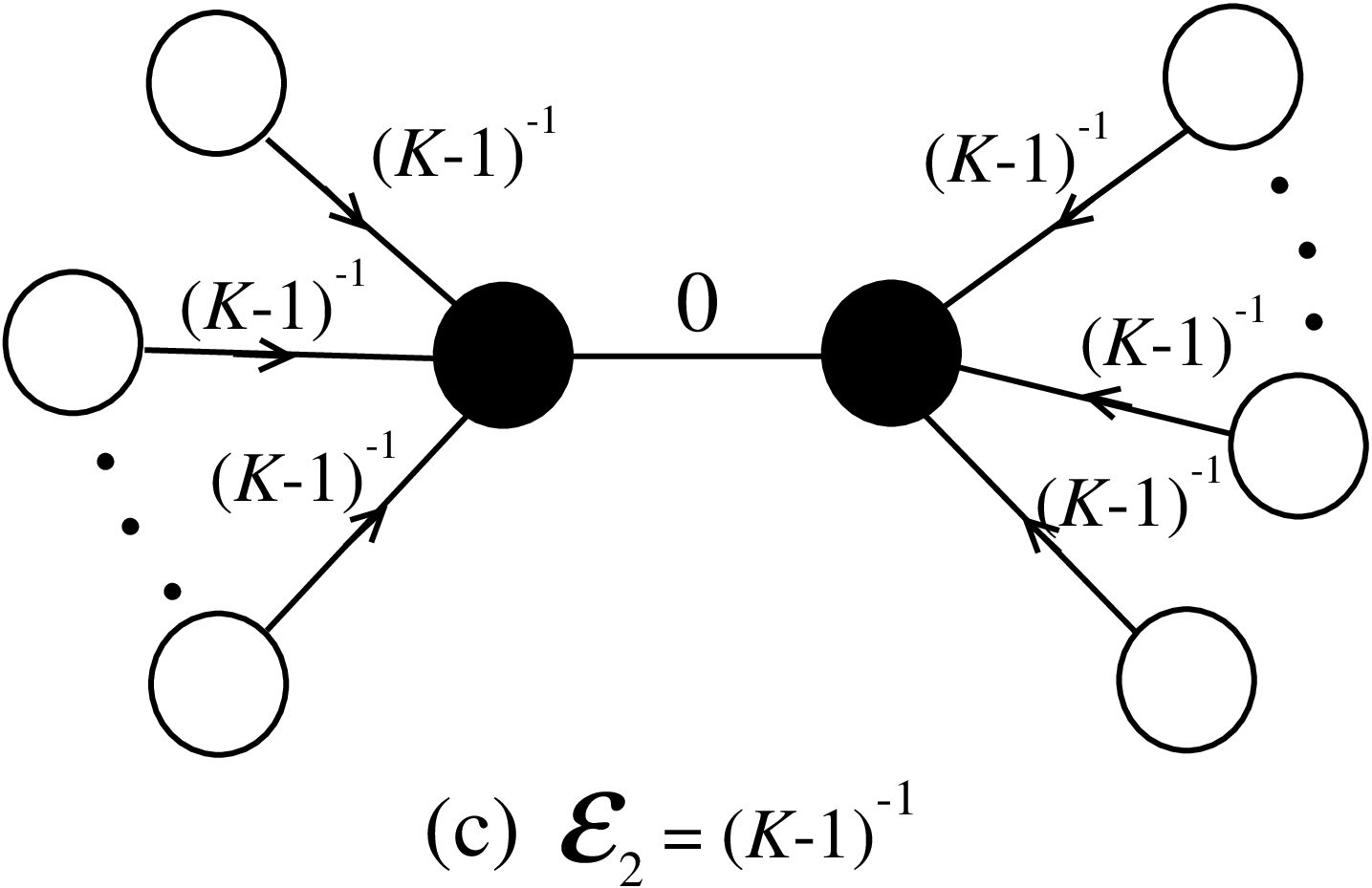}~
\includegraphics[width=0.27\linewidth]{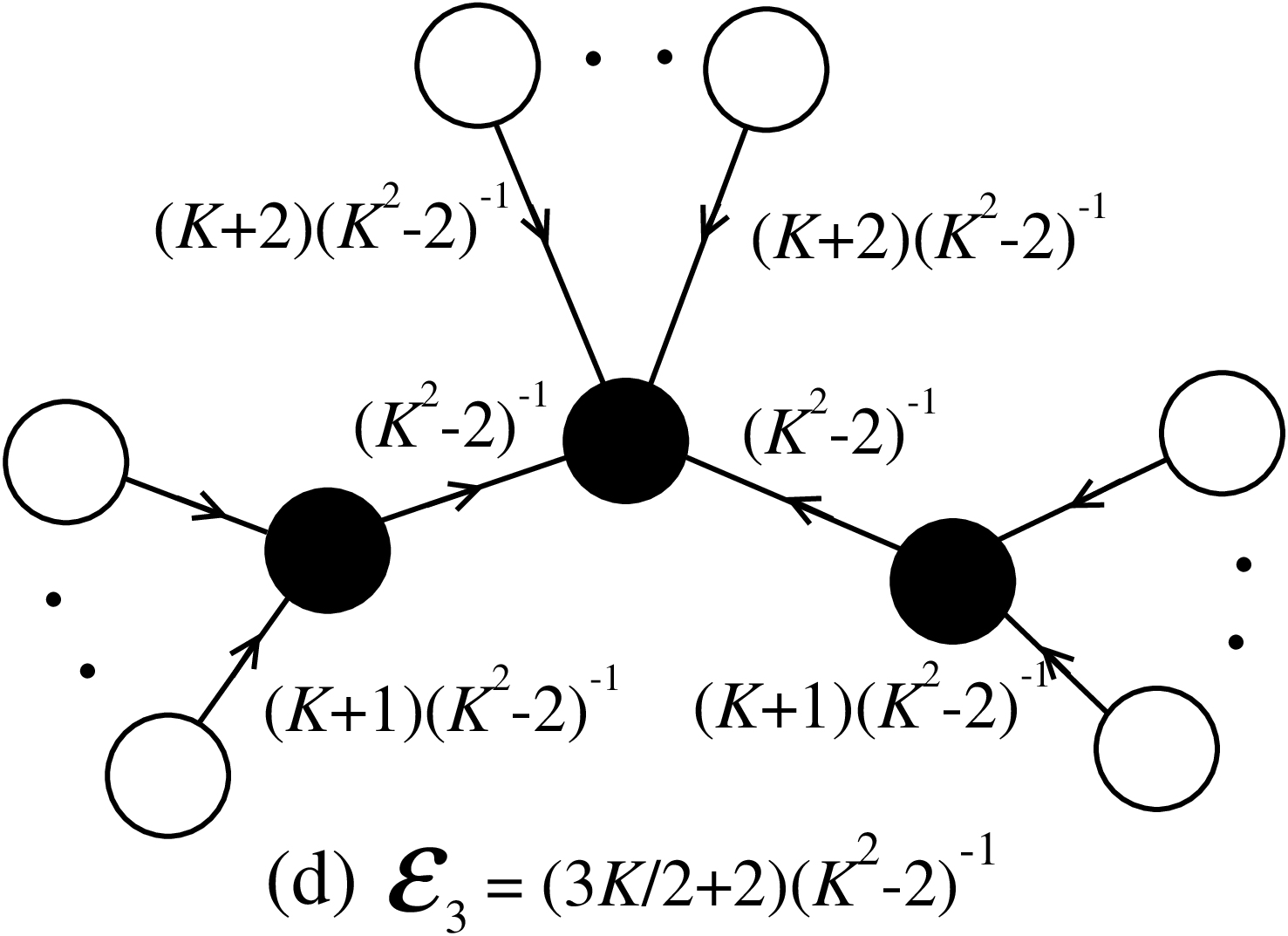}~
\caption{
Cluster of (a) source node,
(b) singly consuming node,
(c) doubly consuming nodes,
and (d) triply consuming nodes.
}
\label{gr_current}
\end{figure*}
%%%%%%%%%%%%%%%%%%%%%%%%%%%%%%%%%%%%%%%%%%%

Measuring the average maximum cluster size of the consumer nodes 
in the samples,
we observe abrupt jumps of the cluster size at the same threshold values.
This indicates that new types of clusters are formed at each jump,
as sketched in the top of \fig{gr_energy}.
The observed threshold values can be calculated by considering the energies of 
consumer clusters surrounded by source nodes as shown in \fig{gr_current},
obtained by the minimization of \req{eq_energy}.
By comparing the energy of different configurations,
we have 
\begin{eqnarray}
	\cE_0 \ge \cE_1 &\quad& \mbox{(singlet)},
	\nonumber\\
	\label{eq_simSingly}
	\cE_0+\cE_1 \ge \cE_2 &\quad& \mbox{(doublet)},
	\\
	\label{eq_simTriply}
	\cE_0+2\cE_1 \ge \cE_3 &\quad& \mbox{(triplet)},
	\nonumber
\end{eqnarray}
resulting in the threshold values in \fig{gr_energy}. 
These results agree with those obtained through the cavity approach 
in Section \ref{sec_RS}.
We call the regime $\sqrt{3/2}<u^{-1}<\sqrt{3}$ with isolated consumer nodes 
the {\it singlet} regime, 
and $\sqrt{21/25}<u^{-1}<\sqrt{3/2}$ the {\it doublet} regime.
The isolated nodes are referred to as {\it singly consuming},
while the paired consumer nodes {\it doubly consuming}.

%%%%%%%%%%%%%%%%%%%%%%%%%%%%%%%%%%%%%%%%%%%
\section{The Replica Symmetric Ansatz}
\label{sec_RS}

\subsection{The RS Recursion at Zero-temperature Limit}
\label{sec_RSgeneral}

We apply the cavity method \cite{mezardBook, nishimoriBook}
assuming that the network has a locally tree-like structure.
We denote as $E_j(y_j)$ the energy of the tree terminated at
node $j$ in the absence of its ancestor node $i$,
when a current $y_j$ is drawn from $j$ to its ancestor.
Relabeling the descendents of $j$ as $k=1,\dots,K-1$,
$E_j(y_j)$ is expressed as
\begin{eqnarray}
\label{eq_recurE}
	&&E_j(y_j) = \CH(E_{k=1},\dots,E_{K-1};\Lambda_j,y_j).
\end{eqnarray}
The functional $\CH$ is given by
\begin{eqnarray}
\label{eq_HE}
	&&\CH(E_{1},\dots,E_{K-1};\Lambda_j,y_j)
	\nonumber\\
	&&\quad\equiv
	\min_{\{y_k\}}\biggl[ \sum_{\kinjmi}E_k(y_k)
	+\frac{u^2}{2}\Theta\biggl(-\Lambda_j-\sum_{\kinjmi}y_k+y_j\biggr)
	 +\frac{y_j^2}{2}\biggr].
\end{eqnarray}
In the absence of node $i$,
there is no supply or demand of resources through the cavity
and the last term $y_{j}^2/2$ should be absent.
However,
the presence of the extra term results in a clear interpretation of $E_j$,
as we will see in the following sections.
Care has to be taken when dealing with the change of the cavity energy,
where $y_{j}$ is taken to be zero 
to eliminate the effect of the extra transportation cost on the dangling bond.

We note that  $E_j$ is an extensive quantity that depends on
size of the tree.
To formulate a recursion of an intensive energy,
we write $E_j(y_{j})$ as a sum of two terms,
\begin{eqnarray}
\label{EV}
	E_j(y_j) = E^V_j(y_j) + E_j(0).
\end{eqnarray}
We call $E^V_j(y_j)$ the {\it cavity energy functions}
which correspond to the cavity fields
in the language of the cavity approach,
and represent the energy variation from $E_j(0)$,
as $y_j$ varies.
In this case, 
$E^V_j(0) = 0$.
$E_j(0)$ corresponds to the energy of 
the tree when no current is drawn from the vertex.
We further define the energy change $\Delta E_j$ 
due to the addition of a vertex,
\begin{eqnarray}
\label{eq_deltaE}
	\Delta E_j &&= E_j(0) - \sum_{\kinjmi} E_k(0)
\end{eqnarray}
which simplifies \req{eq_recurE} to
\begin{eqnarray}
\label{eq_recurEV}
	E^V_j(y_j) = \CH(E^V_{k=1},\dots,E^V_{K-1};\Lambda_j,y_j)
	- \Delta E_j
\end{eqnarray}
where
\begin{eqnarray}
\label{eq_deltaEsim}
	\Delta E_j =
	\CH(E^V_{1},\dots,E^V_{K-1};\Lambda_j,0).
\end{eqnarray}
We have thus separated the energy contribution due to the addition
of a new vertex from the energy variation due to the changes in the 
current drawn from the tree.

The distribution $\sP[E^V]$ of $E^V$ over the vertices of the tree 
is given by the solution of
\begin{eqnarray}
\label{eq_RSPE}
	\sP[E^V_j] =&& \int d\Lambda_j \prod_{k=1}^{K-1}
	\int d E^V_k \sP[E^V_k] 
%	\nonumber\\
	\delta[E^V_j-\CH(E^V_{1},\dots,E^V_{K-1};\Lambda_j,y_j)
	\nonumber\\
	&&+\Delta E(E^V_{1},\dots,E^V_{K-1};\Lambda_j)].
\end{eqnarray}

To elucidate the physical behavior of the system,
we consider a node fed by $K$ trees forming a Bethe lattice.
For instance,
we consider the average energy per node.
The change in energy due to the additional node $i$ is given by 
\begin{eqnarray}
\label{eq_cdeltaE}
	\Delta \cE_{\rm node} = \CH(E^V_{1},\dots, E^V_{K}; \Lambda_j, 0).
\end{eqnarray}
Similarly,
we can consider a link bridging two trees forming a Bethe lattice.
The energy change due to the addition of a link between nodes $L$ and $R$ 
is given by
\begin{eqnarray}
\label{eq_linkDeltaE}
	\Delta{\cal E}_{\rm link}(E^V_L, E^V_R)
	= \min_{y}\bigg[E^V_L(y) + E^V_R(-y) - \frac{y^2}{2}\bigg].
\end{eqnarray}
Denoting $\avg{\dots}$ as the average over the capacities,
the average energy per node is given by 
\begin{eqnarray}
\label{eq_deltaCalE}
	\avg{\Delta{\cal E}} = \avg{\Delta{\cal E}_{\rm node}} 
	- \frac{K}{2}\avg{\Delta{\cal E}_{\rm link}}.
\end{eqnarray}

\subsection{The Piecewise Quadratic Solution}

Due to the quadratic form of the transportation cost 
assumed in \req{eq_energy},
we propose that the cavity energy functions are continuous 
and piecewise quadratic, namely,
\begin{eqnarray}
\label{eq_quadAnsatz}
	E^V_k(y_k) = \min_{n_k}[f_{n_k}^k(y_k)].
\end{eqnarray}
where $n_k=0,1,2,\dots$. 
Indeed, 
the recursive nature of the quadratic
cavity energy functions have been fully employed in deriving the
message-passing approach in \cite{wong2006,wong2007}.
We note in passing
that a similar recursive structure was used in the Gaussian Belief
Propagation algorithm \cite{weiss2001} and applied to processing
continuous signals such as those in CDMA multiuser detection
\cite{bickson2008}.
As a step forward, 
the ansatz in \req{eq_quadAnsatz} further captures 
the multi-valley features in $E^V_k(y_k)$,
which is crucial in formulating the cavity messages for the present model.

\begin{figure}
\centerline{\epsfig{figure=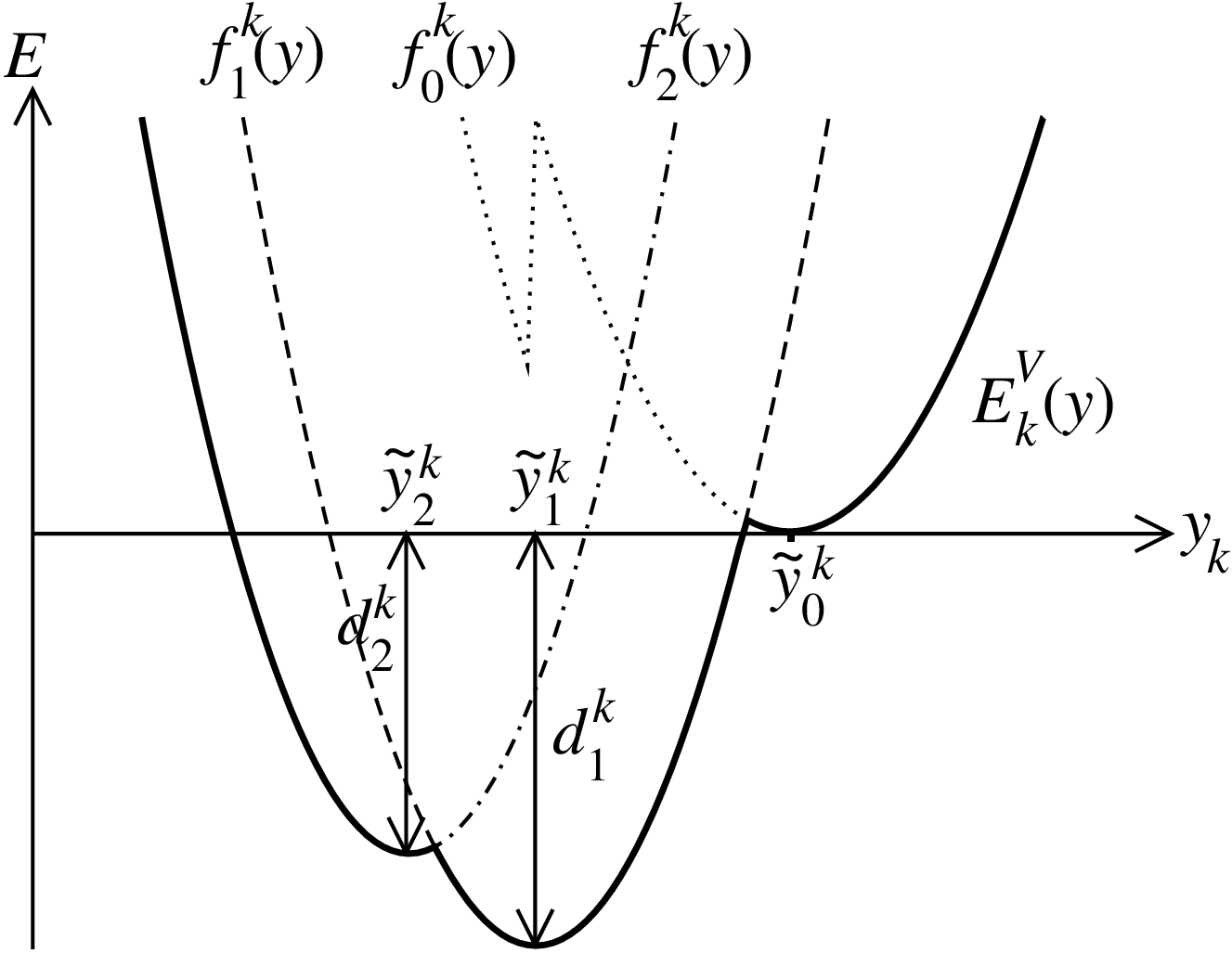, width=0.4\linewidth}}
\caption{
An example of $E_k^V(y)$ composed of three
quadratic functions $f^k_{n_k}(y)$
labelled by $n_k=0,1,2$.
Each composite function is characterized by its minimum position 
$\tilde y^k_{n_k}$,
minimum value $d^k_{n_k}$, and curvature $a^k_{n_k}$.
}
\label{gr_valley}
\end{figure}

We call $f^k_{n_k}$ the $n_k$-th {\it composite function} of $E_k(y_{k})$.
For $n_k>0$,
$f_{n_k}^k(y_k)$ is a quadratic function of the form
\begin{eqnarray}
	f_{n_k}^k(y_k) = a_{n_k}^k(y_k-\tilde y_{n_k}^k)^2+d_{n_k}^k.
\end{eqnarray}
whereas for $n_k=0$,
to takes the form
\begin{eqnarray}
\label{eq_f0}
	f_{0}^k(y_k) = \frac{y_k^2}{2}+c_k+\frac{u^2}{2}\Theta(y_k-\alpha_k).
\end{eqnarray}
The form of $f_0$ is relevant when node $k$ is a source node.
Though $f_0$ is discontinuous,
we will show that the resulting $E^V(y_k)$ is continuous
since the discontinuity at $y_k=\alpha_k$ is masked 
by other quadratic functions.
An example of a cavity energy function $E^V_k(y_k)$ 
composed of three composite functions
is shown in Fig.~\ref{gr_valley}.

%%%%%%%%% Figure 4 %%%%%%%%%%%%%%%%%
%%%%%%%%%%%%%%%%%%%%%%%%%%%%%%%%%%%%

We denote the $f^k_{n_k}$ with the minimum $d^k_{n_k}$ among all 
composite functions in $E^V_k$ as $f^k_{\mnk}$, i.e.,
\begin{eqnarray}
\label{eq_mnk}
	\mnk = \arg \bigg(\min_{n_k} d^k_{n_k}\bigg).
\end{eqnarray}
$\mnk$ is particularly relevant when we evaluate the energy of the system
in the case that the ancestor node of $k$ is a source node.
In this case,
the resource of the ancestor can be freely drawn by node $k$ 
without any consequences to other parts of the network,
and the optimal current $y_k$ takes the value $\tilde y^k_\mnk$.

As will be shown in the next subsection,
only a few composite functions are relevant
in the singlet and the doublet regimes. 
Each composite function has its fixed values of $\tilde y^k_\nk$ and $a^k_\nk$
independent of $k$,
but the constant terms is $k$-dependent.
Thus, the functional form of $E^V_k(y_k)$ 
is effectively parametrized by the constant terms $d^k_{n_k}$ 
of the composite functions, as given by 
\begin{eqnarray}
\label{eq_Ekd}
	E^V_k(y_k) \doteq (d_{0}^k, d_{1}^k, d_{ 2}^k, \dots).
\end{eqnarray}
The relevant composite functions in the singlet and
doublet regimes are shown in \tab{tab_minima}.

%%%%%%%%%%%%%%%%%%% Table 2 %%%%%%%%%%%%%%%%%%%%%%%%
\begin{table}\centering
\begin{tabular}{|l||c|c|}
\hline
$n$ & $\tilde y_n$ & $a_n$ \\
 \hline \hline
 0 & 0 & $\displaystyle\frac{1}{2}$ \\
 \hline 
 1 & $-K^{-1}$ & $\displaystyle\frac{K}{2(K-1)}$ \\
 \hline
 2 & $-(K-1)^{-1}$ & $\displaystyle\frac{K^2-1}{2(K^2-K-1)}$ \\
 \hline
\end{tabular}
\caption{The table of cavity composite functions 
$f_n(y) \doteq (\tilde y_n, a_n)$ from $n=0$ to $2$.
}
\label{tab_minima}
\end{table}
%%%%%%%%%%%%%%%%%%%%%%%%%%%%%%%%%%%%%%%%%%%%%%%%%%%%

\subsection{The Recursion of $E^V$}

Using the piecewise quadratic ansatz in \req{eq_quadAnsatz},
the recursion of $E^V$ in \req{eq_recurEV} becomes
\begin{eqnarray}
\label{eq_recurEVsim}
	E^V_j(y_j) = \min_{\{n_k\}}\CH(f^{k=1}_{n_1},\dots,f^{K-1}_{n_{K-1}};
	\Lambda_j,y_j)
%	\nonumber\\
	-\min_{\{n_k\}}\Delta E_j(f^{k=1}_{n_1},\dots,f^{K-1}_{n_{K-1}};
	\Lambda_j).
	\nonumber\\
\end{eqnarray}

We first consider the recursions for a deficient node $j$. 
Suppose the node is assigned to be a source node, i.e.,
$\Lambda_j+\sum_{\kinjmi}y_k-y_j<0$.
In this case,
\begin{eqnarray}
\label{eq_minHEunsat}
	&&\CH(f^{1}_{n_1},\dots,f^{K-1}_{n_{K-1}};\Lambda_j,y_j)
	|_{\Lambda_j+\sum_{\kinjmi}y_k-y_j< 0}
	\nonumber\\
	&& =\frac{y_j^2}{2} 
	+ \frac{u^2}{2}\Theta\bigg(y_j-\Lambda_j
	-\sum_{\kinjmi}\tilde y^k_{n_k}\bigg)
	+\sum_{\kinjmi}d^k_{n_k},
%	\nonumber\\
\end{eqnarray}
which is exactly the form of $f_0$ in \req{eq_f0}.
The combination $\{\mnk\}$ of composite function minimizes the last
term in \req{eq_minHEunsat}.
Subject to a vertical shift by $\Delta E_j$,
\req{eq_minHEunsat} is taken to be $f^j_0$ in \req{eq_f0} characterized by
\begin{eqnarray}
\label{eq_recurD0}
	d^j_0 = \frac{u^2}{2} + \sum_{\kinjmi}d^k_\mnk - \Delta E_j
\end{eqnarray}
and
\begin{eqnarray}
\label{eq_alpha}
	\alpha_j = \Lambda_j + \sum_{\kinjmi}\tilde y^k_\mnk.
\end{eqnarray}
For the moment,
we ignore the possibility that combinations other than ${\{\mnk\}}$
may result in further discontinuity in $f_0$,
as we will show that composite functions $f_n$ with $n>0$ eliminate the effect 
of the discontinuities.

Next we suppose the node remains as a consumer node.
In this case,
$\Lambda_j+\sum_{\kinjmi}y_k-y_j=0$,
and $\min_\snk\CH(f^{k=1}_{n_1},\dots,f^{K-1}_{n_{K-1}};\Lambda_j,y_j)$ 
is computed subject to this equality constraint,
\begin{eqnarray}
\label{eq_minHEsat}
	&&\CH(f^{k=1}_{n_1},\dots,f^{K-1}_{n_{K-1}};\Lambda_j,y_j)
	|_{\Lambda_j+\sum_{\kinjmi}y_k-y_j= 0}
	\nonumber\\
	&&=A_{\{n_k\}}(y_j-\tilde Y_{\{n_k\}})^2+D_{\{n_k\}}
	+\sum_{\kinjmi}d_{n_k}^k.
\end{eqnarray}
where
\begin{eqnarray}
\label{eq_justSat}
	A_{\{n_k\}} =&& \frac{1}{2}\biggl[
	1+\frac{1}{\sum_{\kinjmi}(2a_{n_k}^k)^{-1}}\biggr],
	\nonumber\\
	\tilde Y_{\{n_k\}} =&& \frac{\Lambda_j+\sum_{\kinjmi}\tilde y_{n_k}^k}
	{1+\sum_{\kinjmi}(2a_{n_k}^k)^{-1}},
	\\
	D_{\{n_k\}} =&&
	\frac{(\Lambda_j+\sum_{\kinjmi}\tilde y_{n_k}^k)^2}
	{2\left[1+\sum_{\kinjmi}(2a_{n_k}^k)^{-1}\right]},
	\nonumber
\end{eqnarray}
and the optimal currents drawn from the descendents are
\begin{eqnarray}
	y^*_{k,\{n_k\}} 
	=&& \frac{y_j-\Lambda_j-\sum_{\linjmi}\tilde y_{n_l}^l}
	{2a_k\sum_{\linjmi}(2a_{n_l}^l)^{-1}}
	+ \tilde y_k.
\end{eqnarray}

Next, we consider the recursions for a source node $j$. 
In this case,
the cavity energy function consists of the composite function 
with $n_j=0$ only, with 
\begin{eqnarray}
	d^j_0=\sum_{\kinjmi}d^k_\mnk-\Delta E_j.
\end{eqnarray}

Table \ref{tab_EVrecurC}
summarizes the combinations $\{n_k\}$ which lead to  
the composite functions $n_j=1,2$ in Table \ref{tab_minima},
enabling us to analyze the singlet and doublet regimes.
We represent these relations of $\snk$ with $n_j$ by the mapping $\cM$.
As an illustration, the relations in 
Table \ref{tab_EVrecurC} can be expressed as
\begin{eqnarray}
\label{eq_M}
	{\cal M}(0,\dots, 0) = 1,
	\nonumber\\
	{\cal M}(1,0, \dots, 0) = 2.
\end{eqnarray}
The constant term $d^j_{n_j}$ is taken to be the minimum constant
term in all combination of $f^k_{n_k}$ with $\cM(\snk)=n_j$,
yielding 
\begin{eqnarray}
\label{eq_recurD}
	d_{n_j}^j = &&\min_{(\{n_k\}|n_j
	={\cal M}(\{n_k\})}\bigg[D_{\{n_k\}} 
	+ \sum_{\kinjmi}d_{n_k}^k\bigg] -\Delta E_j.
%	\nonumber\\
\end{eqnarray}
Remarkably,
the functional recursion of $E^V$ in \req{eq_recurEV} is now 
simplified to a recursion of the constant terms 
in Eqs. (\ref{eq_recurD0}) and (\ref{eq_recurD}).

%%%%%%%%%%%%%%%%%%% Table 3 %%%%%%%%%%%%%%%%%%%%%%%%
\begin{table*}\centering
\begin{tabular}{|c||c|c|c|c|c|}
\hline
$\{n_k\}=$ & $\tilde Y_{\{n_k\}}=\tilde y_{n_j}$ & $A_{\{n_k\}}
=a_{n_j}$ & $\displaystyle D_{\{n_k\}}=D_{n_j}$ & 
$(y^\ast_1, y^\ast_2, \dots,y^\ast_{c-1})$ & $n_j$\\
$(n_1, n_2, \dots, n_{K-1})$ & & & & & \\
 \hline \hline
 $(0, 0, \dots, 0)$ & $\displaystyle-\frac{1}{K}$ &
 $\displaystyle\frac{K}{2(K-1)}$  &
 $\displaystyle\frac{1}{2K}$ &
 $\displaystyle{\Big(\frac{1}{K}, \frac{1}{K}, \dots, \frac{1}{K}\Big)}$ &1 \\
 \hline
 $(1, 0, \dots, 0)$ &
 $\displaystyle-\frac{1}{K-1}$ &
 $\displaystyle\frac{K^2-1}{2(K^2-K-1)}$ &
 $\displaystyle\frac{(K+1)}{2K(K-1)}$ &
 $\displaystyle{\Big(0, \frac{1}{K-1}, \dots, \frac{1}{K-1}\Big)}$ & 2\\
 \hline
\end{tabular}
\caption{Combinations $\{n_k\}$ from descendents 
that lead to the composite functions $n_j=1,2$ given by Table \ref{tab_minima}.
Permutation of $(n_1, n_2, \dots, n_{K-1})$ result in the same $n_j$.
}
\label{tab_EVrecurC}
\end{table*}
%%%%%%%%%%%%%%%%%%%%%%%%%%%%%%%%%%%%%%

The physical interpretation of the composite functions $n_j=0,1,2$ 
is revealed by considering the patterns of optimal currents around node $j$.
For $n_j=0$,
no resources are drawn from the ancestor, 
i.e., $\tilde y_0=0$.
Moreover,
the optimal currents drawn from the descendents are $\tilde y_\mnk$,
which are non-positive according to \tab{tab_EVrecurC}.
This shows that $n_j=0$ corresponds to a resource providing state 
(see \fig{gr_current}(a)).

For $n_j=1$,
currents of $K^{-1}$ are drawn from the ancestor and all descendents,
i.e., $y^\ast_{k=1}=\dots=y^\ast_{K-1}=K^{-1}$, 
as shown in \tab{tab_EVrecurC}.
Hence $n_j=1$ corresponds to the singly consuming state.
(see \fig{gr_current}(b)).

For $n_j=2$,
currents of $(K-1)^{-1}$ are drawn from the ancestor and $K-2$ descendents,
leaving the link to the remaining descendent idle.
This corresponds to the doubly consuming state (see \fig{gr_current}(c)).
Nodes with $n_j>1$ correspond to other modes of resource consumption.

%%%%%%%%% Figure 5 %%%%%%%%%%%%%%%%%
\begin{figure}
\centerline{\epsfig{figure=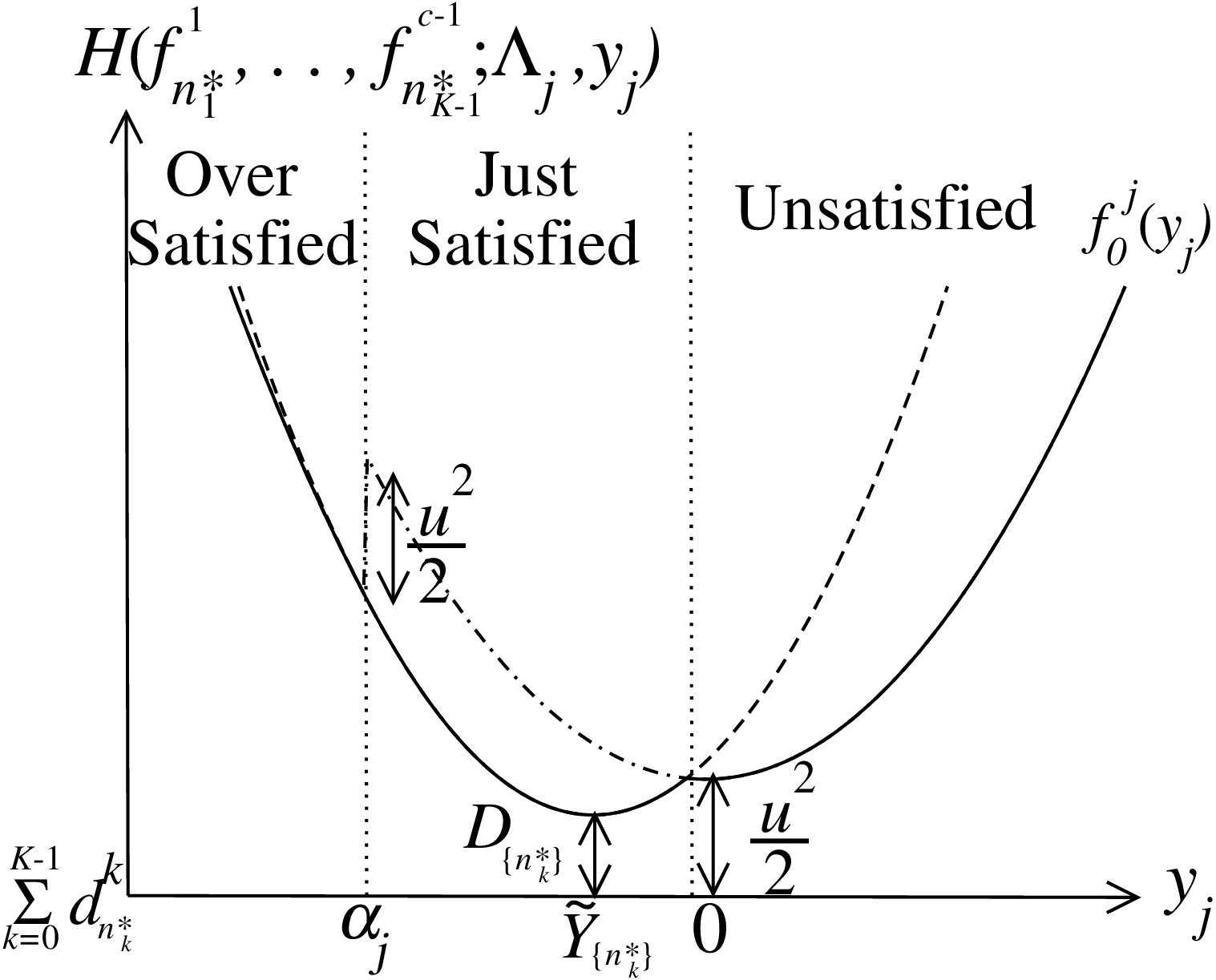, width=0.45\linewidth}}
\caption{
An example of minimization by $\CH(f^{k=1}_{n^\ast_1},\dots,
f^{K-1}_{n^\ast_{K-1}};\Lambda_j,y_j)$
with the combination $\snk=(n^*_1, n^*_2, \dots, n^*_{K-1})$.
The labels Over Satisfied, Just Satisfied, and Unsatisfied 
refer to the conditions $\xi_j>0$, $\xi_j=0$ and $\xi_j<0$ respectively. 
The just-satisfied portion of $\CH$ intersects with 
$f^j_0$ at $y_j=\alpha_j$ as shown.
}
\label{gr_min}
\end{figure}
%%%%%%%%%%%%%%%%%%%%%%%%%%%%%%%%%%%%

We return to verify that $E^V$ is continuous piecewise quadratic,
despite the discontinuity of $f^k_0$ in \req{eq_f0}.
This can be done by substituting $y_j=\alpha_j$
in \req{eq_alpha} into the composite function \req{eq_minHEsat} characterized
by  $A_{\{\mnk\}}$, $Y_{\{\mnk\}}$ and $D_{\{\mnk\}}$ in \req{eq_justSat}.
The result shows that $\CH = \alpha_j^2/2+\sum_{\kinjmi}d^k_\mnk$.
Hence it intersects $f^j_0$ at the lower end of the discontinuity.
As shown in \fig{gr_min},
the discontinuity is masked.

We finally derive a simplified expression of $\Delta E_j$ 
from \req{eq_deltaEsim}.
Since $\CH(f^1_{n_1},\dots,f^{K-1}_{n_{K-1}}; \Lambda_j, 0)$ 
is already obtained in Eqs. (\ref{eq_minHEunsat}) and (\ref{eq_minHEsat}) 
with $y_j=0$,
\begin{eqnarray}
\label{eq_deltaED2}
	\Delta E_j=\min\bigg[&&\min_\snk\bigg(
	A_\snk\tilde Y_\snk^2+D_\snk
%	\nonumber\\
	+\sum_\kinjmi d_\nk^k\bigg),
	\nonumber\\
	&&\frac{u^2}{2}\Theta(-\Lambda_j)+\sum_\kinjmi d_\nk^k\bigg].
\end{eqnarray}

%%%%%%%%%%%%%%%%%%%%%%%%%%%%%%%%%%%%%%%%%%%%
\section{Closed Sets of Cavity Energy Functions: The Intense Simplifications}
\label{sec_RSresult}

\subsection{The Intense Simplification}
\label{sec_closeSet}

\begin{figure}\centering
%\hspace{0.35cm}
\includegraphics[width=0.18\linewidth]{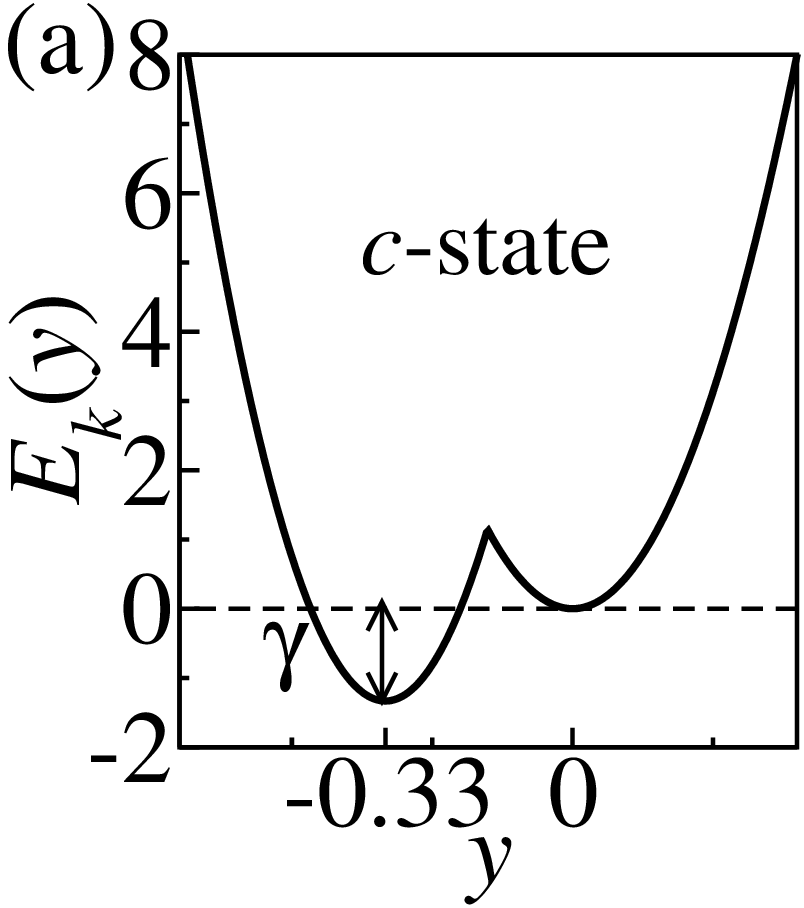}~
\includegraphics[width=0.18\linewidth]{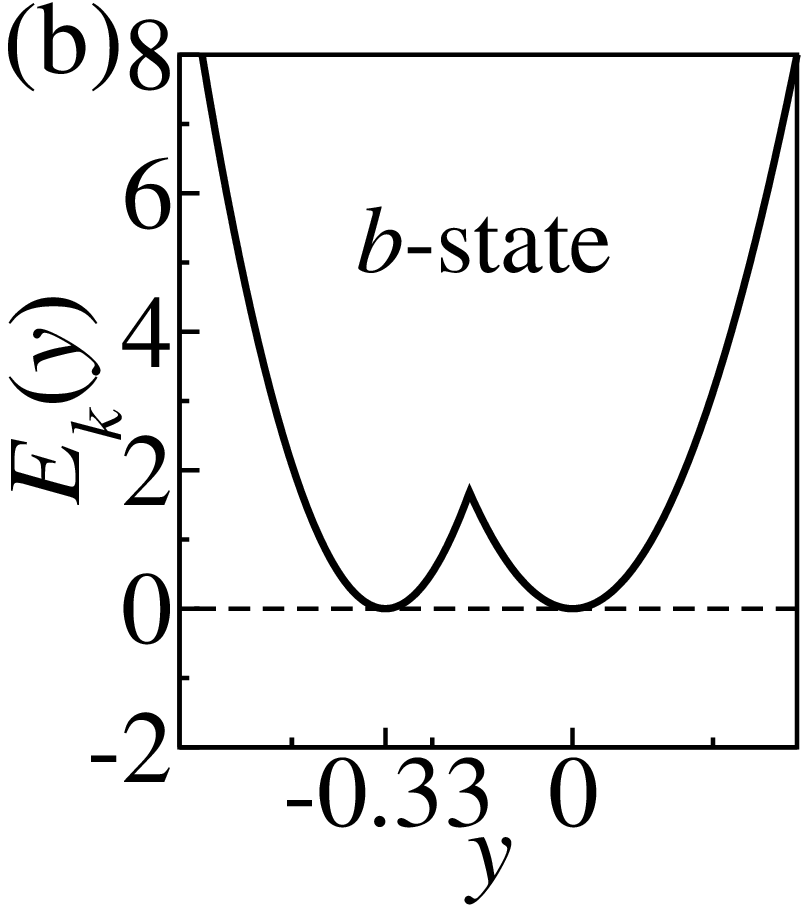}~
\includegraphics[width=0.18\linewidth]{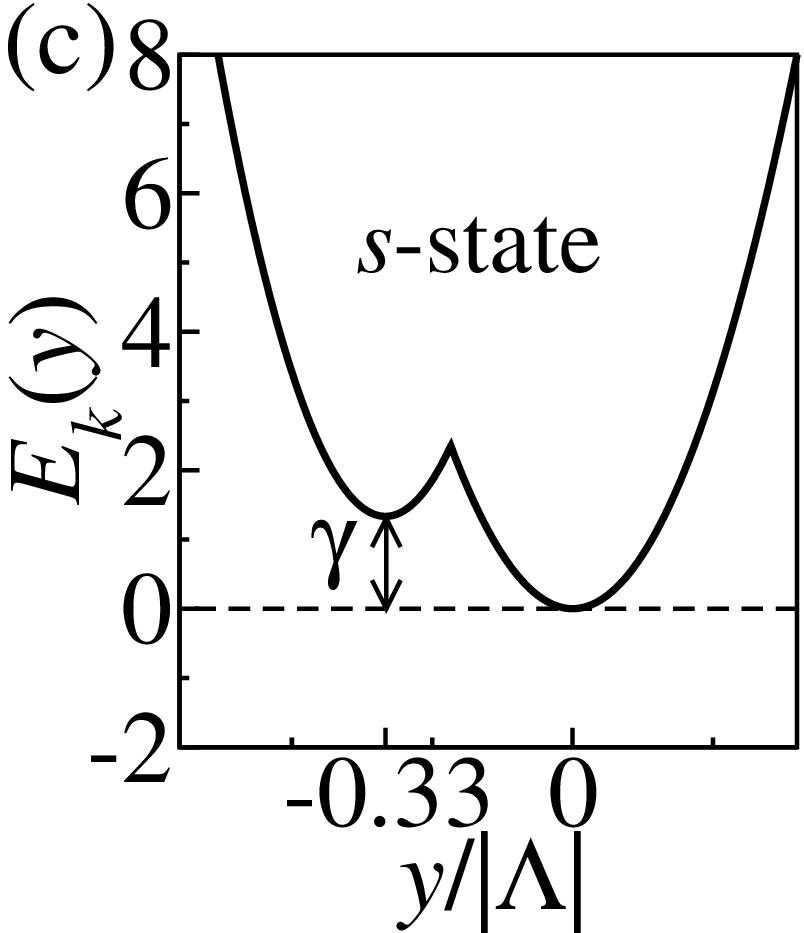}~
\caption{
The closed set of $E_k(y)$ at $K=3$ and $u^{-1}=5/3$, 
corresponding to 
(a) the $c$ state $E^V_c$,
(b) the $b$ state $E^V_b$ and
(c) the $s$ state $E^V_s$.
}
\label{gr_usb}
\end{figure}

In this subsection,
we consider networks with no surplus nodes ($\phi_s=0$).
In the singlet regime,
the energetically stable configurations consist of 
only the source nodes and singly consuming nodes.
Hence we consider $E^V$with only $f_0$ and $f_1$ as composite functions.
$E^V$ is thus given by 
\begin{eqnarray}
	E^V_k(y_k)  \doteq (d_{0}^k, d_{1}^k),
\end{eqnarray}
as a simplification of \req{eq_Ekd}.
Composite functions with $n_k\ge 2$ have 
$d^k_{n_k} > d^k_\mnk$ in this regime,
and their corresponding configurations are not stable.
The recursion relations of $d_0$ and $d_1$ 
in Eqs. (\ref{eq_recurD0}) and (\ref{eq_recurD}) are simplified to
\begin{eqnarray}
	d^j_0&=&\frac{u^2}{2}+\sum_{\kinjmi}\min(d^k_0,d^k_1)-\Delta E_j,
	\\
	d^j_1&=&\frac{1}{2K}+\sum_{\kinjmi}d^k_0-\Delta E_j,
\end{eqnarray}
where 
\begin{eqnarray}
	\Delta E_j=\min\left[\frac{u^2}{2}+\sum_{\kinjmi}\min(d^k_0,d^k_1), 
%	\right.
%	\nonumber\\
%	&&\left.
	\frac{1}{2(K-1)}+\sum_{\kinjmi}d^k_0\right].
\end{eqnarray}
To determine the pattern of current flow, 
it is sufficient to consider the recursion of $\epsilon_j\equiv d^j_1-d^j_0$,
given by
\begin{eqnarray}
\label{eq_singSatCE}
	\epsilon_j=-\gamma-\sum_{\kinjmi}\min(0,\epsilon_k),
\end{eqnarray}
where
\begin{equation}
	\gamma\equiv\frac{u^2}{2}-\frac{1}{2K}.
\end{equation}
The simple recursion leads to a closed set of $K$ cavity energy functions
\begin{eqnarray}
	E^V_q(y) \doteq (d^j_0, d^j_0+q\gamma)
	\quad\mbox{with $q=-1,0,\dots,K-2$}.
\end{eqnarray}
These functions are classified to be {\it consuming} for $q=-1$,
{\it bistable} for $q=0$,
and {\it resource providing} for $q=1,\dots,K-2$.
Their absolute minima are located at $y=-K^{-1}$,
both $y=0$ and $-K^{-1}$,
and $y=0$ respectively.
We call their states $c$ state,
\bst and \sst respectively;
examples for the case of $K=3$ are shown in \fig{gr_usb}.
The \bst behaves in the same way as the \sst in the recursion relation,
but physically they correspond to different cavity states.
The differentiation between $b$ and $s$ states is required 
only when the entropy of the ground state is calculated.
For most other purposes,
grouping $b$ and $s$ states together further simplifies the analyses.
We thus denote the $b$ and $s$ states as \Sst,
and the \cst as \Cst in subsequent analyses.
Their recursion relations are summarized in \tab{tab_recurUSB}.
(The source state ($s$ state) in this paper should not be confused 
with the satisfied state (also denoted as $s$ state) in \cite{yeung2009b}. 
In fact, the $s$, $u$, $b$ states in \cite{yeung2009b} 
have the same cavity energy functions as the $c$, $s$, $b$ states 
in this paper respectively.)

%%%%%%%%% Figure 6 %%%%%%%%%%%%%%%%%
%%%%%%%%%%%%%%%%%%%%%%%%%%%%%%%%%%%%

%%%%%%%%%%%%%%%%%%% Table 4 %%%%%%%%%%%%%%%%%%%%%%%%
\begin{table}\centering
\begin{tabular}{|c|c|c|c|c|}
\hline
Descendent states & 
$\Delta E_j $ & $\cE_j$ & 
\multicolumn{2}{c|}{Vertex state} \\
\hline \hline
$(S, \dots, S)$ &  $\displaystyle\frac{u^2}{2}$ & $-\gamma$ & $c$ & $C$ \\
\hline
$(C, S, \dots, S)$ & $\displaystyle\frac{u^2}{2}-\gamma$ & 0 & $b$ & $S$ \\
\hline
$(C, C, S, \dots, S)$ & $\displaystyle\frac{u^2}{2}-2\gamma$ & $\gamma$ & 
$s$ & $S$ \\
$\vdots$ & $\vdots$ & $\vdots$ & $\vdots$ & $\vdots$ \\
$(C, \dots, C)$ & $\displaystyle\frac{u^2}{2}-(K-1)\gamma$ & 
$(K-2)\gamma$ & $s$ & $S$ \\
\hline
\end{tabular}
\caption{The recursion relations given by \req{eq_singSatCE}.
The column $\Delta E_j$ is valid for $u^{-1}\ge\sqrt{K-1}$ 
in the singlet regime;
a constant $(K-1)[u^2/2-1/2(K-1)]$ has to be added to $\Delta E_j$ 
for $u^{-1}<\sqrt{K-1}$ in the singlet regime.
}
\label{tab_recurUSB}
\end{table}
%%%%%%%%%%%%%%%%%%%%%%%%%%%%%%%%%%%%%%%%%%%%%%%%%%%%%%

The recursion relation in \tab{tab_recurUSB} 
can be summarized by the symbolic equations
\numparts
\begin{eqnarray}
\label{eq_recurSymbol}
\underbrace{S+\dots+S}_{K-1}\rightarrow C \label{eq_recura}.
\\
\mbox{all other combinations}\rightarrow S \label{eq_recurb}.
\end{eqnarray}
\endnumparts
To calculate the average energy per node,
we can apply the same simplification to $\Delta\cE_{\rm node}$ 
and $\Delta\cE_{\rm link}$ in 
Eqs. (\ref{eq_cdeltaE}) and (\ref{eq_linkDeltaE}) respectively. 
We denote the {\it full} states of a node as $\cC$, $\cB$ and $\cS$,
respectively representing the consuming, bistable 
and resource providing states.
They are obtained symbolically via
\numparts
\begin{eqnarray}
\label{eq_recurSymbolFull}
\underbrace{S+\dots+S}_{K}\rightarrow \cC \label{eq_recuraF}
\\
\underbrace{C+S+\dots+S}_{K}\rightarrow \cB \label{eq_recurbF}
\\
\mbox{all other combinations}\rightarrow \cS \label{eq_recurcF}.
\end{eqnarray}
\endnumparts
The energy changes are 
\begin{eqnarray}
\label{eq_deltaCE}
	\Delta \cE_{\rm node}&=&\frac{u^2}{2}
	+\min\left[-\gamma,\sum^K_{j=1}\min(0,\epsilon_j)\right],\\
	\Delta \cE_{\rm link}&=&\min(0,\epsilon_L,\epsilon_R).
\end{eqnarray}
These expressions are valid for $u^{-1}\ge\sqrt{K-1}$ in the singlet regime;
$N_c[u^2/2-1/2(K-1)]$ has to be added to the expressions 
for $u^{-1}<\sqrt{K-1}$ in the singlet regime,
where $N_c$ is the number of vertices in the \Cst.

Numerical iterations of \req{eq_singSatCE} starting from random $\epsilon_k$
show that the closed set of $E^V$ is stable.
The closed set corresponds to the integer cavity fields 
in the language of the cavity approach.
Another example of a closed set of $E^V$ is found in the doublet regime 
as described in Appendix.

We return to discuss the range of the singlet regime. 
From \req{eq_singSatCE},
it becomes apparent that when $\gamma<0$,
we would have $\epsilon_j$ always positive,
implying that the singly consuming state is always unstable.
Hence a necessary condition of the singlet regime is $\gamma\ge 0$,
or $u^{-1}<\sqrt{K}$.

Similarly,
in the doublet regime, 
we obtain the recursion relations
\begin{eqnarray}
	d^j_1-d^j_0&=&
	-\gamma-\sum_\kinjmi\min(0,d^k_1-d^k_0,d^k_1-d^k_0),
	\\
	d^j_2-d^j_0&=&-\kappa+\min_{\{k\}}(d^k_1-d^k_0)
%	\nonumber\\
	-\sum_\kinjmi\min(0,d^k_1-d^k_0,d^k_1-d^k_0),
\end{eqnarray}
where
\begin{equation}
	\kappa\equiv\frac{u^2}{2}-\frac{K+1}{2K(K-1)}.
\end{equation}
Consider the difference $d^j_2-d^j_1=-\kappa+\min(d^k_1-d^k_0)+\gamma$.
In the singlet regime described by \req{eq_singSatCE}, 
$d^j_2-d^j_1$ is always positive only if $\kappa<0$,
implying that $u^{-1}>\sqrt{K(K-1)/(K+1)}$.
Combining the two results,
the range of the singlet regime is $\sqrt{K(K-1)/(K+1)}<u^{-1}<\sqrt{K}$,
agreeing with the result reported in \fig{gr_energy}.

It is convenient to represent the recursion relations 
in the probabilistic framework of BP algorithm \cite{pearl1988}.
We denote as $\psiji_c$ the probability that node $j$ is in the \Cst,
in the absence of the ancestor node $i$.
The probability that node $j$ is in the \Sst in the absence of $i$
is described as $\psiji_s$.
We call $\psiji_c$ and $\psiji_s$ the {\it cavity probabilities}.
In this framework,
Eq. (\ref{eq_recurSymbol}) in the singlet regime can be written as 
\begin{eqnarray}
\label{eq_recurDisorder}
	\psiji_c=\prod_\kinjmi (1-\psikj_c),
\end{eqnarray}
and $\psiji_s+\psiji_s=1$.
Note that the recursion relation (\ref{eq_recurE}) 
with the functional form of $E^V$ 
is now represented by a recursion of a single scalar $\psiji_c$,
greatly simplifying the analysis. 
This framework will be useful 
in establishing a connection with the RSB ansatz.

\subsection{Networks with Surplus Nodes}
\label{sec_disorder}

In this subsection,
we consider networks with surplus nodes ($\phi_s>0$) 
in the singlet regime.
The recursion relations of the deficient nodes 
follow those in \tab{tab_recurUSB},
whereas the $E^V$ of the surplus nodes are always in the $s$ state.
The symbolic equations \req{eq_recurSymbol} are thus extended to
\numparts
\begin{eqnarray}
\label{eq_recurSymbol3}
\underbrace{S+\dots+S}_{K-1}\overset{\Lambda=-1}{\rightarrow} C \label{eq_recura3}.
\\
\mbox{all other combinations}\overset{\Lambda=-1}{\rightarrow} S \label{eq_recurb3}.
\\
\mbox{all combinations}\overset{\Lambda=A}{\rightarrow} S \label{eq_recurc3}.
\end{eqnarray}
\endnumparts
The recursion of $\psiji_c$ is given by 
\begin{eqnarray}
	\psiji_c=\delta_{\Lambda_j,-1}\prod_\kinjmi (1-\psikj_c),
\end{eqnarray}
and $\psiji_c+\psiji_s=1$.

The {\it full} energy change is given by 
\begin{eqnarray}
\label{eq_cdeltaESingSat}
	\Delta{\cal E}^i_{\rm node}
%	\\
	=\cases{
	\displaystyle\frac{u^2}{2} +
	\min\bigg[-\gamma, \sum_{\jini}\min(0,\epsilon^j_1)\bigg], 
	& $\Lambda_i=-1$
	\\
	\displaystyle
	\sum_{\jini}\min(0,\epsilon^j_1), & $\Lambda_i = A$
	}
%	\nonumber
\end{eqnarray}
and $\Delta \cE_{\rm link}$ is still given by \req{eq_deltaCE}.
For deficient nodes, 
the optimized state is either consuming, bistable, or resource providing.
Surplus nodes are fixed to be source nodes. 
$\Delta \cE_{\rm node}$ and the full states of node $i$ 
are shown in \tab{tab_Enode}
with the corresponding combination of descendent states.
The full states of the node $i$ are thus described 
by the probabilities $\psi^i_\cC$, 
$\psi^i_\cB$ and $\psi^i_\cS$ given by
\begin{eqnarray}
	\psi^i_\cC &=& \delta_{\Lambda_i,-1}\prod_{\jini}(1-\psiji_c),
	\nonumber\\
	\psi^i_\cB &=& \delta_{\Lambda_i,-1}\sum_{\jini}
	\psiji_c\prod_{k\in\cL_j\backslash\{i\}}(1-\psi^{k\rightarrow i}_c),
\end{eqnarray}
with $\psi^i_\cC + \psi^i_\cB + \psi^i_\cS=1$.

We derive the fraction of nodes with different full states 
by assuming the independence between descendent branches of a tree.
From \req{eq_recurDisorder}, 
the average probability $\avg{\psi_c}$ of a node in the cavity \Cst 
is thus given by 
\begin{eqnarray}
\label{eq_avgPsi}
	\avg{\psi_c}=\phi_d[1-\avg{\psi_c}]^{K-1},
\end{eqnarray}
where $\avg{\dots}$ represents averaging over nodes.
Iteration of \req{eq_recurDisorder} on Cayley trees 
reveals that $\avg{\psi_c}$ does not approach 
the stable fixed point of \req{eq_avgPsi} when $\phi_d$ is high.
At $\phi_d=1$,
a period-two solution of $\avg{\psi_c}=0$ and 1 emerges. 
Physically, 
this corresponds to alternating layers of consumer 
and source nodes on Caylay trees.
This happens in similar problems such as the Bethe glass \cite{rivoire2004} 
and is referred to as the modulation mode.
On real networks,
nodes are randomly connected,
rendering whole layers of consumer and source nodes highly unlikely.
The period-two situations are suppressed, 
making a fixed point solution of \req{eq_avgPsi} possible in random networks.
The fraction of nodes with full states $\cC$, $\cB$ and $\cS$ are thus given by
\begin{eqnarray}
\label{eq_fSfB}
	f_\cC &=& \phi_d(1-\avg{\psi_c})^K,
	\nonumber\\
	f_\cB &=& K\phi_d\avg{\psi_c}(1-\avg{\psi_c})^{K-1},
\end{eqnarray}
with $f_\cC+f_\cB+f_\cS=1$.
We leave the discussion of their physical interpretation 
to Section \ref{sec_freeNode}.

%%%%%%%%%%%%%%%%%%% Table 5 %%%%%%%%%%%%%%%%%%%%%%%%
\begin{table}\centering
\begin{tabular}{|c|c|c|c|c|}
\hline
& \multicolumn{2}{c|}{$\Lambda_i=-1$} &  \multicolumn{2}{c|}{$\Lambda_i=A$} \\
\cline{2-5}
Descendent states & $\Delta{\cal E}^j_{\rm node}$ & Full state & 
$\Delta{\cal E}^j_{\rm node}$ & Full state \\
\hline \hline
$(S, \dots, S)$ & $\displaystyle\frac{u^2}{2}-\gamma$ & $\cC$ & $0$ & $\cS$ \\
\hline
$(C, S, \dots, S)$ & $\displaystyle\frac{u^2}{2}-\gamma$ & $\cB$ & 
$-\gamma$ & $\cS$ \\
\hline
$(C, C, S, \dots, S)$ & $\displaystyle\frac{u^2}{2}-2\gamma$ & $\cS$ 
& $-2\gamma$ & $\cS$ \\
$\vdots$ & $\vdots$ & $\vdots$ & $\vdots$ & $\vdots$ \\
$(C, \dots, C)$ & $\displaystyle\frac{u^2}{2}-K\gamma$ & $\cS$ & 
$-K\gamma$ & $\cS$ \\
\hline
\end{tabular}
\caption{$\Delta{\cal E}_{\rm node}$ as given by \req{eq_cdeltaESingSat}.
}
\label{tab_Enode}
\end{table}
%%%%%%%%%%%%%%%%%%%%%%%%%%%%%%%%%%%%%%%%%%%%%%%%%%%%

\section{The Average Energy}
\label{sec_RSenergy}

We evaluate the average energy by considering 
the energy change due to the addition of new nodes and links.
Summarizing \tab{tab_Enode},
$\avg{\Delta \cE_{\rm node}}$ becomes
\begin{eqnarray}
\label{eq_EnodeSim}
	\avg{\Delta{\cal E}_{\rm node}} 
	= \phi_d\Big[\frac{u^2}{2}-(1-\avg{\psi_c})^K\gamma\Big]
	-K\avg{\psi_c}\gamma.
\end{eqnarray}
Similarly,
$\avg{\Delta{\cal E}_{\rm link}}$ is given in \tab{tab_Elink} by
\begin{eqnarray}
\label{eq_ElinkSim}
	\avg{\Delta\cE_{\rm link}} 
	= -\gamma[2\avg{\psi_c} - \avg{\psi_c}^2].
\end{eqnarray}
After some algebra,
Eqs. (\ref{eq_deltaCalE}), (\ref{eq_avgPsi})
(\ref{eq_EnodeSim}) and (\ref{eq_ElinkSim}) lead to
\begin{eqnarray}
\label{eq_ERS}
	\cE_{\rm RS} = \avg{\Delta{\cal E}} 
	= \phi_d\frac{u^2}{2} 
	- \left[\avg{\psi_c}+\left(\frac{K}{2}-1\right)
	\avg{\psi_c}^2\right]\gamma.
\end{eqnarray}

%%%%%%%%%%%%%%%%%%% Table 6 %%%%%%%%%%%%%%%%%%%%%%%%
\begin{table}\centering
\begin{tabular}{|c|c|}
\hline
Vertex states & $\Delta{\cal E}_{\rm link}$\\
\hline \hline
$(S, S)$ & 0 \\
\hline
$(C, S)$ & $-\gamma$ \\
\hline
$(C, C)$ & $-\gamma$  \\
\hline
\end{tabular}
\caption{$\Delta{\cal E}_{\rm link}$ as given by \req{eq_linkDeltaE}.
}
\label{tab_Elink}
\end{table}
%%%%%%%%%%%%%%%%%%%%%%%%%%%%%%%%%%%%%%%%%%%%%%%%%%%%

When a deficient node changes from a source to a consumer, 
the energy of the node reduces by $\gamma$ from $u^2/2$.
We thus identify the coefficient of $\gamma$ in \req{eq_ERS}
to be the fraction of consumer nodes.
The fraction $f^{\rm RS}_s$ of source nodes is then given by
\begin{eqnarray}
\label{eq_fuRS}
	f_s^{\rm RS}=1-\avg{\psi_c}-\left(\frac{K}{2}-1\right)\avg{\psi_c}^2.
\end{eqnarray}
Note that $f_s^{\rm RS}$ is distinguished from $f_\cS^{\rm RS}$,
since $f_s=f_\cS+f_\cB/2$, i.e., 
$f_s$ also counts those bistable nodes that become source nodes 
in the network configuration.
Through the linear relationship $\avg{\cE}=\phi_d u^2/2-f_c^{\rm RS}\gamma = 1/2K-(1-\phi_d)u^2/2+f_s^{\rm RS}\gamma$,
we consider $f_s$ as a measure of the average energy $\avg{\cE}$.

\begin{figure}
\centerline{\epsfig{figure=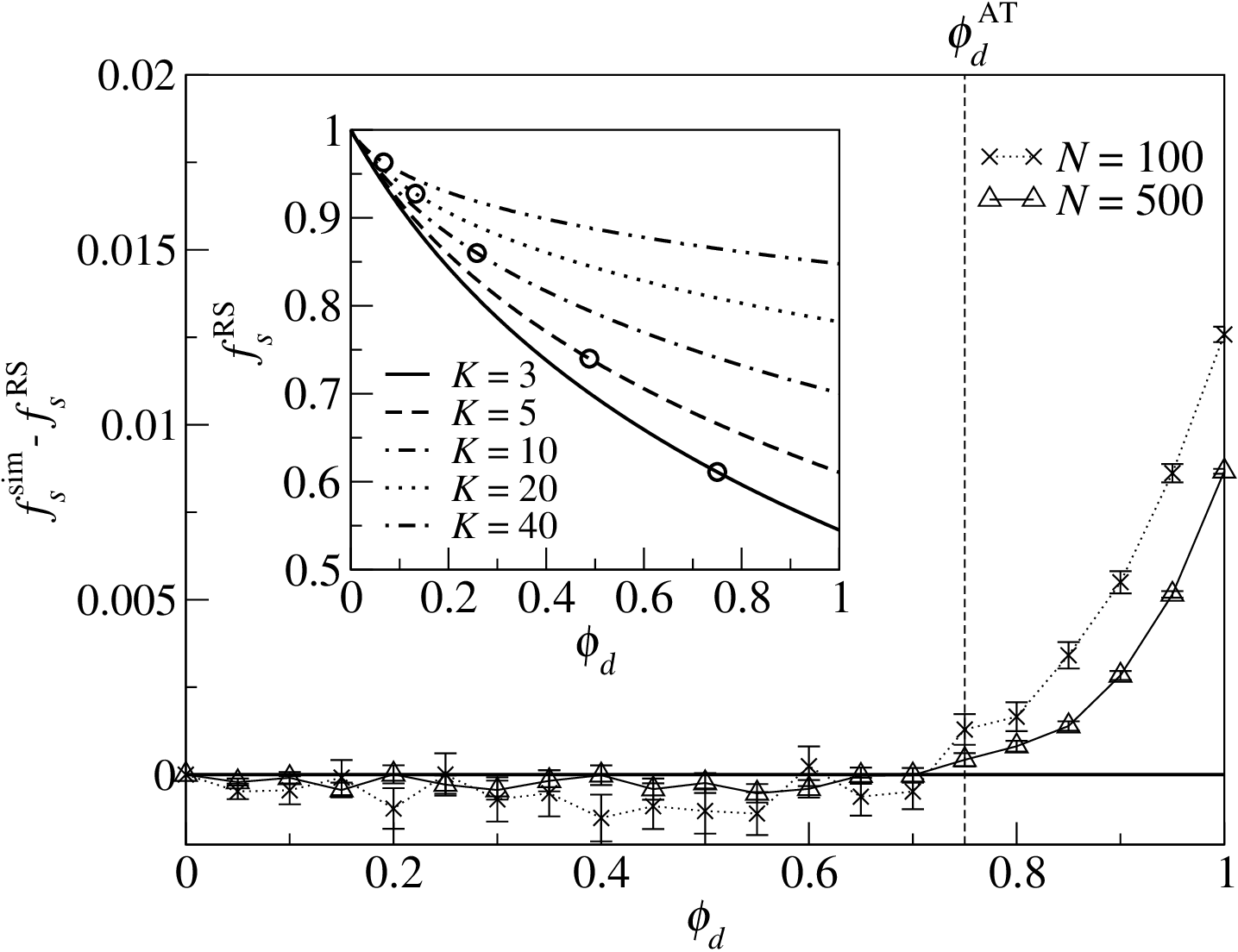, width=0.65\linewidth}}
\caption{
The average fraction of source nodes in simulations with
$K=3$, $N=100, 500$ and 3000$N$ steps with $N_{\rm flip}=4$,
as compared with the $f_s^{\rm RS}$ obtained from the 
RS ansatz.
Inset: $f_s^{\rm RS}$ for different values of $K$.
The circles indicate the values of $\phi_d$ above which the 
RS assumption is not stable.
}
\label{gr_RSenergy}
\end{figure}

The inset of \fig{gr_RSenergy} shows the fraction of source nodes 
as a function of $\phi_d$ derived from Eqs.~(\ref{eq_avgPsi})
and (\ref{eq_fuRS}).
For all connectivities $K$,
$f_s^{\rm RS}$ decreases with $\phi_d$.
A higher connectivity leads to an increase in $f_s^{\rm RS}$ since
more nodes are required to convert to source nodes 
to satisfy the demand of a consumer node.

Figure~\ref{gr_RSenergy} shows the difference between $f_s^{\rm RS}$ and 
$f_s^{\rm sim}$ obtained from numerical simulations,
in which the energy of
real instances is minimized by the GSAT algorithm
as described in Section \ref{sec_simulation}.
The differences between $f_s^{\rm RS}$ and 
$f_s^{\rm sim}$ are roughly zero when $\phi_d$ 
is below some critical value.
Above the critical value,
$f_s^{\rm RS}$ is significantly lower than $f_s^{\rm sim}$.
Hence the energy $\cE_{\rm RS}$ is lower than the simulated energy.
This discrepancy is related to the instability of the RS ansatz, 
which will be discussed below.

%%%%%%%%% Figure 7 %%%%%%%%%%%%%%%%%
%%%%%%%%%%%%%%%%%%%%%%%%%%%%%%%%%%%%

\subsection{The Soft Nodes in the Ground States}
\label{sec_freeNode}

%%%%%%%%% Figure 8 %%%%%%%%%%%%%%%%%
\begin{figure}
\centerline{\epsfig{figure=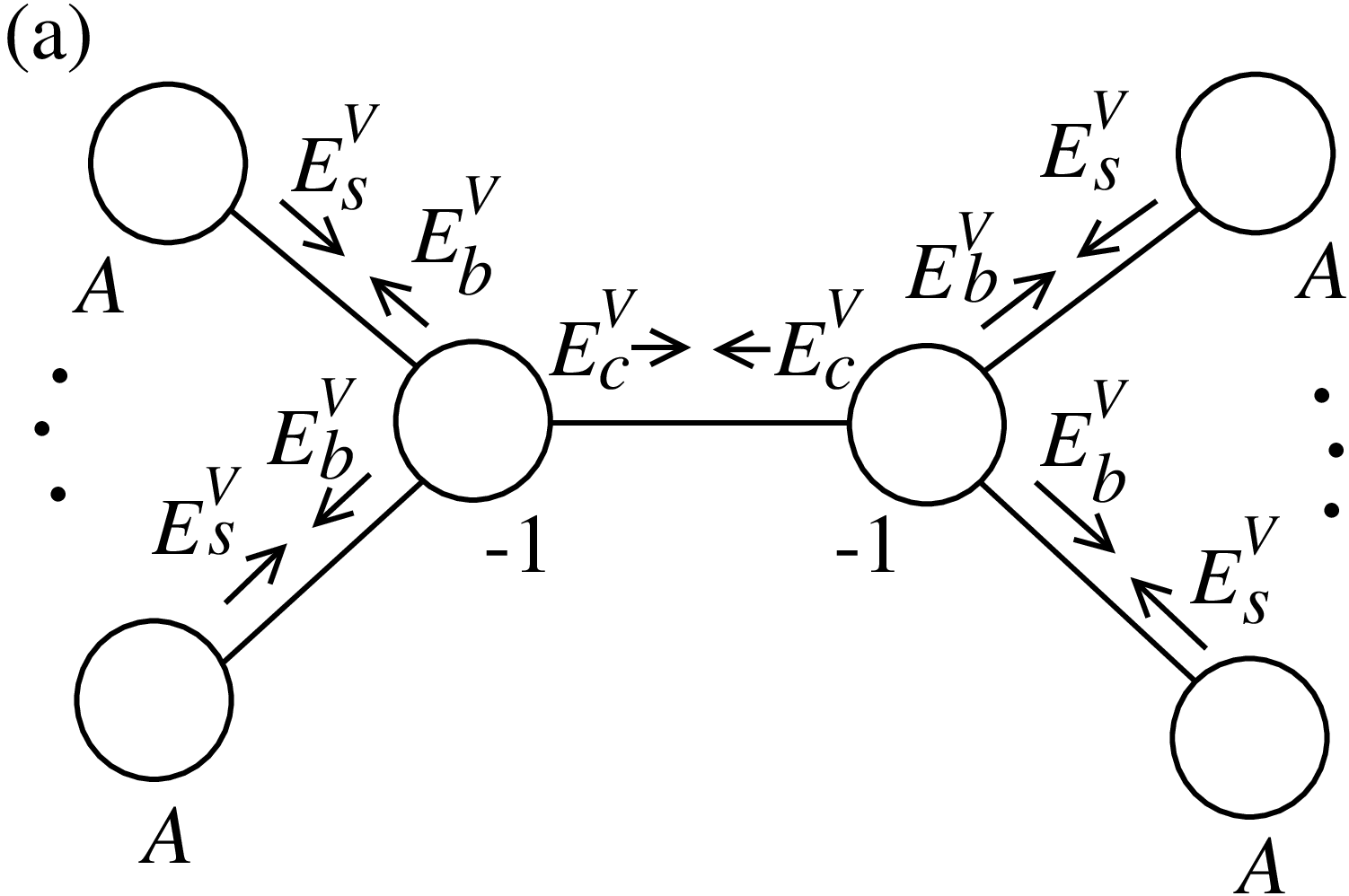, width=0.35\linewidth}}
\hspace{2cm}
\leftline{
\epsfig{figure=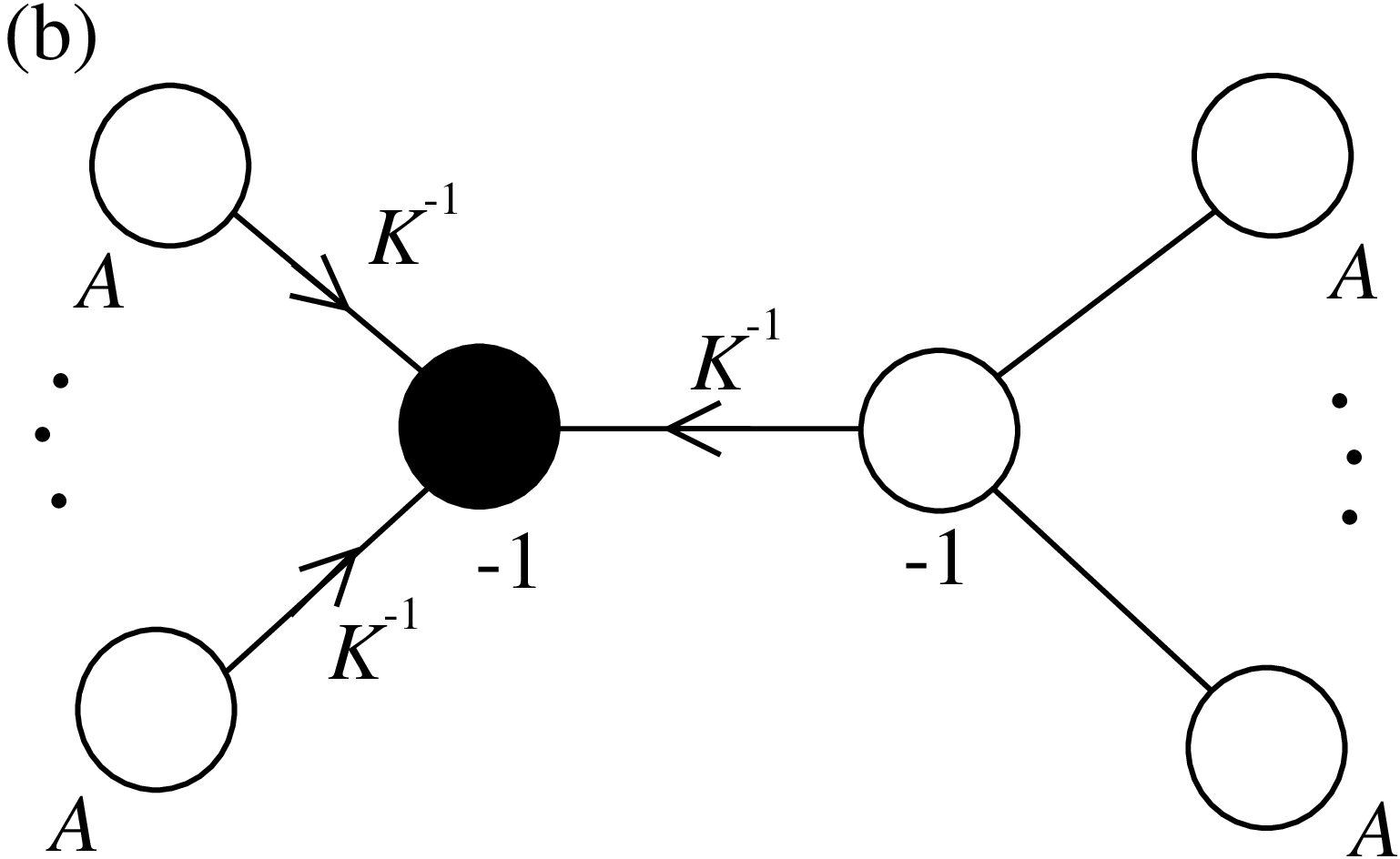, width=0.3\linewidth}
\hspace{1cm}
\leftline{\epsfig{figure=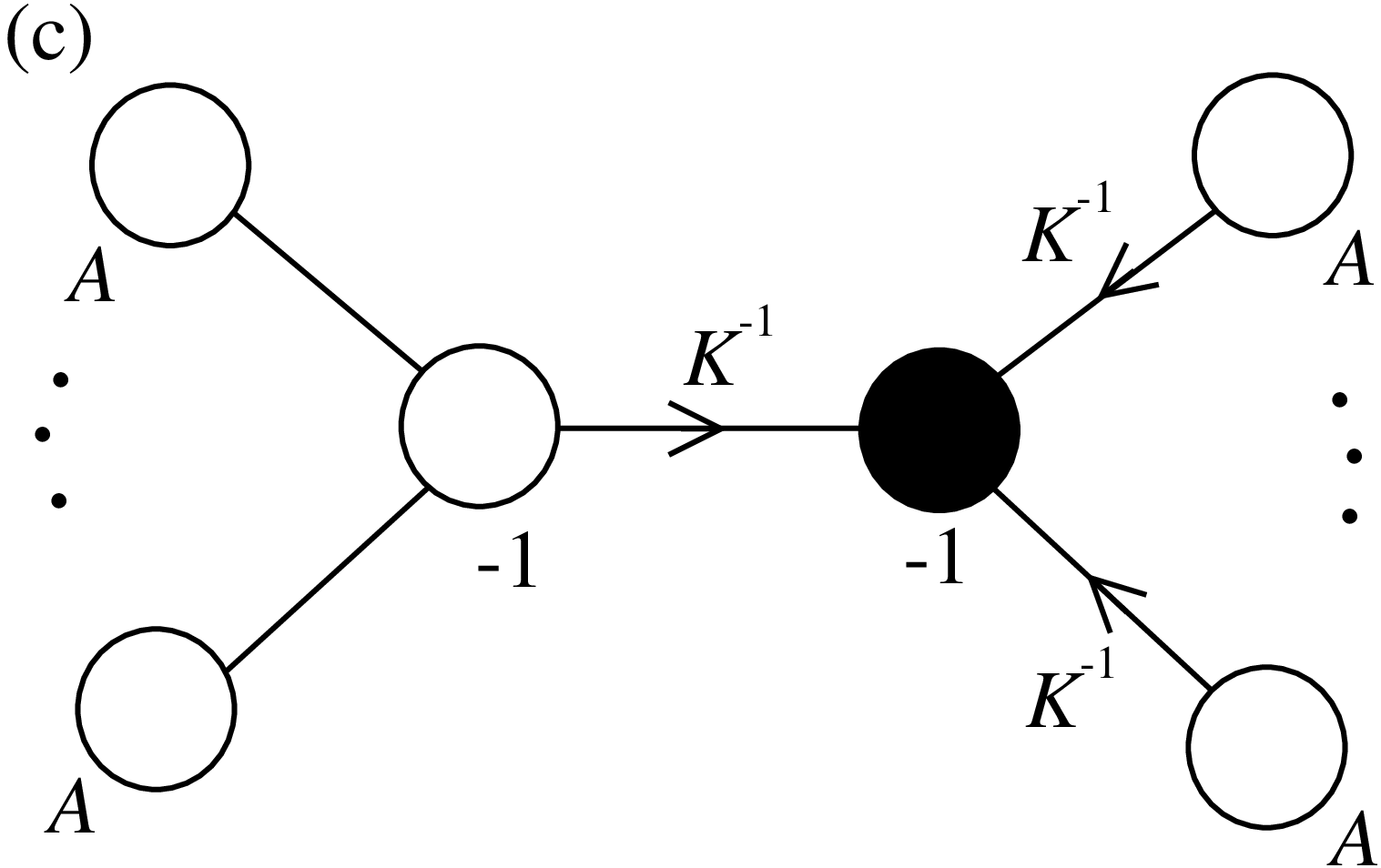, width=0.3\linewidth}}}
\caption{
(a) An example of a configuration with two central soft nodes,
annotated with cavity energy functions passed among them.
(b-c) The two degenerate states of configuration in (a).
}
\label{gr_freeNode}
\end{figure}
%%%%%%%%%%%%%%%%%%%%%%%%%%%%%%%

Nodes with the full state $\cB$
are bistable between the consumer and source states.
Figure~\ref{gr_freeNode} shows an example with two central nodes 
in the $\cB$-state.
Using \tab{tab_recurUSB},
the cavity energy functions passed among them are worked out.
In particular,
$E^V_b$ is sent from the bistable nodes 
to all its resource-providing neighbors,
implying a zero current or a current of $-K^{-1}$ would have 
no effect on the optimized energy.
Hence,
the central pairs correspond to the {\it soft} nodes which 
can be a consumer or a source node in different degenerate configurations.
These degenerate configurations are connected in the configuration space,
since transitions among them only involve 
the flipping of states of the soft nodes.
The RS ansatz assumes that the configuration space 
is dominated by a single cluster
whose states can be obtained from each other 
by flipping the states of the soft nodes \cite{zdeborova2007}.
In other words,
using the notions of $\cS$, $\cC$ and $\cB$ states,
the single cluster is described by a unique set of 
$\cS$, $\cC$ and $\cB$ labels for each node.
On the other hand,
if the network enters the RSB phase,
the configuration space is dominated by numerous clusters,
each with its own set of $\cS$, $\cC$ and $\cB$ states of the nodes.
Transitions among the clusters involve flipping the {\it hard} (non-soft)
nodes as well.
In this section,
we discuss the RS case.

The fraction of soft nodes is given by the fraction of 
nodes with full state $\cB$.
From \req{eq_fSfB},
we have
\begin{eqnarray}
\label{eq_freeNode}
 	f_{\rm soft}^{\rm RS}=f_\cB=K\phi_d\avg{\psi_c}(1-\avg{\psi_c})^{K-1}.
\end{eqnarray}
On the other hand, 
the hard nodes are either consumer or source nodes 
in all degenerate states.
They correspond to nodes in the {\it backbone} 
in a vertex cover \cite{weigt2000,weigt2001,zhou2005}.
Surplus nodes are certainly in the source backbone.
Deficient nodes can be found either in the consumer or source backbone.
From \tab{tab_Enode}, 
the fraction of nodes in the consumer backbone is given in \req{eq_fSfB} by
the probability of finding a node with full state $\cC$, namely,
\begin{eqnarray}
\label{eq_satBackbone}
	f_{\rm con-bone}^{\rm RS} = f_\cC = \phi_d(1-\avg{\psi_c})^{K},
\end{eqnarray}
Remarkably,
it can be shown easily $f_{\rm soft}^{\rm RS}/2 + f_{\rm con-bone}^{\rm RS} 
= 1-f_s^{\rm RS}$
which implies that exactly half of the soft nodes are consumers.

%%%%%%%%% Figure 9 %%%%%%%%%%%%%%%%%
\begin{figure}
\centerline{\epsfig{figure=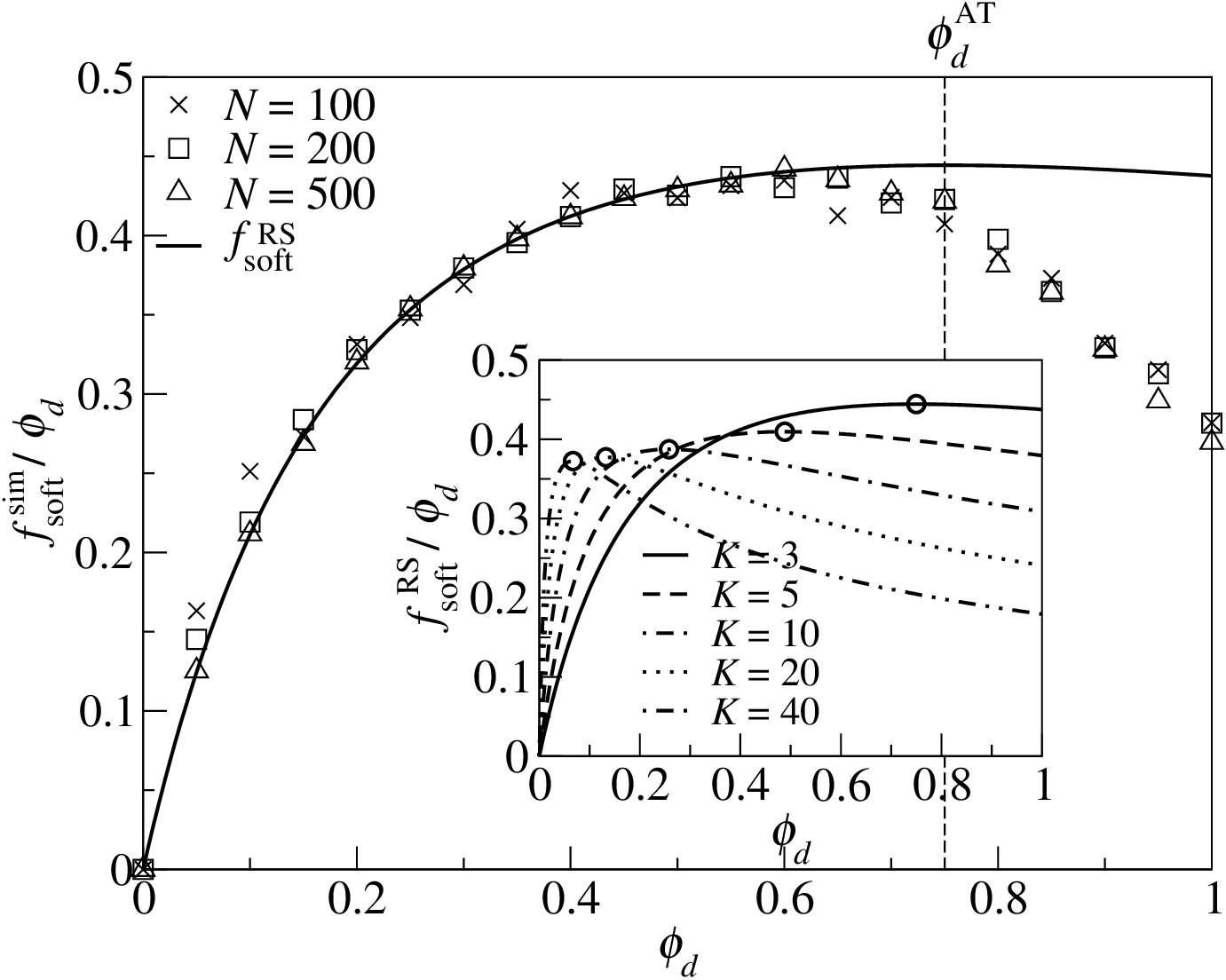, width=0.65\linewidth}}
\caption{
The average fraction of soft nodes in simulations of
$K=3$ and 3000$N$ steps with $N_{\rm flip}=4$,
as compared with $f^{\rm RS}_{\rm soft}$.
Inset: $f_s^{\rm RS}/\phi_d$ for different values of $K$.
The circles indicates the values of $\phi_d$ 
above which the RS assumption is not stable.
}
\label{gr_nBackbone}
\end{figure}
%%%%%%%%%%%%%%%%%%%%%%%%%%%%%%%%%%%%

The fraction of soft nodes relative to the fraction of deficient nodes 
obtained from \req{eq_freeNode} 
is shown in the inset of \fig{gr_nBackbone}.
When the fraction $\phi_d$ of deficient nodes increases from 0,
the fraction increases until it reaches a maximum value 
at $\avg{\psi_c}=K^{-1}$ and $\phi_d=K^{K-2}/(K-1)^{K-1}$.
When $\phi_d$ is small, 
most deficient nodes are surrounded by surplus nodes
and hence are found in the consumer backbone,
leading to a small fraction of soft nodes.
When $\phi_d$ increases,
the probability of finding contiguous deficient nodes increases,
which leads to an increase in $f^{\rm RS}_{\rm soft}$.

The fraction of soft nodes in simulations 
is compared with $f_{\rm soft}^{\rm RS}$ in \fig{gr_nBackbone}.
In simulations,
we use an algorithm similar to the GSAT \cite{selman1996}
to lower the energy of the system
until it becomes steady.
Then we allow further cluster flips and identify the nodes
which can be flipped with no change in global energy, 
such as the example in Fig. \ref{gr_freeNode}(b) and (c).
Similar to \fig{gr_RSenergy},
the simulation results have an excellent agreement with the RS result
when $\phi_d$ is below the value at the peak of 
$f_{\rm soft}^{\rm RS}/\phi_d$.
Above this critical value,
the fraction of soft nodes in simulations is less than the RS prediction,
which will be shown to be due to RS instability.

Algorithmically,
a high fraction of soft nodes leads to an unfavorable consequence.
After the convergence of BP on real instances,
nodes with $\cB$ state are bistable 
and a further determination of the final optimal
configuration is required on the resulting sub-graph of $\cB$ nodes.
A random assignment of $\cB$ nodes 
to either consuming or resource-providing does not generally
result in an optimal configuration.
When $\phi_d$ is low, 
the sub-graph of $\cB$ nodes are disconnected and the assignment is easy.
When $\phi_d$ is high, 
the sub-graphs of $\cB$ nodes are connected 
and the assignment is more difficult.

\subsection{The Instability of the RS Ansatz}
\label{sec_RSinstability}

The discrepancy between the simulation results and
the predicted average energy ${\cal E}_{\rm RS}$ and $f_{\rm soft}^{\rm RS}$ 
suggests that the RS ansatz is unstable at high $\phi_d$.
In the RS formalism,
a single ground state is assumed.
We thus examine the stability of this assumption against the picture 
of multiple ground states,
by relaxing the constraints of $\psiji_s=0,1$.
In other words,
$0\le\psiji_c\le 1$.
To study the stability of the integer ansatz of $\psiji_c$,
we define the variables
\begin{eqnarray}
	\ecji &=& \delta_{\psiji_c,1}
	\nonumber\\
	\esji &=& \delta_{\psiji_s,1}
\end{eqnarray}
and $\egji=1 -\ecji-\esji$.
Thus, the recursion rule (\ref{eq_recurDisorder}) is extended to 
include non-integer values of $\psiji_c$.
$\egji = 1$ corresponds to the onset of
non-zero probabilities over the range of $0<\psiji<1$,
indicating the occurrence of {\it glassy} behavior with node $j$
having a probabilistic distribution
of $C$ and $S$ states in the stable states of the configuration space.
Hence the RS solution is stable 
only if $\egji=0$ for all $j\rightarrow i$ is a stable fixed point.

We thus formulate the recursion relations of $\ecji$,
$\esji$ and $\egji$ by considering \tab{tab_recurUSB},
namely,
\begin{eqnarray}
\label{eq_etaRecur}
	\ecji=
	&&\delta_{\Lambda_j, -1}\prod_{k=1}^{K-1}\eskj
	\nonumber\\
	\esji=
	&&\delta_{\Lambda_j, A}+\delta_{\Lambda_j, -1}
	\biggl[1-\prod_{k=1}^{K-1}(\eskj+\egkj)\biggr]
	\nonumber\\
	\egji=
	&&\delta_{\Lambda_j, -1}
	\biggl[\prod_{k=1}^{K-1}(\eskj+\egkj)-\prod_{k=1}^{K-1}\eskj\biggr].
\end{eqnarray}
Denoting the site average of $\ecji$, $\esji$ and $\egji$ 
as $\avg{\eta_c}$, $\avg{\eta_s}$ and $\avg{\eta_g}$ respectively,
we obtain their stable fixed points by solving
\begin{eqnarray}
\label{eq_etaRecurStable}
	\avg{\eta_c}=
	&&\phi_d\avg{\eta_s}^{K-1}
	\nonumber\\
	\avg{\eta_s}=
	&&1-\phi_d(\avg{\eta_s}+\avg{\eta_g})^{K-1}
	\nonumber\\
	\avg{\eta_g}=
	&&\phi_d[(\avg{\eta_s}+\avg{\eta_g})^{K-1}-\avg{\eta_s}^{K-1}]
\end{eqnarray}
For all $\phi_d$,
$\avg{\eta_g}=0$
is a trivial solution of the last line of \req{eq_etaRecurStable},
and the above recursions reduce to the RS recursions (\ref{eq_recurDisorder}).
By introducing a small perturbation $\delta\egkj$ to $\egkj=0$,
we obtain the corresponding $\delta\egji$.
The solution of $\avg{\eta_g}=0$ is stable under the perturbation if
\begin{eqnarray}
\label{eq_etaInstability}
	\left| \frac{\avg{\delta \egji}}{\avg{\delta\egkj}} \right| 
	=(K-1)\phi_d\avg{\eta_s}^{K-2}\le 1,
\end{eqnarray}

%%%%%%%%% Figure10 %%%%%%%%%%%%%%%%%
\begin{figure}
\centerline{\epsfig{figure=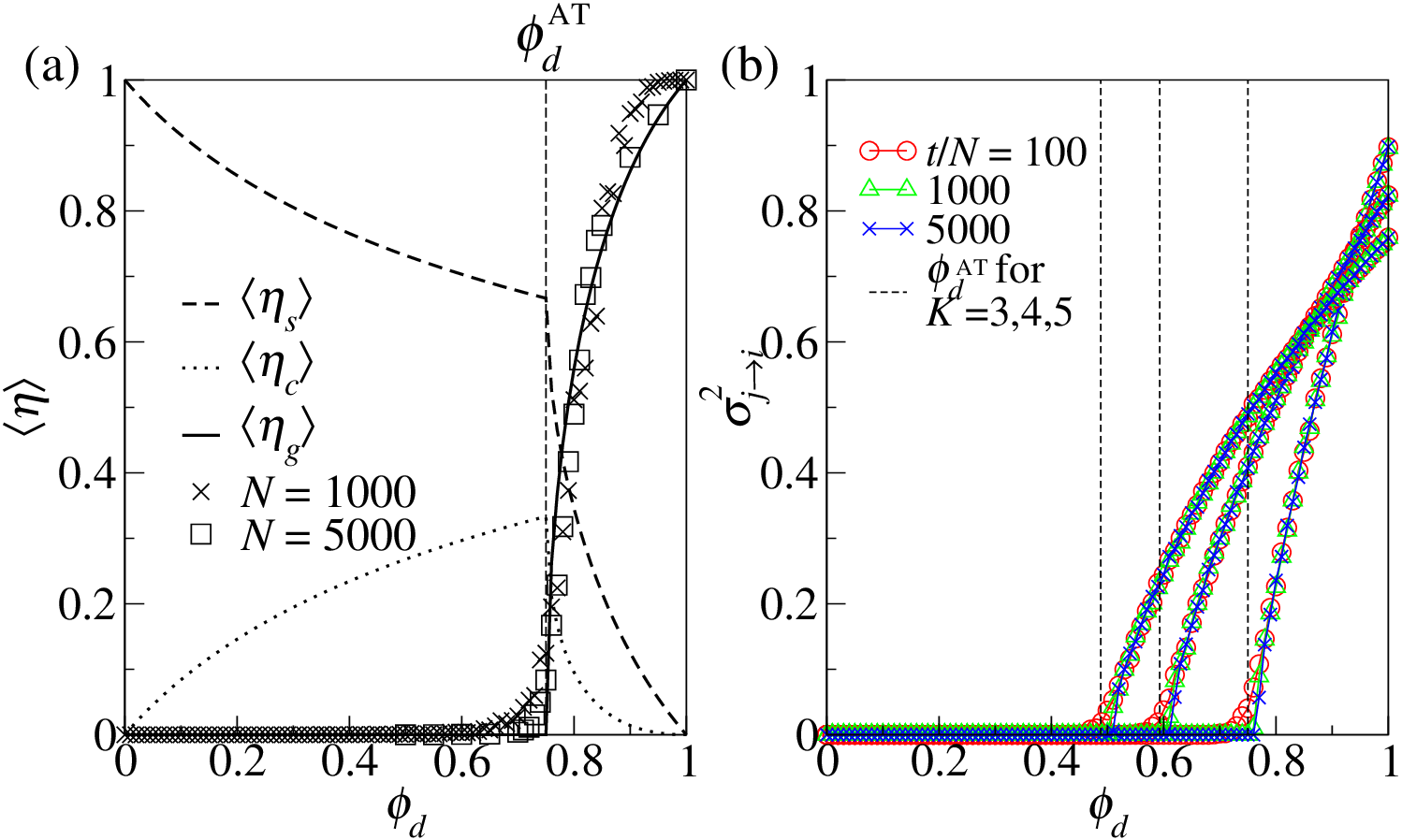, width=0.7\linewidth}}
\caption{
(a) The stable solution of \req{eq_etaRecurStable} for $K=3$.
The symbols represent the simulated fraction of non-converging BP messages.
(b) The dependence of the variance $\sigma^2_{j\to i}$ on $\phi_d$. 
The dotted lines indicate the corresponding values of $\phi_d^{\rm AT}$.
}
\label{gr_BP}
\end{figure}
%%%%%%%%%%%%%%%%%%%%%%%%%%%%%%%%%%%%

Alternatively,
the stability of the RS solution can be studied by considering
the propagation of fluctuations $\avg{(\delta\psikj_c)^2}$ under the 
recursion relation \req{eq_recurDisorder} \cite{thouless1986}.
This leads to the Almeida-Thouless (AT) stability condition,
\begin{eqnarray}
\label{eq_AT}
	\frac{\avg{(\delta\psiji_c)^2}}{\avg{(\delta\psikj_c)^2}}
	= (K-1)\phi_d\avg{(\psi_s)^2}^{K-2}\le 1.
\end{eqnarray}
In the RS regime,
$\avg{(\psi_s)^2}=\avg{\eta_s}$ since $\psiji_s=0$ or 1.
The AT stability condition is thus equivalent to \req{eq_etaInstability}.

\begin{figure}
\centerline{\epsfig{figure=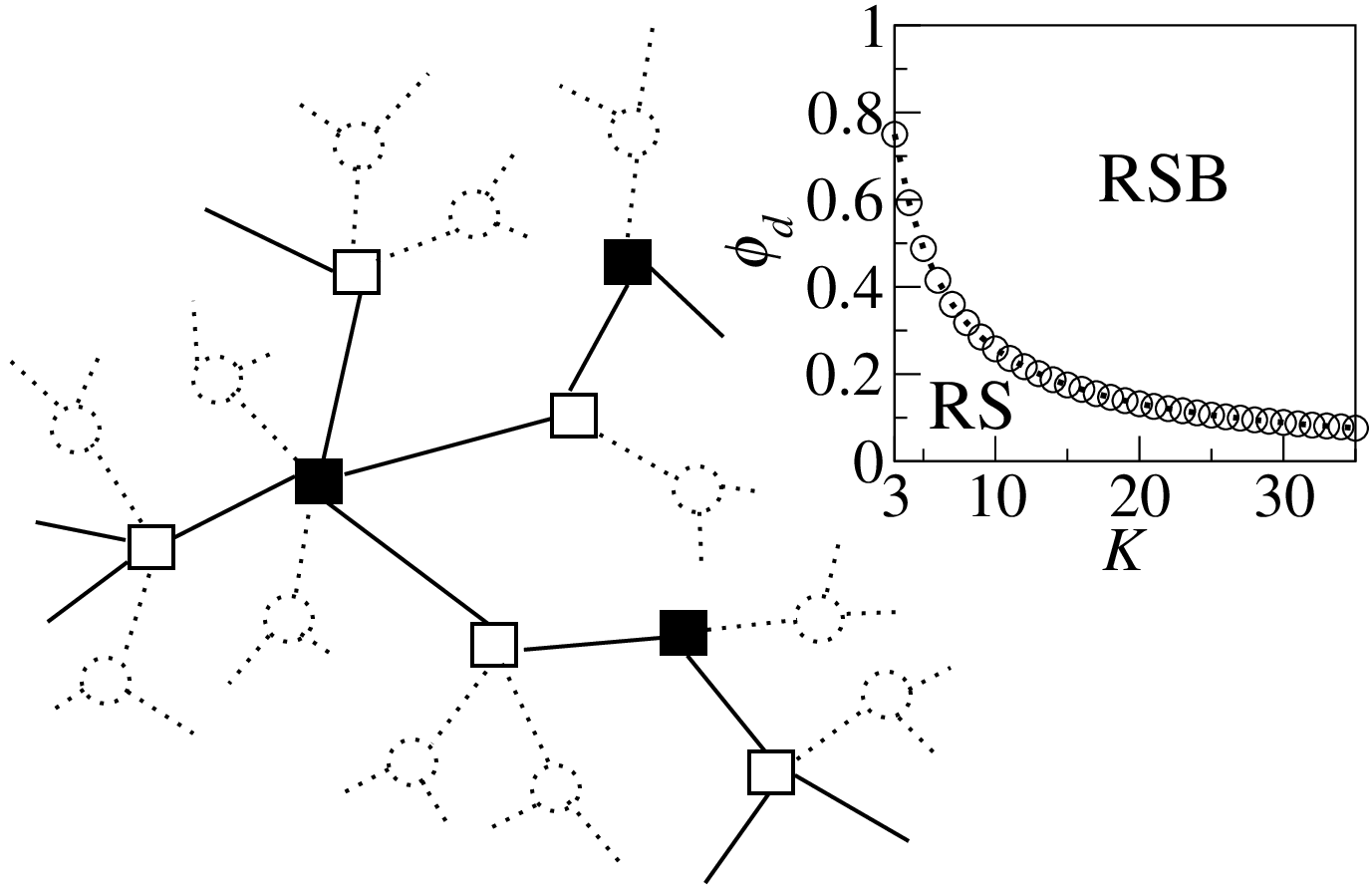, width=0.58\linewidth}}
\caption{
Sketch of the Poissonian subnetwork of deficient nodes
in the large connectivity limit with only a fraction $O(K^{-1})$
of deficient nodes (squares) and the rest being surplus nodes (circles).
Filled and unfilled symbols represent consumer or source states respectively.
Inset:
The RS and RSB phases in the $K$-$\phi_d$ space.
}
\label{gr_phaseDiagram}
\end{figure}

The stable solution of $\avg{\eta_c}$, 
$\avg{\eta_s}$ and $\avg{\eta_g}$ for $K=3$
is shown in \fig{gr_BP}(a).
The RS solution becomes unstable when $\phi_d>0.75$
for $K=3$.
This critical value agrees with those found 
in the simulation of the average energy and the fraction of soft nodes.
For general values of $K$, simple algebra leads to the AT line 
\begin{eqnarray}
\label{eq_ATline}
	\phi_d^{\rm AT}=\frac{K^{K-2}}{(K-1)^{K-1}}
\end{eqnarray}
which separates the RS and the RSB phases in the space
as shown in the inset of \fig{gr_phaseDiagram}.
From Eqs. (\ref{eq_avgPsi}), 
(\ref{eq_freeNode}) and (\ref{eq_AT}), 
it can be shown that 
$\avg{\psi_s}=f_{\rm soft}^{\rm RS}=K^{-1}$ on the AT line.

In the large $K$ limit,
$\phi_d^{\rm AT}$ approaches $e/K$.
This result has an interesting connection with the vertex cover problem. 
Considering the covered set as the set of source nodes, 
all links involving surplus nodes are covered. 
The remaining links are those among the deficient nodes.
These deficient nodes have at least one neighbor being a deficient node.
Hence from \tab{tab_Enode}, 
their surplus node neighbors do not affect the states of the deficient nodes.
Rather,
their states are determined by the states of their
deficient node neighbors.
Thus,
the problem of minimizing the covered set size 
reduces to one that minimizes the subset size of covered nodes 
in the subnetwork of deficient nodes as sketched in \fig{gr_phaseDiagram}. 
In the large $K$ limit, 
this subnetwork has a Poissonian connectivity distribution 
with a mean $K\phi_d$. 
The result $K\phi_d^{\rm AT}=e$ agrees with the point of RS instability 
derived in~\cite{weigt2000, weigt2001}.

%%%%%%%%% Figure 11 %%%%%%%%%%%%%%%%%
%%%%%%%%%%%%%%%%%%%%%%%%%%%%%%%

Comparisons between the simulation results 
and the RS analytical results 
from Figs. \ref{gr_RSenergy} and \ref{gr_nBackbone}
have yielded evidence of an AT transition in real instances 
at $\phi_d=\phi_d^{\rm AT}$.
Here we provide two more supporting numerical experiments.
In the first experiment,
we consider the BP algorithm initialized with $\psiji_c=0,1$ 
for all $j\rightarrow i$.
As shown in Fig.~\ref{gr_BP}(a),
effectively all messages 
converge to their steady states in the RS regime.
However, 
a significant fraction of messages 
fluctuates between 0 and 1 when $\phi_d$ rises above $\phi_d^{\rm AT}$,
indicating the breakdown of the RS ansatz. 
It leads to the non-convergence of the BP algorithm on real instances.
As shown in \fig{gr_BP}(a), 
the fraction of non-converging BP messages 
has an excellent agreement with the theoretical values of $\avg{\eta_g}$. 
Consequently, improved algorithms are needed for ground state searching. 
Algorithmically,
decimation procedures, 
such as those used in the survey propagation (SP)
algorithm~\cite{mezard2002}, are required.
We have tested the BP algorithm with decimation
and obtained results with simulated energy lower than the GSAT algorithm,
which will be reported elsewhere.

In the second experiment,
we consider numerical iterations of \req{eq_recurDisorder}
using {\it population dynamics} \cite{wong2006,wong2007}.
We start with different sets of initial values of $\psiji_c=0,1$,
each set following the same sequence of random connections
among the nodes.
When the dynamics reaches the steady state,
we measure the variance
$\sigma^2_{j\rightarrow i}=\overline{(\psiji_c)^2}
-\left(\overline{\psiji_c}\right)^2$,
where the overline denotes the average over random initial conditions.
$\sigma_{j\rightarrow i}\approx 0$ implies that the cavity state of node $j$
is {\it independent} of the boundary condition of the tree represented
by the population dynamics,
and is always frozen in either $S$ or $C$ states.
$\sigma_{j\rightarrow i}>0$ implies that the cavity state of node $j$
is {\it dependent} on the boundary conditions
and shows a long range correlation.
The numerical results of $\sigma_{j\rightarrow i}$ averaged over
nodes are shown in \fig{gr_BP}(b) for $K=3,4,5$.
In the figure
$\avg{\sigma_{j\rightarrow i}}>0$ when $\phi_d>\phi_d^{\rm AT}$
with almost no dependence on the number of iterations in the population 
dynamics.

%%%%%%%%%%%%%%%%%%%%%%%%%%%%%%%%%%%%%%%%%
\section{The One-step Replica Symmetry Breaking Ansatz}
\label{sec_RSB}

\subsection{The 1RSB Formalism}

In the RSB formalism,
the network behavior is explained by the dominance of numerous states
in the configuration space,
instead of a single ground state in the RS formalism.
Here,
we consider the 1RSB ansatz \cite{mezard2001, mezard2003}
where the density of states with energy $\sfe=\cE/N$ per node is assumed
to have the form 
${\cal N}_N(\sfe)=\exp[N\Sigma(\sfe)]$,
for a network of $N$ nodes and total energy $\cE$.
$\Sigma(\sfe)$ is called the {\it complexity} 
or {\it configurational entropy}. 
For small changes in the average energy,
we can write
$\Sigma(\sfe)=x(\sfe-\sfe^R)$
where $\sfe^R$ is the reference energy.
This assumption leads to a recursion 
for the {\it cavity probability functional} $P_j[E^V_j]$
\begin{eqnarray}
\label{eq_recurPEV}
	&&P_j[E^V_j(y_j)] = \frac{1}{\Xi}\prod_{\kinjmi}
	\bigg(\int \cD E^V_k(y_k) P_k[E^V_k(y_k)]\bigg)
	\nonumber\\
	&&\times\prod_{y_j}\left(\delta[E^V_j(y_j) 
	- \CH(E^V_{1},\dots,E^V_{K-1};\Lambda_j,y_j)\right.
%	\nonumber\\
	+\Delta E_j(E^V_{1},\dots,E^V_{K-1};\Lambda_j)]
	\nonumber\\
	&&\left.\times\exp[-x\Delta E_j(E^V_{1},\dots,E^V_{K-1};\Lambda_j)]
	\right)
\end{eqnarray}
where $\Xi$ is the normalization constant.
We now define the right hand side of \req{eq_recurPEV} to be $\CH_P$ and
the recursion can be represented by 
$P_j[E^V_j] = \CH_P(P_{k=1},\dots,P_{c-1}, E^V_j; \Lambda_j,x)$.
Solving the recursion in \req{eq_recurPEV} by population dynamics 
is equivalent to solving for a stable functional distribution $\cQ$ in
\begin{eqnarray}
\label{eq_rsbpp}
	&&\cQ[P_j] = \int d\Lambda_j\rho(\Lambda_j)\prod_{k=1}^{K-1}
	\int \cD P_k\cQ[P_k]
	\\
	&&\times\prod_{E^V_j}\delta\{P_j[E^V_j] 
	- \CH_P(P_{k=1},\dots,P_{c-1}, E^V_j; \Lambda_j, x)\},
	\nonumber
\end{eqnarray}
which is an analogy to the RS case of solving for $\sP[E^V]$ in \req{eq_RSPE}.
Note that the RS recursions in \req{eq_recurEV} correspond to 
the recursions of the cavity energy functions $E^V$
yielding a solution of the functional $P[E^V]$,
while the 1RSB recursions in \req{eq_recurPEV} correspond to recursions of 
the functional probability $P[E^V]$ 
yielding a solution of the probability functional $Q[P]$.

To analyze the physical properties of the network,
we write the partition function when $K$ cavity probability functional 
feeds a central node.
\begin{eqnarray}
\label{eq_XiNode}
	\Xi_{\rm node} = \prod_{\jini}
	\bigg(\int \cD E^V_j P_j[E^V_j]\bigg)
%	\nonumber\\
	\exp[-x\Delta 
	\cE_{\rm Node}(E^V_{1},\dots, E^V_{K};\Lambda_i)].
\end{eqnarray}
with $\Delta \cE_{\rm node}$ given by \req{eq_cdeltaE}.
Similarly,
the partition function obtained by bridging two trees with a link is
\begin{eqnarray}
\label{eq_XiLink}
	\Xi_{\rm link}
	=\int \cD E^V_L \cD E^V_R P_L[E^V_L]  P_R[E^V_R]
%	\nonumber\\
	\exp[-x\Delta\cE_{\rm link}(E^V_L, E^V_R)].
\end{eqnarray}
The average configuration free energy is given by 
\begin{eqnarray}
\label{eq_phi}
	\Phi(x) = -\frac{1}{x}\bigg(\avg{\ln\Xi_{\rm node}} 
	- \frac{K}{2}\avg{\ln\Xi_{\rm link}}\bigg)
\end{eqnarray}
The averages $\avg{\dots}$ are taken over $\Lambda_i$ and $P_j$ from the
distribution $\cQ[P_j]$.
$\Phi$ is related to the complexity $\Sigma$ and the average energy $\sfe$ by
\begin{eqnarray}
	x\Phi=x\sfe-\Sigma,
\end{eqnarray}
with $\partial\Sigma/\partial\sfe=x$.
$\Sigma$ and $\sfe$ are parametrically dependent on $x$ via
\begin{eqnarray}
\label{eq_sigma}
	\Sigma = x^2\frac{\partial\Phi}{\partial x},
	\\
	\sfe = \frac{\partial (x\Phi)}{\partial x}.
\end{eqnarray}

While solving for $P[E^V]$ in \req{eq_recurPEV} is in general difficult,
simple solutions can be obtained 
if a closed set of countably many $E^V$ is sufficient to 
describe the recursions of $E^V$.
The singlet regime in which the $S$ and $R$ states
form a closed set is a good example.
We emphasize,
however,
that the techniques are generally applicable to regions 
beyond the singlet regime where closed sets of $E^V$ are found,
such as the commensurate point in the doublet regime 
discussed in \ref{sec_closeSetDoubleSat}.

\subsection{The 1RSB Solution}
\label{sec_1RSBsing}

In the singlet regime.
there are only two representative states,
$C$ and $S$ states,
we  parametrize $P[E^V]$ as
\begin{eqnarray}
\label{eq_Pev}
	P_j[E^V_j] = \psi^j_c\delta(E^V_j - E^V_c) 
	+ \psi^j_s\delta(E^V_j - E^V_s).
\end{eqnarray}
Using \tab{tab_recurUSB} to obtain $\Delta E$ for different combinations
of $C$ and $S$ states,
\req{eq_recurPEV} can be simplified to
\begin{eqnarray}
\label{eq_recurSing1RSB}
	\psiji_c =&& \delta_{\Lambda_j, -1}\frac{e^{{-xu^2}/{2}}}
	{\cZ^{j\to i}}\prod_{\kinjmi}\psikj_s,
	\nonumber\\
	\psiji_s =&& 
	\delta_{\Lambda_j, -1}\frac{e^{{-xu^2}/{2}} }
	{\cZ^{j\to i}}\bigg[\prod_{\kinjmi}(\psikj_c e^{x\gamma}+\psikj_s)
	%\nonumber\\
	-\prod_{\kinjmi}\psikj_s\bigg]+\delta_{\Lambda_j, A},
	\nonumber\\
	\cZ^{j\to i} =&& e^{{-xu^2}/{2}}\prod_{\kinjmi}
	(\psikj_c e^{x\gamma}+\psikj_s).
\end{eqnarray}
The above recursions of $\psiji_c$ are gross simplifications of 
the recursions of the functional probabilities in \req{eq_recurPEV}.
These equations can be solved by population dynamics involving 
a pool of values of $\psiji_c$ with $0\le\psiji_c\le 1$.
Alternatively,
the solution to the 1RSB recursion can be found by directly
solving for the distribution $Q(\psi_c)$ in \req{eq_rsbpp}
which is isomorphic to $\cQ[P]$ in \req{eq_rsbpp}.

%%%%%%%%% Figure 12 %%%%%%%%%%%%%%%%%
\begin{figure}
\centerline{\epsfig{figure=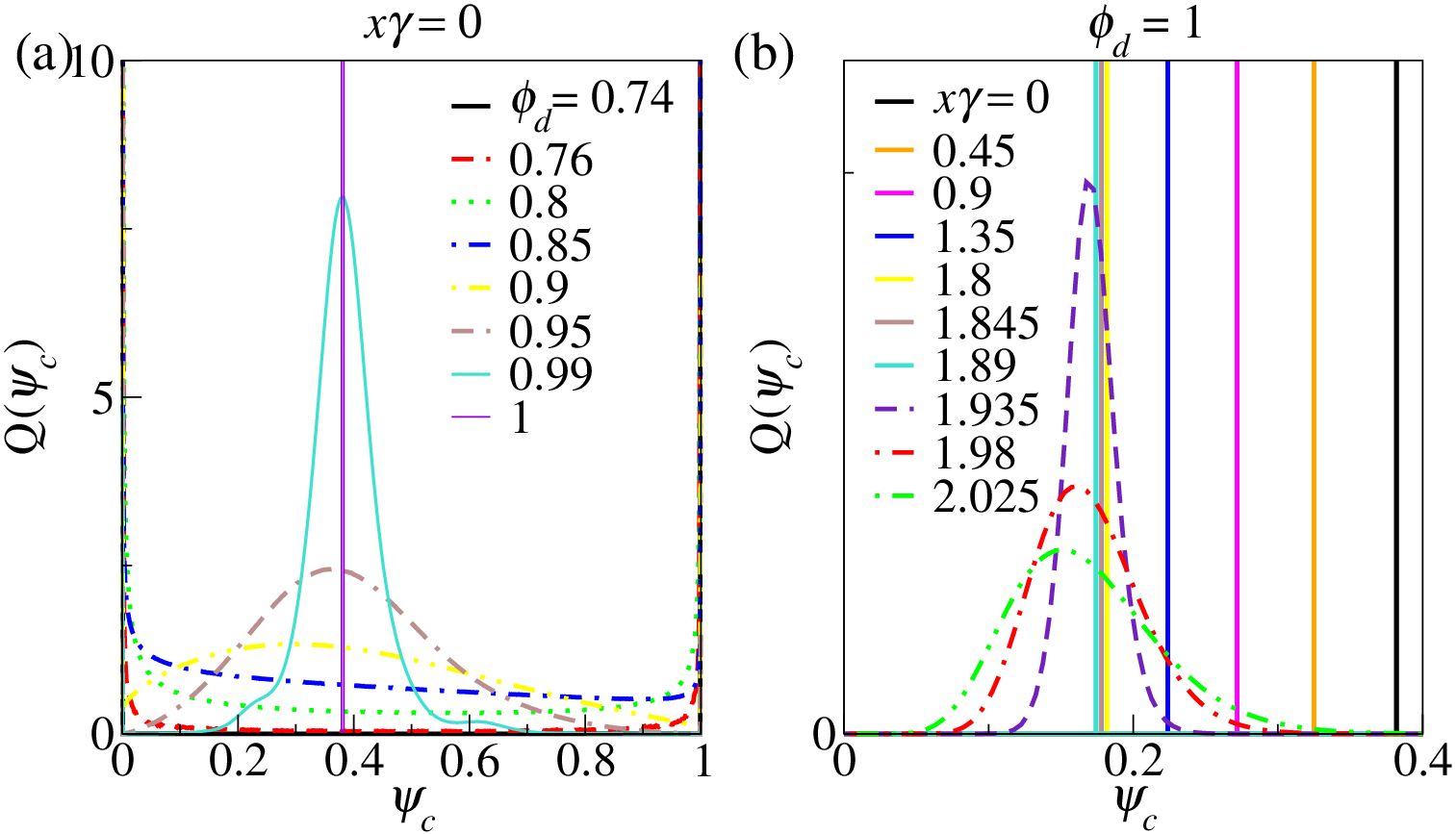, width=0.7\linewidth}}
\caption{
(a) The stable solution of $Q(\psi_c)$ obtained by solving \req{eq_rsbpp}
in the singlet regime using \req{eq_recurSing1RSB}
with $K=3$ and $x=0$.
(b) The stable solution of $Q(\psi_c)$ 
with $K=3$ and $\phi_d=1$.
}
\label{gr_Qps}
\end{figure}
%%%%%%%%%%%%%%%%%%%%%%%%%%%%%%%%%%%%%

The stable solution of $Q(\psi_c)$ for $K=3$ and $x=0$ is shown in 
\fig{gr_Qps}(a).
When $\phi_d\le\phi_d^{\rm AT}$,
there are no fractional components of $\psi_c$,
and $Q(\psi_c) = \avg{\psi_c}\delta(\psi_c-1)+(1-\avg{\psi_c})\delta(\psi_c)$.
When $\phi_d>\phi_d^{\rm AT}$,
non-zero components of $0<\psi_c<1$ exist.
This agrees with the result in Section \ref{sec_RSinstability} that in this
regime,
the RS solution is an unstable solution of \req{eq_rsbpp}.
When $\phi_d=1$,
there is no disorder in the capacities. 
All vertices are identical and $Q(\psi_c)=\delta(\psi_c-\avg{\psi_c})$ where 
$\avg{\psi_c}$ is given by the RS equation (\ref{eq_avgPsi}).
This means that among the different states of the system,
all vertices are equally probable to be in $C$ state with probability
$\avg{\psi_c}$.
The stable solutions of $Q(\psi_c)$ are dependent on $x$ 
via the factor $x\gamma$.
In the case of $\phi_d=1$,
the dependence of $Q(\psi_c)$ on $x$ is shown in \fig{gr_Qps}(b).
The position of the delta peak at 
$\avg{\psi_c}=(1-\avg{\psi_c})^{K-1}/[(e^{x\gamma}-1)\avg{\psi_c}+1]^{K-1}$ 
shifts to left from $x\gamma=0$ to $x\gamma=\xei\gamma=1.92$.
For $x>\xei$ in \fig{gr_Qps}(b), 
$Q(\psi_c)$ becomes a continuous distribution,
indicating the instability of the 1RSB ansatz
to be discussed in the next subsection.

\begin{figure}
\centerline{\epsfig{figure=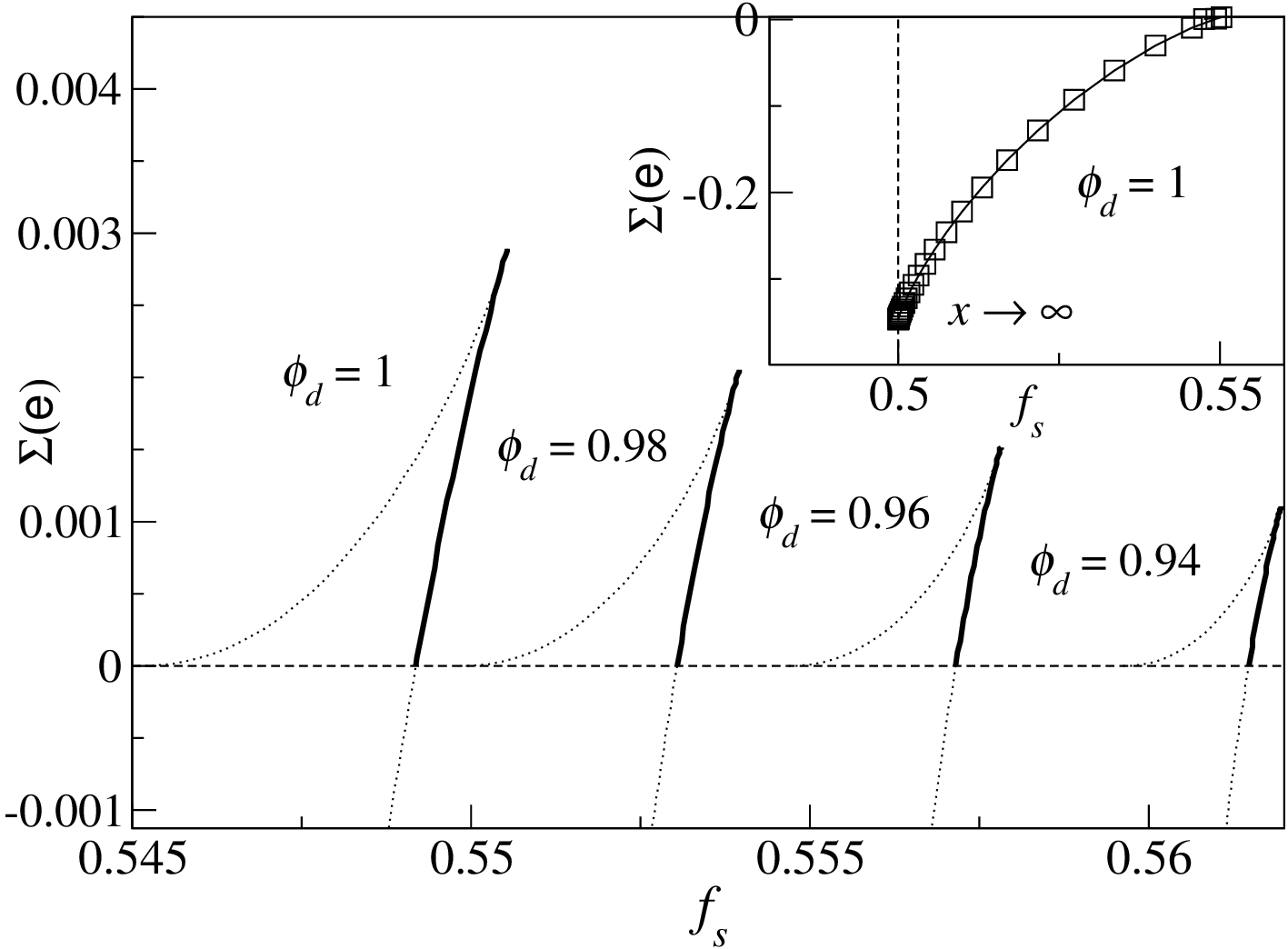, width=0.65\linewidth}}
\caption{
The 1RSB complexity $\Sigma(\sfe)$ obtained by numerically
solving for $Q(\psi_c)$ with $K=3$ and $\phi_d=0.94, 0.96, 0.98, 1$.
Inset: $\Sigma(\sfe)$ for $K=3$ and $\phi_d=1$ 
obtained from the ansatz $Q(\psi_c)=\delta(\psi_c-\avg{\psi_c})$.
It approaches the modulation limit at $f_s=0.5$ as $x\rightarrow\infty$.
Symbols are spaced at intervals of 0.9 in $x\gamma$.
}
\label{gr_complexity}
\end{figure}

With the stable solution of $Q(\psi_c)$,
we derive the complexity $\Sigma(\sfe)$ 
in the singlet regime.
Obtaining $\Delta \cE_{\rm node}$
in \tab{tab_Enode},
we write the partition function $\Xi_{\rm node}$ in \req{eq_XiNode} as
\begin{eqnarray}
\label{eq_XiNodeSingSat}
	\Xi_{\rm node}
	=&& \delta_{\Lambda_j, -1}\bigg[\prod_{\jini}
	(\psiji_c e^{x\gamma} + \psiji_s) 
%	\nonumber\\
	+ (e^{x\gamma}-1)\prod_{\jini}\psiji_s\bigg]e^{-xu^2/2}
	\nonumber\\
	&&+\delta_{\Lambda_j, A}\bigg[\prod_{\jini}
	(\psiji_c e^{x\gamma} + \psiji_s)\bigg].
\end{eqnarray}
With $\Delta E_{\rm link}$ from \tab{tab_Elink},
we write $\Xi_{\rm link}$ as
\begin{eqnarray}
\label{eq_XiLinkSingSat}
	\Xi_{\rm link}= \psi^{L\rightarrow R}_s \psi^{R\rightarrow L}_s
	+(\psi^{L\rightarrow R}_s \psi^{R\rightarrow L}_c
%	\nonumber\\
	+ \psi^{L\rightarrow R}_c \psi^{R\rightarrow L}_s)e^{x\gamma} 
	+ \psi^{L\rightarrow R}_c \psi^{R\rightarrow L}_c e^{x\gamma}.
	\nonumber\\
\end{eqnarray}
The configurational free energy $\Phi$ is given by \req{eq_phi}
with $\psiji_s$ averaged over $Q(\psi_c)$.
The complexity $\Sigma(\sfe)$ obtained is shown in \fig{gr_complexity}.
Again,
$\sfe$ is expressed in terms of the fraction of source nodes 
through the relation $f_s = [\sfe-1/2K+(1-\phi_d) u^2/2]/\gamma$.

%%%%%%%%% Figure 13 %%%%%%%%%%%%%%%%%
%%%%%%%%%%%%%%%%%%%%%%%%%%%%%%%%%%%%%

Generally,
we identify three segments on the complexity curve: 
(i) the unphysical segment (the dotted segment with $\Sigma\ge 0$),
(ii) the physical segment (the solid segment with $\Sigma\ge 0$), and
(iii) the highly unlikely segment (the dotted segment with $\Sigma< 0$).
The physical segment of $\Sigma$ is related to the 
number $\cN$ of metastable states with energy $\sfe$.
We denote the values of $x$ when $\Sigma=0$ and $\Sigma$ 
is maximum as $x_s$ and $x_d$ respectively.
$\sfe(x_s)$ corresponds to the lowest energy 
among the states with non-vanishing complexity,
which is considered as the ground state in the picture of 1RSB.
$\sfe(x_d)$ corresponds to the energy of the states 
with the highest complexity,
which is believed to be the states where search algorithms get trapped,
giving rise to dynamical transitions.
However,
recent work on the coloring problem shows 
that the efficacy of the BP algorithm
is not affected by the dynamical transition \cite{zdeborova2007}.
We leave this issue for future studies.

The segment of negative complexity  corresponds to states
with vanishing number in the thermodynamic limit.
Its physical meaning is clear in the limit $x\rightarrow\infty$,
which corresponds to a single state with lowest possible energy,
since the reweighting process allows only one state.
We show $\Sigma$ in the inset of \fig{gr_complexity} as $x\rightarrow\infty$.
The result is obtained from the solution of $Q(\psi_c)$ 
restricted to be the the 1RSB solution in 
the form of $Q(\psi_c)=\delta(\psi_c-\avg{\psi_c})$,
i.e. the unstable solution when $x>\xei$.
The complexity curve approaches
the limit of $\psi_c=0.5$,
which corresponds to the highly unlikely structure of networks with
alternating layers of consumer and source nodes.

\subsection{Comparison with Real Instances}
\label{sec_decimatedBP}

Though successful in finding low lying states on real instances,
the GSAT algorithm we described in Section \ref{sec_simulation} 
requires long computation time for large systems.
To compare the predictions of RS and 1RSB approximations with real instances, 
we employ the BP with decimation.
As the BP does not converge in the RSB phase,
we measure the time average of the fluctuating messages $\psiji_c$
and evaluate for each node the time average probability 
of the full $\cS$-states.
The node with the highest $\psi^i_\cS$ is fixed to be resource providing.
Only source nodes are decimated, 
as bistable nodes should be left undecimated, 
and decimating a consumer node will fix its neighbors simultaneously, 
which may hinder the convergence of the BP.
By repeating the above procedure,
the BP messages finally converge and the full state of all nodes
are determined.

%%%%%%%%% Figure 14 %%%%%%%%%%%%%%%%%
\begin{figure}
\centerline{\epsfig{figure=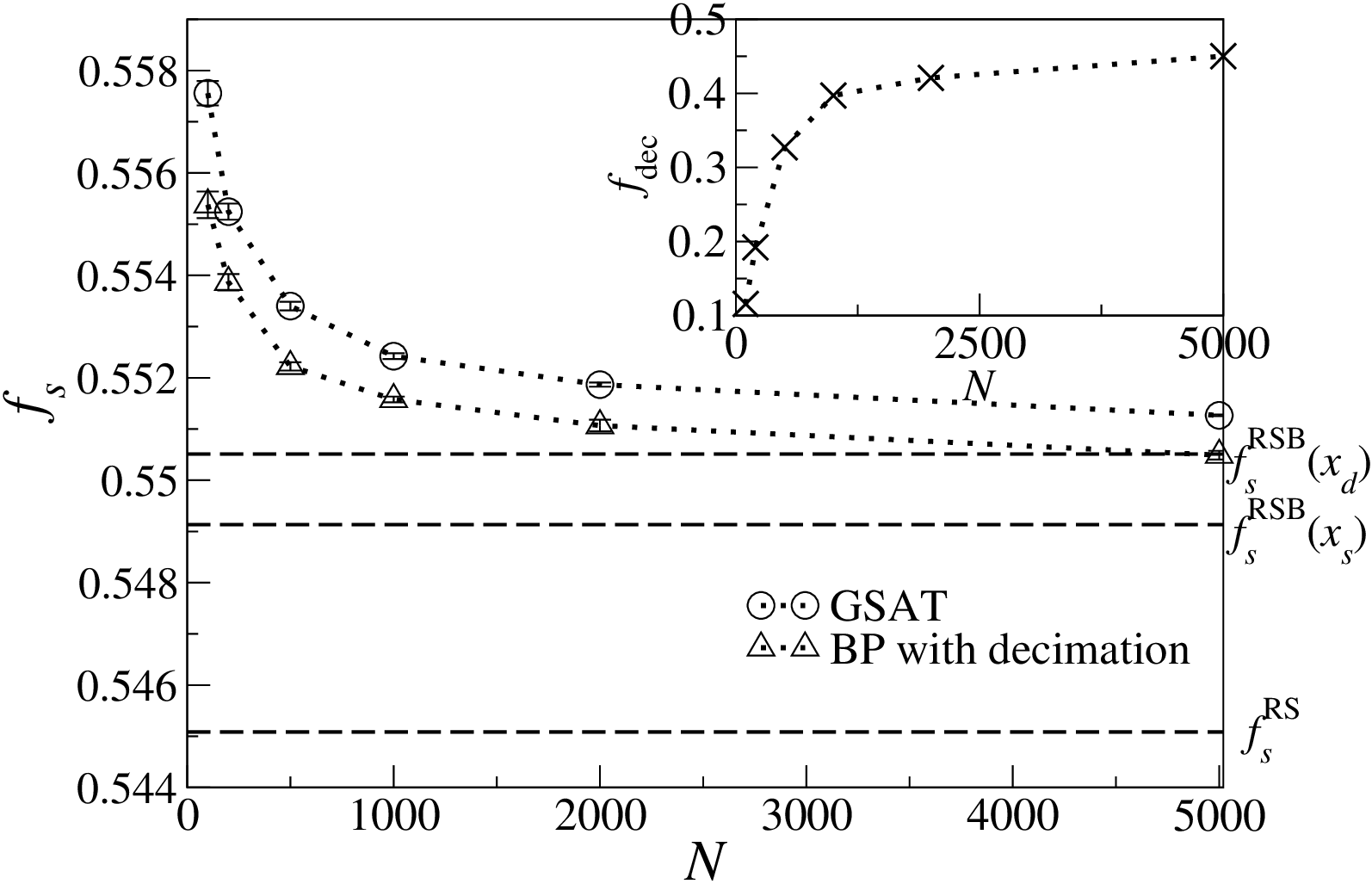, width=0.65\linewidth}}
\caption{
The optimized $f_s$ on real instances as a function of $N$
with $K=3$,
as obtained by the GSAT algorithm ($\circ$) and the BP algorithm with decimation ($\triangle$).
The horizontal dashed lines show the analytical results of the RS ansatz,
$f_s(x_s)$ and $f_s(x_d)$ in the 1RSB ansatz.
Inset: the fraction $f_{\rm dec}$ of decimated nodes before convergence.
}
\label{gr_decimatedBP}
\end{figure}
%%%%%%%%%%%%%%%%%%%%%%%%%%%%%%%%%%%%%

Figure \ref{gr_decimatedBP} shows
that lower simulated energy can be obtained 
from the BP with decimation as compared with the GSAT algorithm.
The simulated energy approaches the $\sfe(x_d)$ as obtained by the 1RSB ansatz.
However,
the fraction of decimated nodes before convergence 
increases with system size as shown in the inset of \fig{gr_decimatedBP}.
This greatly increases the computational time for large systems 
if only a single node is decimated at a time.
Several nodes can be decimated simultaneously
to shorten the computation time,
with a tradeoff in energy.

\subsection{The instability of 1RSB formalism}
\label{sec_1RSBvalidity}

\begin{figure}
\centerline{\epsfig{figure=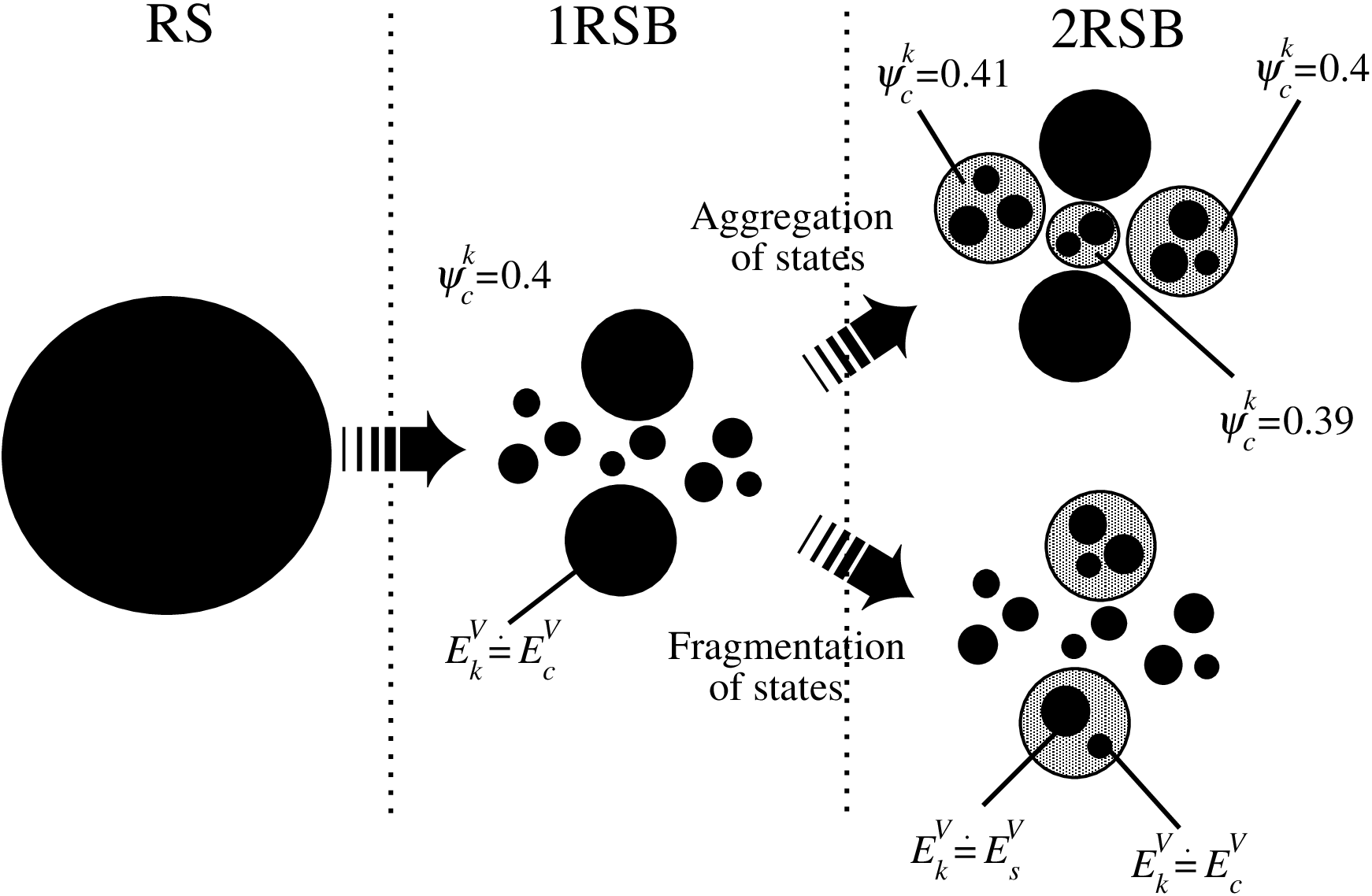, width=0.7\linewidth}}
\caption{
The schematic picture of the state space in the RS,
1RSB and 2RSB phase.
Black circles represent singly connected states.
Grey circles represent clusters of states.
$k$ is an arbitrary vertex in the network.
Its probabilities $\psi^k_c$ of the $C$ states in the 1RSB phase,
and several clusters in the 2RSB phase,
are shown in the figure
to illustrate the aggregation of states.
Its cavity energy $E^V_k$ in an arbitrary state in 
the 1RSB phase,
and several of its fragmented states in the 1RSB phases,
are also shown in the figure to illustrate the fragmentation of states.
The values of $\psi^k_c$ are assigned for the purpose of illustration only.
}
\label{gr_2RSB}
\end{figure}

To test the stability of the 1RSB ansatz against further steps of RSB,
we consider two kinds of instability that leads to the two-step RSB (2RSB)
formalism \cite{montanari2003}.
They are the so-called type I and type II instabilities,
corresponding to the aggregation of states and fragmentation of states 
respectively,
as shown schematically in \fig{gr_2RSB}.
In type I instability,
metastable states aggregate in clusters and the 1RSB ansatz is valid
in each cluster.
The whole state space is composed of clusters 
and the 2RSB formalism is required
to describe the state space structure.
In type II instability,
some states split to form clusters of states instead of single states.
The 1RSB ansatz is valid inside the clusters 
while 2RSB ansatz is required to describe the
state space structure.
It is generally believed that once 1RSB is not stable,
the {\it full} RSB is required to describe the system,
as illustrated by the 1RSB instability found in graph coloring
\cite{krzakala2004} and $K$-satisfiability problems \cite{montanari2004}.
Here we focus on the case of $K=3$ in the 1RSB regime 
with $\phi_d>\phi_d^{\rm AT}$,
following the approach of \cite{krzakala2004}.

%%%%%%%%% Figure 15 %%%%%%%%%%%%%%%%%
%%%%%%%%%%%%%%%%%%%%%%%%%%%%%%%%%%%%%%

\subsubsection{Type I instability: Aggregation of states}

In the 1RSB phase,
each vertex $k$ is characterized 
by the cavity probability functional $P_k[E^V]$.
This characterization is much simplified 
due to the small closed set of states in the singlet regime,
rendering it isomorphic to single values of $\psi^k_c$.
In the 2RSB formalism,
each vertex $k$ is characterized by the probability functional of $P_k$.
For the closed set of states in the singlet regime,
this characterization is isomorphic to the distribution $Q_k(\psi^k_c)$.
In the example illustrated in \fig{gr_2RSB},
the probability $\psi^k_c=0.4$ is originally uniform in the state
space of the 1RSB phase.
When it enters the 2RSB phase,
$\psi^k_c$ starts to take up diversified values of $\psi^k_c=0.39,0.40,0.41$
in three different clusters of states.

To examine type I instability,
we test for the possible spreading in $\psiji_c$ 
by introducing small fluctuations in $\psikj_c$ of the descendents.
We define $T_{k\rightarrow j}(x)$ to be the derivative
\begin{eqnarray}
\label{eq_1RSB1stab}
	T_{k\rightarrow j}(x)
	=\bigg|\frac{\partial \psiji_c}{\partial \psikj_c}\bigg|_{\rm 1RSB}
\end{eqnarray}
from \req{eq_recurSing1RSB},
which is evaluated at the values of $\psikj_c$ from 
the 1RSB solution.
The propagation of noise is thus given by
\begin{eqnarray}
	(\delta \psiji_c)^2 
	= \sum_{\kinjmi}[T_{k\rightarrow j}(x)]^2(\delta \psikj_c)^2.
\end{eqnarray}
In the thermodynamic limit,
we consider a tree structure with $L$ generations.
The noise at the $L^{\rm th}$ generation vanishes if
$(T_{L-1\rightarrow L}T_{L-2\rightarrow L-1}\dots T_{0\rightarrow 1})^2 
\rightarrow 0$ as $L\rightarrow\infty$.
Hence we measure the quantity
\begin{eqnarray}
	\li_L(x) 
	= (K-1)^L(T_{L-1\rightarrow L}T_{L-2\rightarrow L-1}\dots 
	T_{0\rightarrow 1})^2,
\end{eqnarray}
averaged over the quenched disorder and boundary conditions of the trees.
This can be implemented by iterating \req{eq_recurSing1RSB} 
using population dynamics 
and evaluating the corresponding $\li_L(x)$ in each iteration.
Alternatively,
we solve for the distribution $\cP_L[\li_L, \psi_c^L]$ 
at layer $L$ by the recursion relation
\begin{eqnarray}
\label{eq_calPI}
	&&\cP_{L+1}[\li_{L+1}, \psi_c^{L+1}] 
%	\nonumber\\
	= \int d\li_L \int d \psi_c^{L} \cP_L[\li_L, \psi_c^L] 
	\prod_{k=2}^{K-1}\int d\psi_c^k Q(\psi_c^k)
	\nonumber\\
	&&\bigg\{\phi_d\delta[\li_{L+1} - (T_{L\rightarrow L+1})^2\li_L]
%	\nonumber\\
	\delta\bigg[\psi_c^{L+1} - \frac{e^{-xu^2/2}}{\cZ^{j\to i}}
	(1-\psi_c^L)\prod_{k=2}^{K-1}(1-\psi_c^k)\bigg] 
	\nonumber\\
	&&+\phi_s\delta(\li_{L+1})\delta(\psi_c^{L+1})\bigg\}
\end{eqnarray}
with the initial condition
\begin{eqnarray}
	P_1[\li_1, \psi^1_c]\equiv Q(\psi^1_c)\delta(\li_1-1) 
\end{eqnarray}
for $L=1$.
We note that the case of $\phi_d=1$ 
is a special case where no disorder is present
and $T_{0\rightarrow 1}=\dots =T_{L-1\rightarrow L}$ 
since $Q(\psi^k_c)=\delta(\psi^k_c-\avg{\psi^k_c})$ for all $k$.
Thus for $\phi_d=1$,
$\avg{\lambda^{(I)}_1(x)}\ge 1$ is sufficient 
to show that 1RSB solution is unstable at $x$.
To evaluate $\avg{\li_L(x)}$ for general values of $\phi_d<1$,
solving \req{eq_rsbpp} gives more reliable results than population dynamics.
This is because for large $L$ and nonvanishing values 
of $\phi_s$,
such as those close to $\phi_s^{\rm AT}$,
the presence of the factor 
$\prod_{l=1}^{L}\delta_{\Lambda_l,A}$ in $\li_L(x)$
requires an extremely large population for a finite fraction of nonzero $\li_L(x)$
in the pool of population dynamics.

We show $\avg{\li_L(x)}$ as a function of $L$ in \fig{gr_eigen}(a)
for $\phi_d=0.9$ from $x\gamma=0$ to $x\gamma=4.5$,
and define $\xei$ by 
\begin{eqnarray}
\label{eq_xei}
	\lim_{L\rightarrow\infty}
	\frac{d \log\avg{\li_L(x)}}{d L}\bigg|_{x=\xei} = 1.
\end{eqnarray}
From \fig{gr_eigen}(a),
the 1RSB solution for $\phi_d=0.9$ is stable against type I instability 
when $x\gamma<\xei\gamma\approx 3.07$.

\subsubsection{Type II instability: Fragmentation of states} 

In the 1RSB phase,
vertex $k$ is characterized by a cavity energy function $E^V_k$
in each single state.
In the 2RSB phase,
some 1RSB single states split into different states in which
some vertices are characterized by more than one cavity energy functions.
In the example illustrated in \fig{gr_2RSB},
a state with $E^V_k=E^V_c$ is fragmented to states with 
$E^V_k=E^V_c$ and $E^V_k=E^V_s$ respectively
on entering the 2RSB phase.

As the cavity energy functions of some vertices are modified 
during fragmentation,
we examine the probability of changes in $E^V_i$ of a
node $i$ due to changes in $E^V_j$
among its descendent nodes $j$,
the so-called proliferation of bugs \cite{krzakala2004}.
In the singlet regime,
we denote $\pi_{s\rightarrow c}^{k\rightarrow j}$
as the joint probability that vertex $k\rightarrow j$ is in the \Sst 
in the absence of bugs,
and in the \Cst in the presence of a small number of bugs.
From the recursion \req{eq_recurSing1RSB},
we note that contributions to $\pi_{c\rightarrow s}^{j\rightarrow i}$
come from the case that all descendents of $j$ are in the \Sst,
and one of them changes to \Cst in the presence of bugs.
Hence
\begin{eqnarray}
\label{eq_recurPi}
	\pi^{j\rightarrow i}_{c\rightarrow s} 
	&&= \delta_{\Lambda_j, -1}\frac{e^{{-xu^2}/{2}}}{\cZ^{j\to i}}
	\sum_{\kinjmi}\pi^{k\rightarrow j}_{s\rightarrow c}
	\prod_{\linjmik}\psi^{l\rightarrow j}_s e^{x\gamma}.
\end{eqnarray}
Similarly,
contributions to $\pi_{s\rightarrow c}^{j\rightarrow i}$ come from the case
that only one descendent of $j$ is in the \Cst 
which changes to \Sst in the presence of bugs.
Hence
\begin{eqnarray}
	\pi^{j\rightarrow i}_{s\rightarrow c} 
	&&= \delta_{\Lambda_j, -1}\frac{e^{{-xu^2}/{2}} }{\cZ^{j\to i}}
	\sum_{\kinjmi}\pi^{k\rightarrow j}_{c\rightarrow s}
	\prod_{\linjmik}\psi^{l\rightarrow j}_s.
\end{eqnarray}
We define the matrix $V_{k\rightarrow j}(x)$ to be
\begin{eqnarray}
\label{eq_V}
	V_{k\rightarrow j}(x)
	= \delta_{\Lambda_j, -1}\frac{e^{{-xu^2}/{2}}}{\cZ^{j\to i}}
%	\\
	\left(\begin{array}{cc}
	0 & \displaystyle\prod_{\linjmik}\psi^{l\rightarrow j}_s e^{x\gamma} \\
	\displaystyle\prod_{\linjmik}\psi^{l\rightarrow j}_s & 0
	\end{array}\right).
\end{eqnarray}
The instability of the 1RSB solution against fragmentation of states
can be thus examined by considering the maximum eigenvalue of 
the products of matrices $V$.
We measure the quantity
\begin{eqnarray}
\label{eq_calPII}
	\lii_L(x) = (K-1)^L \cI(V_{L-1\rightarrow L}V_{L-2\rightarrow L-1}
	\dots V_{0\rightarrow 1}),
\end{eqnarray}
averaged over quenched disorders,
where $\cI(\dots)$ is defined as 
the maximum eigenvalue of the matrix.
The 1RSB solution is stable against the type II instability 
if $\avg{\lii_L(x)}\rightarrow 0$ as $L\rightarrow\infty$. 
$\avg{\lii_L(x)}$ can be solved 
by population dynamics of \req{eq_recurSing1RSB}
or by solving for the distribution on $\cP_L[\lii_L, \psi_c^L]$
analogous to \req{eq_calPI}.

\begin{figure}
\centerline{\epsfig{figure=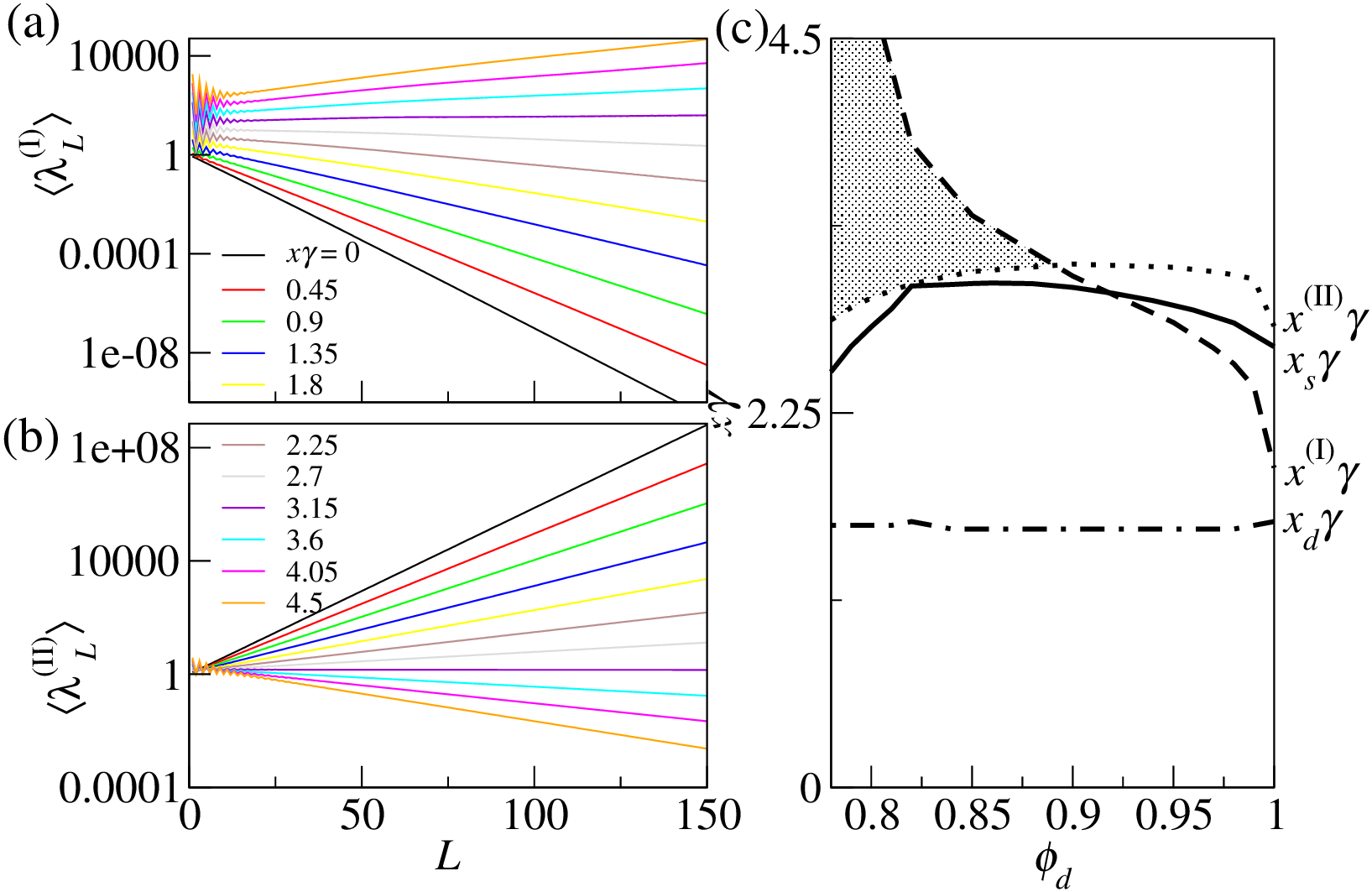, width=0.7\linewidth}}
\caption{
(a) $\avg{\li_L(x)}$ and (b) $\avg{\lii_L(x)}$ 
as a function of $L$ for systems with $K=3$ and $\phi_d=0.9$,
from $x\gamma=0$ to $x\gamma=4.5$.
(c) $\xei\gamma$, $\xeii\gamma$, $x_s\gamma$ and $x_d\gamma$ as a function of $\phi_d$ for $K=3$.
The shaded region corresponds to the stable range of the 1RSB ansatz.
}
\label{gr_eigen}
\end{figure}

We show $\avg{\lii_L(x)}$ as a function of $L$ in \fig{gr_eigen}(b)
for $\phi_d=0.9$ from $x\gamma=0$ to $x\gamma=4.5$,
and define $\xeii$ similarly as \req{eq_xei}.
From \fig{gr_eigen}(b),
the 1RSB solution for $\phi_d=0.9$ is stable against type II instability 
when $x\gamma<\xeii\gamma\approx 2.93$.

%%%%%%%%% Figure 16 %%%%%%%%%%%%%%%%%
%%%%%%%%%%%%%%%%%%%%%%%%%%%%%%%%%%%%%

$\xei$ and $\xeii$ from $\phi_d=0.78$ to $1$,
together with $x_s$ and $x_d$ obtained from the complexity curve,
are plotted in \fig{gr_eigen}(c) as a function of $\phi_d$.
Reliable results for $\phi_d\approx \phi_d^{\rm AT}=0.75$ 
are difficult to obtain
as the continuous component in $Q(\psi_c)$ becomes extremely small 
(see for instance, \fig{gr_Qps})
and extremely high precision is required.
For $\phi_d$ lower than $\approx 0.88$,
the 1RSB ansatz is stable in the range of $\xeii<x<\xei$,
corresponding to the shaded region in \fig{gr_eigen}(c).
Both $x_s$ and $x_d$ are found below the shaded region,
indicating the instability of the physical segment of $\Sigma$.
The stable range of $x$ lies on the negative segment of $\Sigma$,
which implies that states with vanishing probability
are stable in the 1RSB ansatz.
For $\phi_d$ higher than $\approx 0.88$,
the shaded region disappears and all 
complexity curves are unstable.
We thus conclude that the 1RSB quantities 
evaluated at $x_s$ and $x_d$ are unstable.
Nevertheless,
agreement with simulation  results show that the 1RSB ansatz
is a good approximation of the physical picture of the system.

%%%%%%%%%%%%%%%%%%%%%%%%%%%%%%%%%%%%%%%
\section{Conclusion}
\label{sec_conclusion}

In this paper,
we have studied the source location problem on transportation networks.
As the formulation involves continuous variables,
the cavity fields are represented by the cavity energy functions
which satisfy the piecewise quadratic ansatz.
The ansatz decomposes the cavity energy functions into composite
functions,
and effectively parametrizes them by
the energy minimum of each composite function.

This enables us to obtain a small closed set of cavity energy functions
in the singlet regime,
which greatly simplifies the functional RS recursions
to simple recursions of probabilities.
Physical results such as the average energy and the 
fraction of soft nodes are obtained
and have excellent agreement with simulations
when the fraction of deficient nodes is small.
We examined the stability of the RS solution and
derive the AT-line for the transition to the RSB phase.
In the high connectivity limit,
such results are consistent with the RS instability 
obtained in the vertex cover problem
for Poissonian graphs.

Though the solution of the 1RSB ansatz 
corresponds to a stable distribution of functionals,
which is in general infeasible to solve,
the closed sets of cavity energy functions greatly simplify
the 1RSB recursions which make the 1RSB solution feasible.
We remark that the analysis is applicable to regimes other than the singlet
regime of the system,
such as the commensurate points of the doublet regime
being considered in \ref{sec_closeSetDoubleSat}.
In other cases,
closed sets with a large number of functions are found
and the 1RSB solution may once again become computational
infeasible.
We expect that the present techniques are applicable
to other problems where closed sets of cavity fields exist.

\appendix

\section{The doublet Regime}
\label{sec_closeSetDoubleSat}

\subsection{The Closed Set of Cavity Energy Functions 
and the Simplified RS Recursion}

Apart from the closed set of cavity energy functions $E^V$ 
in the singlet regime,
a closed set with countable elements of $E^V$ can also be found 
in other regimes.
In the doublet regime,
we find that a small number of $E^V$ spans a closed set
at the commensurate points
\begin{eqnarray}
\label{eq_comensurate}
	\gamma = m\kappa,
\end{eqnarray}
with the rational number $m\ge 2$.
These commensurate points correspond to the 
values of 
\begin{eqnarray}
	u^{-1}=-\sqrt{\frac{K(K-1)(m-1)}{(m-1)K+m+1}},
\end{eqnarray}
at which abrupt jumps in the fraction of source nodes 
are found in the range $\sqrt{21/25}<u^{-1}<\sqrt{3/2}$ in \fig{gr_energy} 
for $K=3$.
Each value of rational $m$ corresponds to a switch of energetic stability
from one configuration of consumer nodes to another.

%%%%%%%%% Figure 17 %%%%%%%%%%%%%%%%%
\begin{figure}
%\leftline{\epsfig{figure=1.eps, width=0.31\linewidth}
%\leftline{\epsfig{figure=7.eps, width=0.31\linewidth}
%\leftline{\epsfig{figure=4.eps, width=0.31\linewidth}
%}}}
%\leftline{
%\epsfig{figure=2.eps, width=0.31\linewidth}
%\leftline{\epsfig{figure=3.eps, width=0.31\linewidth}
%\leftline{\epsfig{figure=5.eps, width=0.31\linewidth}
%}}}
%\leftline{\epsfig{figure=6.eps, width=0.31\linewidth}
%\leftline{\epsfig{figure=8.eps, width=0.31\linewidth}
%}}
\leftline{
\hspace{1cm}
\epsfig{figure=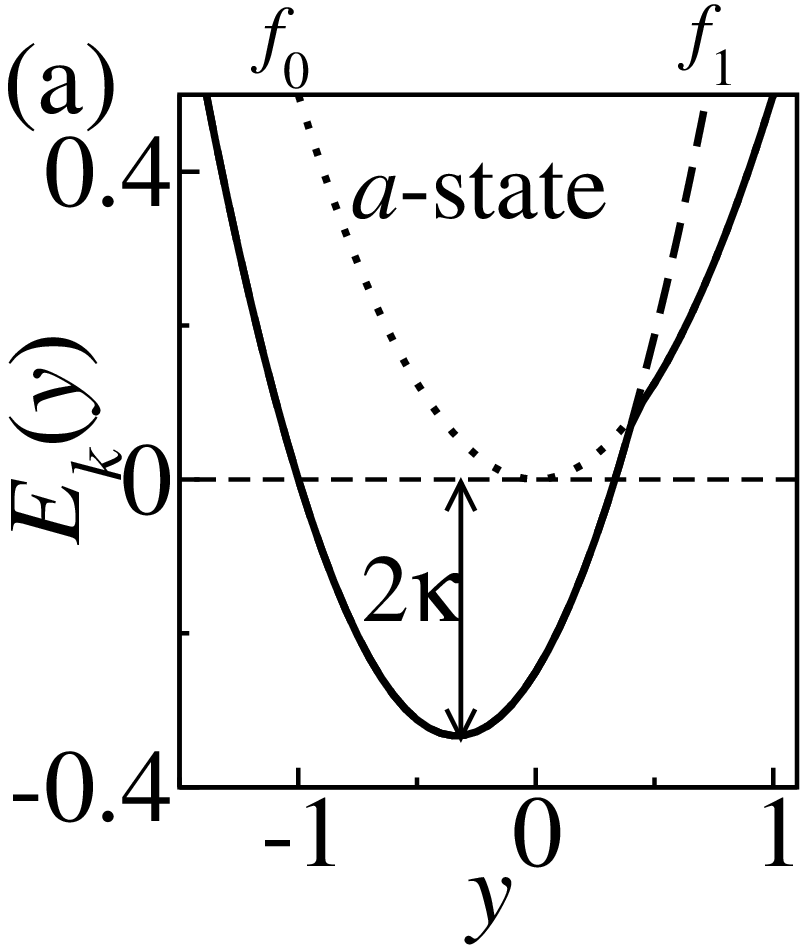, width=0.21\linewidth}
\leftline{\epsfig{figure=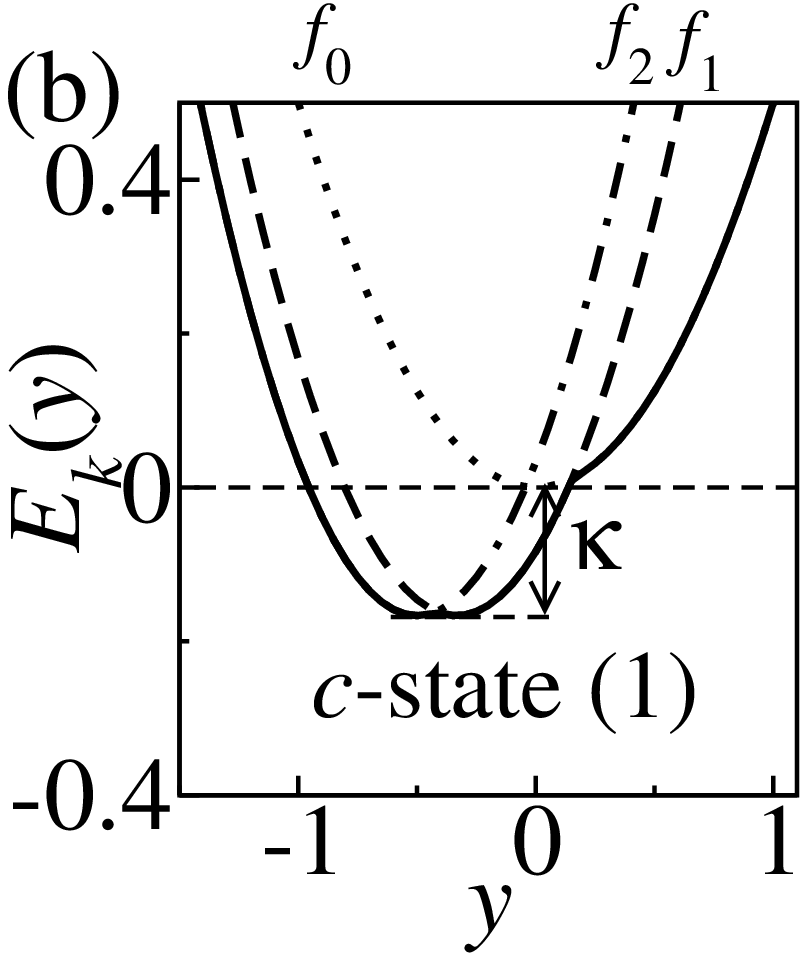, width=0.21\linewidth}
\leftline{\epsfig{figure=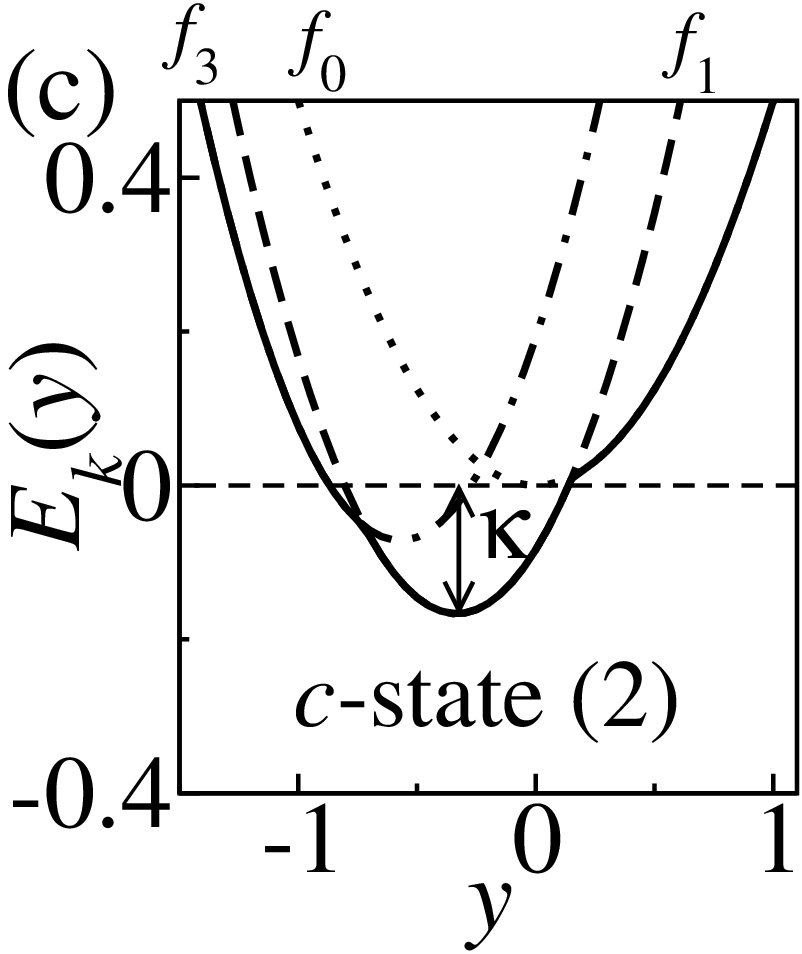, width=0.21\linewidth}
\leftline{\epsfig{figure=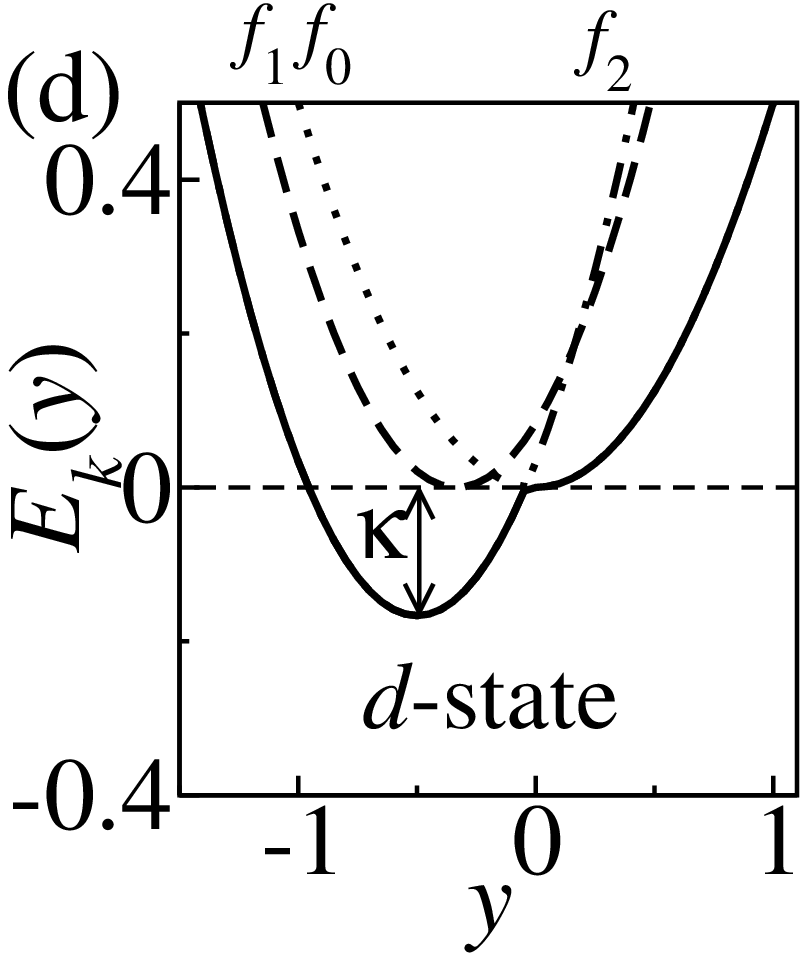, width=0.21\linewidth}
}}}}

\leftline{
\hspace{1cm}
\epsfig{figure=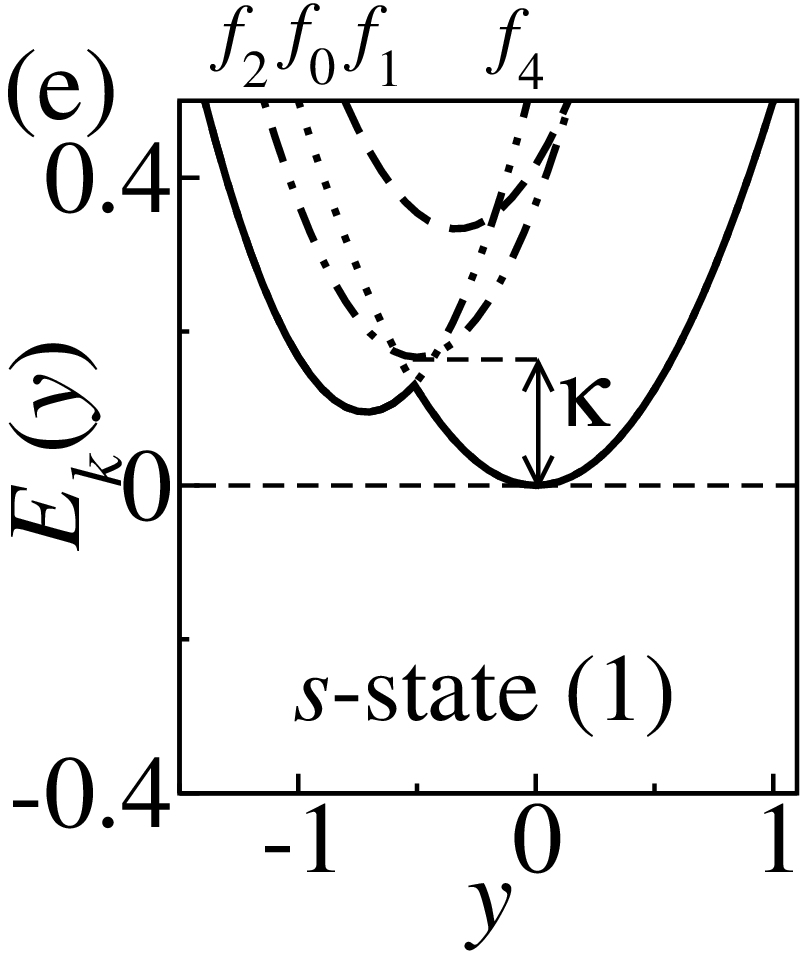, width=0.21\linewidth}
\leftline{\epsfig{figure=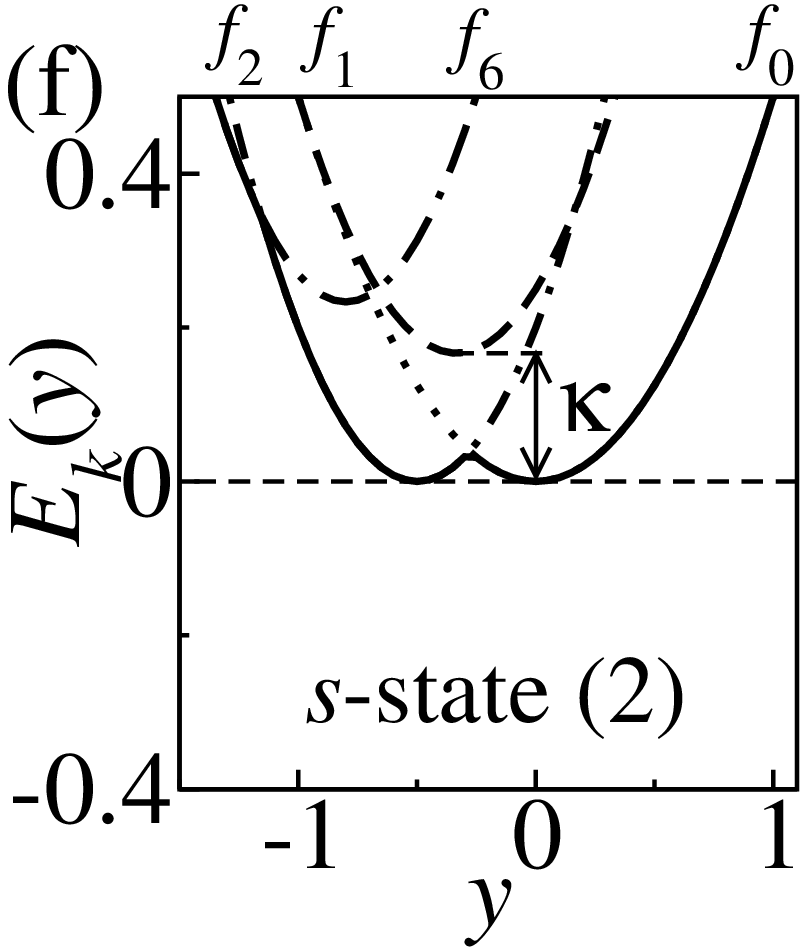, width=0.21\linewidth}
\leftline{\epsfig{figure=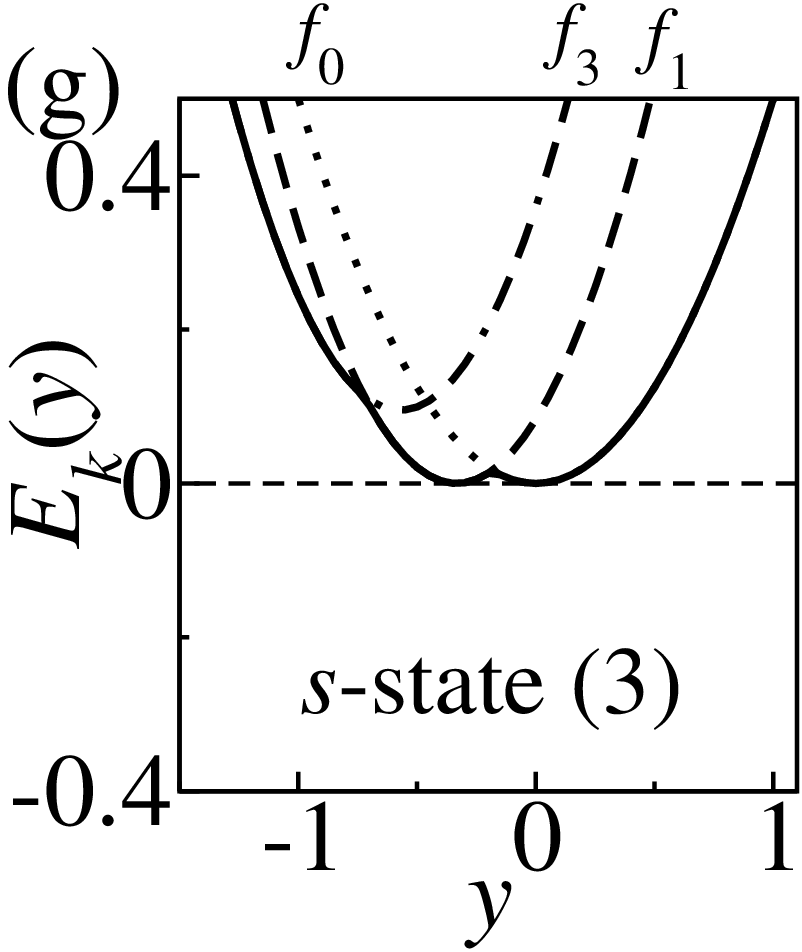, width=0.21\linewidth}
\leftline{\epsfig{figure=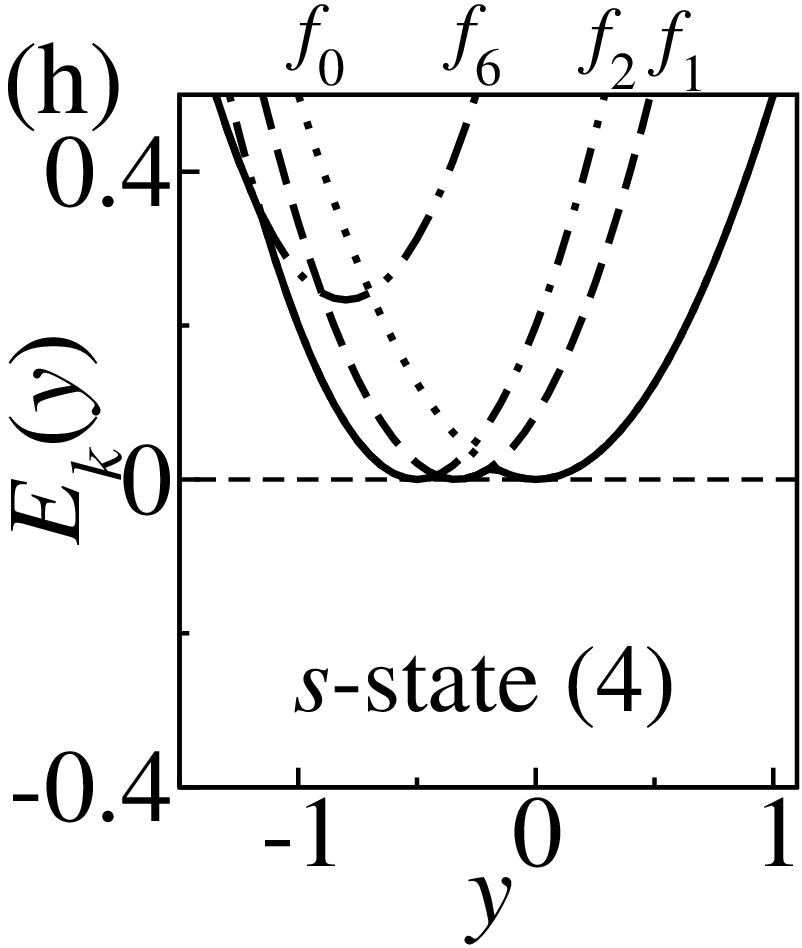, width=0.21\linewidth}
}}}}

\caption{A closed set of cavity energy functions $E_k(y)$ 
at $K=3$, $u^{-1}=1$ in the doublet regime.
The forms of $E^V$ corresponds to 
(a) the $a$ state,
(b-c) the $c$ state,
(d) the $d$ state and
(e-h) the $s$ state.
}
\label{gr_closeSetDou}
\end{figure}
%%%%%%%%%%%%%%%%%%%%%%%%%%%%%%%

To find the closed set of $E^V$ in the doublet regime,
we only have to consider the composite functions $f_0$, 
$f_1$ and $f_2$.
$E^V_k(y_k)$ is thus denoted as 
\begin{eqnarray}
	E^V_k(y_k) \doteq (d^k_0, d^k_1, d^k_2).
\end{eqnarray}
All other composite functions $f_{n_k}$ with $n_k\ge 3$
have $d^k_{n_k}>d^k_{\mnk}$ and their corresponding configurations
are not stable in the doublet regime.
We thus consider only the recursion $\cM(0,\dots,0)=1$
and $\cM(1,0,\dots,0)=2$. 
As an illustration,
we consider the case $m=2$ and $\phi_d=1$ 
where closed set of $E^V$ 
is shown in \fig{gr_closeSetDou}.
The cavity energy change $\Delta E_j$ from \req{eq_deltaED2},
the constant terms $d^j_0$, $d^j_1$ and $d^j_2$ from
Eqs. (\ref{eq_recurD0}) and (\ref{eq_recurD}) are simplified to
\begin{eqnarray}
\label{eq_recurDou}
	\Delta E_j &&= \frac{u^2}{2} + \sum_{\kinjmi}\min(0, d^k_1, d^k_2),
	\nonumber\\
	d^j_0 &&= 0,
	\\
	d^j_1 &&= -2\kappa - \sum_{\kinjmi}\min(0, d^k_1, d^k_2),
	\nonumber\\
	d^j_2 &&= -\kappa + \min_{\kinjmi} d^k_1 
	- \sum_{\kinjmi}\min(0, d^k_1, d^k_2),
	\nonumber
\end{eqnarray}
where the zero-point of $E^V$ is made at the minimum values of $f_0$
for convenience of analysis.
The energy change of $\Delta E_j$ of adding vertex $j$
is obtained by comparing the energies of the $S$ states of the 
vertex and its descendents,
and can be shown to produce the same physical results as 
the full energetic comparison.
From the recursion relations,
the full closed set of $E^V$ is found to be
\begin{eqnarray}
\label{eq_closeSetDou}
	E^V_a(y) &&\doteq (0, -2\kappa, (-1+r)\kappa) 
	\nonumber\\
	E^V_c(y) &&\doteq (0, -\kappa, (-1+r)\kappa)
	\nonumber\\
	E^V_d(y) &&\doteq (0, r\kappa, -1\kappa) 
	\nonumber\\
	E^V_s(y) &&\doteq (0, q\kappa, r\kappa) 
\end{eqnarray}
with integers $q,r\ge 0$.
$E^V_a$ corresponds to cavity states with a strong preference
to be singly consuming.
$E^V_c$, $E^V_d$ and $E^V_s$ correspond to cavity states 
which are respectively singly consuming,
doubly consuming and resource providing.
They are denoted as the $a$,
$c$, $d$ and $s$ states.
Note that the integer $r\ge 1$ in \req{eq_closeSetDou}
may correspond to the form of $E^V$ where 
$f_1$ and $f_2$ are not relevant 
(see for instance \fig{gr_closeSetDou} (f) and (g)).
For $K=3$,
there are two forms of $E^V_c$ with $r =0$ and $r\ge 1$ and
four forms of $E^V_s$ with $(q,r)$ = $(\ge 1, 1)$,
$(\ge 1, 0)$, 
$(0, \ge 1)$ and $(0, 0)$.
Hence the closed set of $E^V$ has eight forms of $E^V$ as 
shown in \fig{gr_closeSetDou}.

Next,
we consider the disordered case $\phi_d<1$.
The recursion relations can be simplified in terms of 
$\psi_a$, $\psi_c$, $\psi_d$ and $\psi_s$,
corresponding to the probabilities for a vertex
to be in the $a$, $c$, $d$ and $s$ states,
as given by 
\begin{eqnarray}
\label{eq_recurPDou}
	&& \psiji_a = \delta_{\Lambda_j, -1}\prod_{\kinjmi}\psikj_s,
	\nonumber\\
	&& \psiji_c = \delta_{\Lambda_j, -1}\sum_{\kinjmi}
	(\psikj_c+\psikj_d)\prod_{\linjmik}\psi^{l\rightarrow j}_s, 
	\nonumber\\
	&& \psiji_d = \delta_{\Lambda_j, -1}\sum_{\kinjmi}\psikj_a
	\prod_{\linjmik}\psi^{l\rightarrow j}_s,
\end{eqnarray}
and  $\psiji_s = 1-\psiji_a-  \psiji_c- \psiji_d$
In the RS phase,
we set $\psiji_a,  \psiji_c, \psiji_d, \psiji_s=0,1$.
The average optimized energy in the RS ansatz is obtained by evaluating
the full energy change $\Delta \cE_{\rm node}$ and $\Delta \cE_{\rm link}$
resulting from the addition of new nodes and new links.
The expression for $\Delta\cE_{\rm node}$ and $\Delta \cE_{\rm link}$ 
from \req{eq_deltaE} can be simplified as
\begin{eqnarray}
\label{eq_DeltaCEDou}
	&&\Delta\cE^i_{\rm node} 
	= \delta_{\Lambda_i,-1} \Big\{\frac{u^2}{2}
%	\nonumber\\
	+ \min\bigg[\sum_{\jini}\min(0,d^j_1, d^j_2), -2\kappa, 
	-\kappa+\min_{\jini}d^j_1\bigg]\Big\}
	\nonumber\\
	&&+\delta_{\Lambda_i, A}\sum_{\jini}\min(0,d^j_1, d^j_2).
	\nonumber\\
	&&\Delta\cE_{\rm link}=\min\Big[0, d^L_1, d^R_1,
	\frac{1}{K(K-1)}+d^L_1+d^R_1, d^L_2, d^R_2\Big].
\end{eqnarray}

\subsection{The Instability of the RS Ansatz}

%%%%%%%%% Figure 18 %%%%%%%%%%%%%%%%%
\begin{figure}
\hspace{1cm}
\centerline{\epsfig{figure=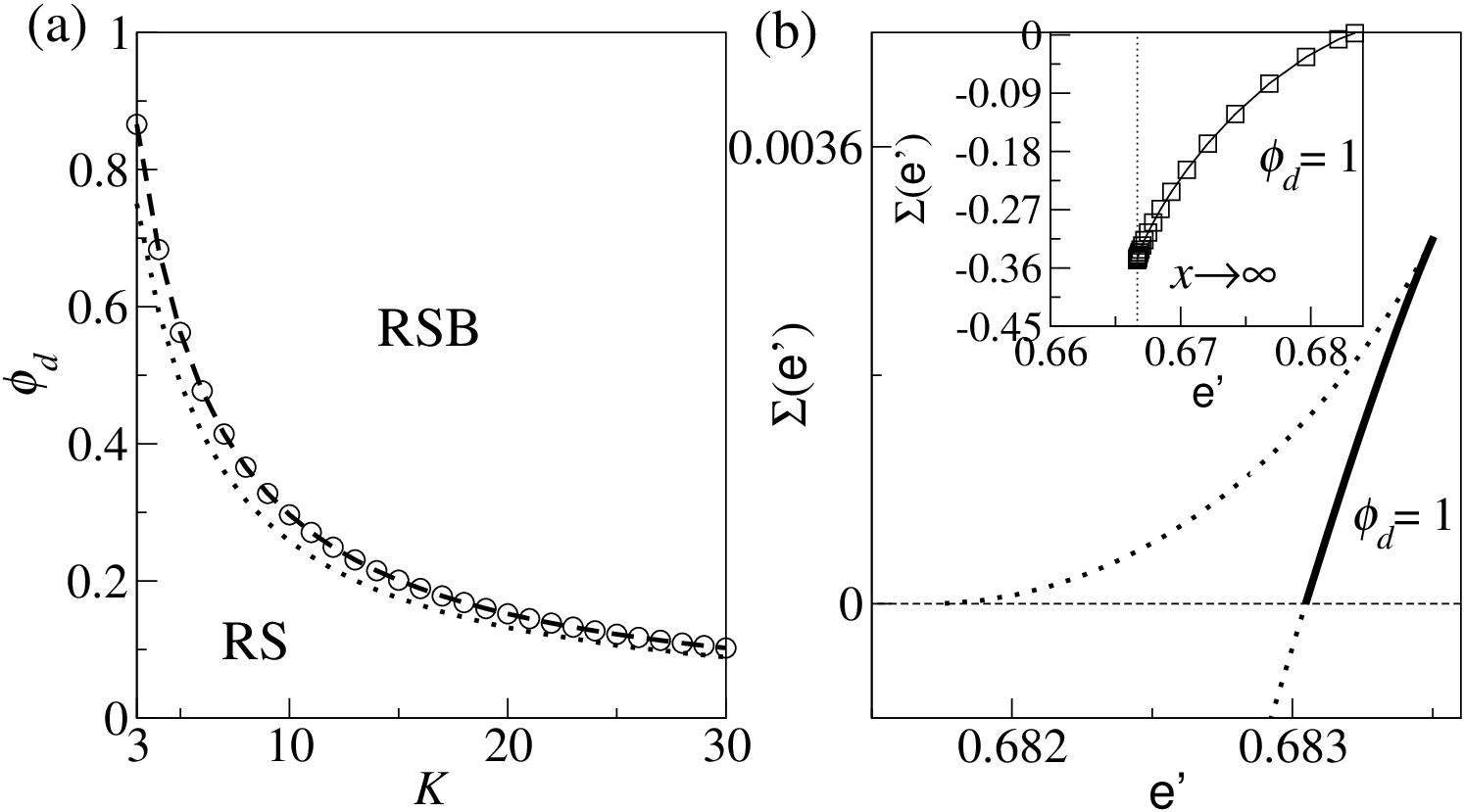, width=0.65\linewidth}}
\caption{
(a) The $K$-$\phi_d$ phase diagram and the AT line (dashed line with circles) 
in the doublet regime with $u^{-1} = \sqrt{K(K-1)/(K+3)}$.
The dotted line is the AT line in the singlet regime.
(b) The complexity function $\Sigma(\sfe)$ from the 1RSB solution 
for $K=3$ with $u^{-1} = 1$. 
$\sfe'=\sfe/(u^2/2)$.
Inset: The negative segment of the complexity function shown 
in the limit $x\rightarrow \infty$.
Each symbol on the line indicates an increase of 1.2 in $x\kappa$.
}
\label{gr_phaseDiagramDou}
\end{figure}
%%%%%%%%%%%%%%%%%%%%%%%%%%%%%%%%%%%%%

To obtain the AT line between the RS and the RSB phases,
we consider the variations $\delta \psikj_a, \delta \psikj_c, 
\delta \psikj_d$ and $\delta \psikj_s$
in the recursions of probabilities in \req{eq_recurPDou},
with $\delta \psikj_a+ \delta \psikj_c+ \delta \psikj_d +\delta \psikj_s=0$.
Though the $c$ and $d$ states lead to different degeneracies 
of the full states, 
the two states play the same role in the recursion relations
and lead to the same $\Delta E$, $\Delta\cE_{\rm node}$ 
and $\Delta\cE_{\rm link}$.
Combining $\psikj_c$ and $\psikj_d$ in the recursions,
the variations $\delta \psikj_a, \delta(\psikj_c +\psikj_d)$ 
and $\delta \psikj_s$ depends only on 
$\delta \psikj_s$ of the descendents.
We thus write down the AT condition for the disordered case 
of the doublet regime,
\begin{eqnarray}
\label{eq_RSinstabDou}
	\frac{\avg{(\delta \psiji_s)^2}}{\avg{(\delta\psikj_s)^2}}
	=(K-1)(K-2)\phi_d(1-\avg{\psikj_s})\avg{\psikj_s}^{K-3},
\end{eqnarray}
where we have applied the relation of $\avg{(\psikj_s)^2}=\avg{\psikj_r}$
in the derivation.
Simple algebra leads to the following form of the AT line
\begin{eqnarray}
\label{eq_ATDou}
	\phi_d^{\rm AT} 
	= \frac{1}{(K-1)(K-2)(1-\avg{\psi_s}^{\rm AT})
	(\avg{\psi_s}^{\rm AT})^{K-3}},
	\nonumber\\
	\avg{\psi_s}^{\rm AT}=\frac{(2K-3)(K-1)-\sqrt{(K-1)(5K-9)}}{2K(K-2)},
\end{eqnarray}
where $\avg{\psi_s}^{\rm AT}$ is the average value of $\psi_s$ on the AT line.
The $K$-$\phi_d$ phase diagram is shown in \fig{gr_phaseDiagramDou}(a),
which is compared to the phase diagram of the singlet regime
in \fig{gr_phaseDiagram} inset.
The comparison suggests that the RSB phase shrinks when $u^{-1}$ decreases
from the singlet regime to the doublet regime.

\subsection{The 1RSB solution}

To obtain the 1RSB solution,
we follow the approach adopted in the singlet regime 
and evaluate
the partition functions $\Xi_{\rm node}$ and $\Xi_{\rm link}$
using the full energy change $\cE_{\rm node}$ and $\cE_{\rm link}$.
We solve the 1RSB solution for the case without disorder (i.e. $\phi_d=1$)
with the 1RSB restriction on delta functional form of $\cQ\{P[E^V]\}$,
i.e. $Q(\psi_a, \psi_c, \psi_d)=\delta(\psi_a-\avg{\psi_a})
\delta(\psi_c-\avg{\psi_c})\delta(\psi_d-\avg{\psi_d})$.
The complexity function $\Sigma$ is obtained from the 1RSB solution for $K=3$,
and is shown in in \fig{gr_phaseDiagramDou}(b).
Compared with the singlet regime,
$\Sigma$ in both cases have similar form and
similar maximum values. 
The physical segment of the curve is shown by the solid segment.
Note that $\sfe=(1-f_\cC-f_\cD)/2 + f_\cC/6 + f_\cD/3$
and the fraction $f_\cC$ and $f_\cD$ of singly and doubly consuming nodes 
are not uniquely determined by the energy $\sfe$.
The negative segment of the complexity function approaches
the limit of $\sfe = 1/3$ as $x\rightarrow\infty$,
as shown in the inset of \fig{gr_phaseDiagramDou}(b).
It corresponds to the lowest possible energy on graphs 
with special structures,
which occur with vanishing probability in the thermodynamic limit.
$\sfe = 1/3$ implies $4 f_\cC + 3 f_\cD = 2$,
suggesting the modulation limit of $f_\cC=1/2$ when $f_\cD=0$
as in the singlet regime,
or the limit of $f_\cD=2/3$ when $f_\cC=0$ where two-third
of nodes are doubly consuming.

%%%%%%%%%%%%%%%%%%%%%%%%%%%%%%%%
\section*{Acknowledgements}

This work is supported
by the Research Grants Council of Hong Kong
(grant numbers HKUST 603607 and HKUST 604008).

%%%%%%%%%%%%%%%%%%%%%%%%%%


\begin{thebibliography}{0}

%\bibitem{doyle1984}
%P. G. Doyle and J. L Snell, 
%{\it Random Walks and Electric Networks} 
%(Mathematical Association of America, Washington D. C., 1984).

%\bibitem{rockafellar1984}
%R. T. Rockafellar, 
%{\it Network Flows and Monotropic Optimization} 
%(Wiley, New York, 1984).

\bibitem{garey1979}
M. R. Garey and D. S. Johnson, 
{\it Computers and Intractability: A Guide to the Theory of NP-Completeness} 
(Freeman, New York, 1979).

\bibitem{mezardBook}
M. M\'ezard, G. Parisi and M. A. Virasoro, 
{\it Spin Glass Theory and Beyond}
(World Scientific, 1987).

\bibitem{nishimoriBook} 
H.~Nishimori, 
{\it Statistical Physics of Spin Glasses and Information Processing}
(Oxford University Press, Oxford, UK, 2001).

\bibitem{mezard2002}
M. M\'ezard and R. Zecchina,
Phys. Rev. E {\bf 66}, 056126 (2002).

\bibitem{mulet2002}
R. Mulet, A. Pagnani, M. Weigt, and R. Zecchina,
Phys. Rev. Lett. {\bf 89}, 268701 (2002); 
A. Braunstein, R. Mulet, A. Pagnani, M. Weigt, and R. Zecchina,
Phys. Rev. E {\bf 68}, 036702 (2003).

\bibitem{weigt2000}
M. Weigt and A.~K. Hartmann,
Phys. Rev. Lett. {\bf 84}, 6118 (2000).

\bibitem{weigt2001}
M. Weigt and A.~K. Hartmann,
Phys. Rev. E {\bf 63}, 056127 (2001).

\bibitem{zhou2005}
H. Zhou,
Phys. Rev. Lett. {\bf 94}, 217203 (2005).

\bibitem{wong2006} 
K.~Y~.M.~Wong and D.~Saad, 
Phys. Rev. E {\bf 74}, 010104 (2006).

\bibitem{wong2007} 
K.~Y~.M.~Wong and D.~Saad, 
Phys. Rev. E {\bf 76}, 011115 (2007).

\bibitem{yeung2009a}
C.~H. Yeung and K.~Y.~M. Wong,
J. Stat. Mech, P03029 (2009).

\bibitem{shenker1996}
C. Kopparapu, 
{\it Load Balancing Servers, Firewalls and Caches}
(Wiley, 2002).

\bibitem{rardin1998}
R. L. Rardin 
{\it Optimization in Operations Research}
(Prentice Hall, Englewood Cliffs, NJ, 1998).

\bibitem{yeung2009b}
C.~H. Yeung and K.~Y.~M. Wong,
Phys. Rev. E {\bf 80}, 021102 (2009).

\bibitem{wong1990}
K. Y. M. Wong and D. Sherrington, 
J. Phys. A {\bf 23}, L175 (1990).

\bibitem{wong1993}
K. Y. M. Wong and D. Sherrington, 
Phys. Rev. E {\bf 47}, 4465 (1993); 
erratum, Phys. Rev. E {\bf 50}, 1727 (1994).

\bibitem{whyte1995}
W. Whyte, D Sherrington and K. Y. M. Wong, 
J. Phys. A {\bf 28}, 7105 (1995).

\bibitem{luo2001}
P. Luo and K. Y. M. Wong, 
Phys. Rev. E {\bf 64}, 061912 (2001).

\bibitem{imry1975}
Y. Imry and S. K. Ma, 
Phys. Rev. Lett. {\bf 50}, 1399 (1975).

\bibitem{bruinsma1983}
R. Bruinsma and G. Aeppli, 
Phys. Rev. Lett. {\bf 50}, 1494 (1983).

\bibitem{bruinsma1984}
R. Bruinsma, Phys. Rev. B {\bf 30}, 289 (1984).

\bibitem{biroli2002}
G. Biroli and M. M\'ezard,
Phys. Rev. Lett. {\bf 88}, 025501 (2002).

\bibitem{rivoire2004}
O. Rivoire, G. Biroli, O.~C. Martin, and M. M\'ezard, 
Eur. Phys. J. B {\bf 37}, 55 (2004).

\bibitem{krzakala2008}
F. Krzakala, M. Tarzia, and L. Zdeborov\'a, 
Phys. Rev. Lett. {\bf 101}, 165702 (2008).

\bibitem{toulouse1977}
G. Toulouse, 
Comm. on Phys. {\bf 2}, 115 (1977).

\bibitem{devaney1989}
R.~L. Devaney, 
{\it An Introduction to Chaotic Dynamical Systems}
(Addison-Wesley, Redwood City, CA, 1989).

\bibitem{weiss2001}
Y. Weiss and W. T. Freeman, Neural Computation {\bf 13}, 2173 (2001).

\bibitem{bickson2008}
D. Bickson, D. Dolev, O. Shental, P. H. Siegel, and J. K. Wolf,
The 2008 International Symposium on Information Theory (ISIT2008),
Toronto, 2008.

\bibitem{pearl1988} 
J. Pearl, 
{\it Probabilistic Reasoning in Intelligent Systems: 
Networks of Plausible Inference} 
(Morgan Kaufmann, San Mateo, CA, 1988).

\bibitem{zdeborova2007}
L. Zdeborov\'a and F. Krzakala,
Phys. Rev. E {\bf 76}, 031131 (2007).

\bibitem{selman1996}
B. Selman, H. Kautz and B. Cohen
{\it DIMACS Series in Discrete Mathematics and 
Theoretical Computer Science} {\bf 26}, 521 (1996).

\bibitem{mezard2001}
M. M\'ezard and G. Parisi,
Eur. Phys. J. B {\bf 20}, 217 (2001).

\bibitem{mezard2003}
M. M\'ezard and G. Parisi,
J. Stat. Phys. {\bf 111}, 112 (2003).

\bibitem{thouless1986}
D. J. Thouless,
Phys. Rev. Lett. {\bf 56}, 1082 (1986).

\bibitem{montanari2003}
A. Montanari and F.Ricci-Tersenghi,
Eur. Phys. J. B {\bf 33}, 339 (2003).

\bibitem{krzakala2004}
F. Krzakala, A. Pagnani and Martin Weigt,
Phys. Rev. E {\bf 70}, 046705 (2004).

\bibitem{montanari2004}
A. Montanari, G. Parisi and F. Ricci-Tersenghi,
J. Phys. A: Math. Gen. {\bf 37}, 2073 (2004).



%\bibitem{banavar2000}
%J. R. Banavar, F. Colaiori, A. Flammini, A. Maritan, and A. Rinaldo, 
%Phys. Rev. lett. {\bf 84}, 4745 (2000); 
%M. Durand, 
%Phys. Rev. Lett. {\bf 98}, 088701 (2007); 
%S. Bohn and M. O. Magnasco, 
%Phys. Rev. Lett. {\bf 98}, 088702 (2007);
%Z. Shao and H. Zhou, 
%Phys. Rev. E {\bf 75}, 066112 (2007).


%\bibitem{wong1993}
%K.~Y.~M. Wong and D.~Sherrington, 
%Phys. Rev. E {\bf 47}, 4465 (1993).


\end{thebibliography}
\end{document}